\documentclass[twocolumn]{aastex631}

\usepackage[T1]{fontenc} 
\usepackage{array}
\usepackage{amsmath}
\usepackage{xspace}
\usepackage{xcolor}
\usepackage{gensymb}

\usepackage{graphicx}
\usepackage{dcolumn}
\usepackage{bm}
\hypersetup{colorlinks=true,citecolor=blue,filecolor=blue,urlcolor=blue,}

\usepackage[noabbrev,capitalise]{cleveref}

\makeatletter
\usepackage{etoolbox}
\patchcmd\H@refstepcounter{\protected@edef}{\protected@xdef}{}{}
\makeatother

\defcitealias{DESI2024.IV.KP6}{\texttt{DESI24-\lya}}
\newcommand{\kplya}{\citetalias{DESI2024.IV.KP6}\xspace}

\newcommand{\lya}{Ly$\alpha$}
\newcommand{\lyaf}{Ly$\alpha$ forest}
\newcommand{\lyb}{Ly$\beta$}

\newcommand{\lyalya}{Ly$\alpha\times$Ly$\alpha$}
\newcommand{\lyaqso}{Ly$\alpha\times$QSO}

\newcommand{\alpac}{Alcock-Paczynski}

\newcommand{\lyacolore}{\texttt{Ly$\alpha$CoLoRe}}
\newcommand{\saclay}{\texttt{Saclay}}
\newcommand{\iminuit}{\texttt{iminuit}}

\newcommand{\apar}{$\alpha_{||}$}
\newcommand{\atrans}{$\alpha_\bot$}

\newcommand{\phip}{$\phi_\mathrm{p}$}
\newcommand{\phis}{$\phi_\mathrm{s}$}
\newcommand{\phif}{$\phi_\mathrm{f}$}
\newcommand{\alphap}{$\alpha_\mathrm{p}$}
\newcommand{\alphas}{$\alpha_\mathrm{s}$}
\newcommand{\alphaf}{$\alpha_\mathrm{f}$}
\newcommand{\fsig}{$f\sigma_8$}

\newcommand{\lyalyalyalya}{Ly$\alpha$(A)$\times$Ly$\alpha$(A)}
\newcommand{\lyalyalyalyb}{Ly$\alpha$(A)$\times$Ly$\alpha$(B)}
\newcommand{\lyalyaqso}{Ly$\alpha$(A)$\times$QSO}
\newcommand{\lyalybqso}{Ly$\alpha$(B)$\times$QSO}

\newcommand{\hMpc}{\ $h^{-1}\text{Mpc}$}
\newcommand{\kmsMpc}{\,{\rm km\,s^{-1}\,Mpc^{-1}}}

\newcommand{\lcdm}{{$\Lambda$CDM}}

\newcommand{\vega}{\texttt{Vega}}

\begin{document}

\title{DESI DR1 Ly$\alpha$ forest: 3D full-shape analysis and cosmological constraints}

\author[0000-0002-2169-0595]{Andrei~Cuceu}
\altaffiliation{NASA Einstein Fellow}
\altaffiliation{\url{acuceu@lbl.gov}}
\affiliation{Lawrence Berkeley National Laboratory, 1 Cyclotron Road, Berkeley, CA 94720, USA}
\affiliation{Center for Cosmology and AstroParticle Physics, The Ohio State University, 191 West Woodruff Avenue, Columbus, OH 43210, USA}

\author[0000-0002-9136-9609]{Hiram~K.~Herrera-Alcantar}
\affiliation{Institut d'Astrophysique de Paris. 98 bis boulevard Arago. 75014 Paris, France}
\affiliation{IRFU, CEA, Universit\'{e} Paris-Saclay, F-91191 Gif-sur-Yvette, France}

\author[0000-0003-2561-5733]{Calum~Gordon}
\affiliation{Institut de F\'{i}sica d’Altes Energies (IFAE), The Barcelona Institute of Science and Technology, Edifici Cn, Campus UAB, 08193, Bellaterra (Barcelona), Spain}

\author{C\'esar~Ram\'irez-P\'erez}
\affiliation{Institut de F\'{i}sica d’Altes Energies (IFAE), The Barcelona Institute of Science and Technology, Edifici Cn, Campus UAB, 08193, Bellaterra (Barcelona), Spain}

\author[0000-0001-7600-5148]{E.~Armengaud}
\affiliation{IRFU, CEA, Universit\'{e} Paris-Saclay, F-91191 Gif-sur-Yvette, France}

\author[0000-0002-3033-7312]{A.~Font-Ribera}
\affiliation{Institut de F\'{i}sica d’Altes Energies (IFAE), The Barcelona Institute of Science and Technology, Edifici Cn, Campus UAB, 08193, Bellaterra (Barcelona), Spain}

\author[0000-0001-9822-6793]{J.~Guy}
\affiliation{Lawrence Berkeley National Laboratory, 1 Cyclotron Road, Berkeley, CA 94720, USA}

\author{B.~Joachimi}
\affiliation{Department of Physics \& Astronomy, University College London, Gower Street, London, WC1E 6BT, UK}

\author[0000-0002-4279-4182]{P.~Martini}
\affiliation{Center for Cosmology and AstroParticle Physics, The Ohio State University, 191 West Woodruff Avenue, Columbus, OH 43210, USA}
\affiliation{Department of Astronomy, The Ohio State University, 4055 McPherson Laboratory, 140 W 18th Avenue, Columbus, OH 43210, USA}
\affiliation{The Ohio State University, Columbus, 43210 OH, USA}

\author[0000-0001-9070-3102]{S.~Nadathur}
\affiliation{Institute of Cosmology and Gravitation, University of Portsmouth, Dennis Sciama Building, Portsmouth, PO1 3FX, UK}

\author[0000-0001-6979-0125]{I.~P\'erez-R\`afols}
\affiliation{Departament de F\'isica, EEBE, Universitat Polit\`ecnica de Catalunya, c/Eduard Maristany 10, 08930 Barcelona, Spain}

\author{J.~Rich}
\affiliation{IRFU, CEA, Universit\'{e} Paris-Saclay, F-91191 Gif-sur-Yvette, France}
\affiliation{Sorbonne Universit\'{e}, CNRS/IN2P3, Laboratoire de Physique Nucl\'{e}aire et de Hautes Energies (LPNHE), FR-75005 Paris, France}

\author{J.~Aguilar}
\affiliation{Lawrence Berkeley National Laboratory, 1 Cyclotron Road, Berkeley, CA 94720, USA}

\author[0000-0001-6098-7247]{S.~Ahlen}
\affiliation{Department of Physics, Boston University, 590 Commonwealth Avenue, Boston, MA 02215 USA}

\author[0000-0003-2923-1585]{A.~Anand}
\affiliation{Lawrence Berkeley National Laboratory, 1 Cyclotron Road, Berkeley, CA 94720, USA}

\author[0000-0003-4162-6619]{S.~Bailey}
\affiliation{Lawrence Berkeley National Laboratory, 1 Cyclotron Road, Berkeley, CA 94720, USA}

\author[0000-0002-9964-1005]{A.~Bault}
\affiliation{Lawrence Berkeley National Laboratory, 1 Cyclotron Road, Berkeley, CA 94720, USA}

\author[0000-0001-9712-0006]{D.~Bianchi}
\affiliation{Dipartimento di Fisica ``Aldo Pontremoli'', Universit\`a degli Studi di Milano, Via Celoria 16, I-20133 Milano, Italy}
\affiliation{INAF-Osservatorio Astronomico di Brera, Via Brera 28, 20122 Milano, Italy}

\author[0000-0002-8934-0954]{A.~Brodzeller}
\affiliation{Lawrence Berkeley National Laboratory, 1 Cyclotron Road, Berkeley, CA 94720, USA}

\author{D.~Brooks}
\affiliation{Department of Physics \& Astronomy, University College London, Gower Street, London, WC1E 6BT, UK}

\author[0000-0002-9553-4261]{J.~Chaves-Montero}
\affiliation{Institut de F\'{i}sica d’Altes Energies (IFAE), The Barcelona Institute of Science and Technology, Edifici Cn, Campus UAB, 08193, Bellaterra (Barcelona), Spain}

\author{T.~Claybaugh}
\affiliation{Lawrence Berkeley National Laboratory, 1 Cyclotron Road, Berkeley, CA 94720, USA}

\author[0000-0002-0553-3805]{K.~S.~Dawson}
\affiliation{Department of Physics and Astronomy, The University of Utah, 115 South 1400 East, Salt Lake City, UT 84112, USA}

\author[0000-0002-1769-1640]{A.~de la Macorra}
\affiliation{Instituto de F\'{\i}sica, Universidad Nacional Aut\'{o}noma de M\'{e}xico,  Circuito de la Investigaci\'{o}n Cient\'{\i}fica, Ciudad Universitaria, Cd. de M\'{e}xico  C.~P.~04510,  M\'{e}xico}

\author[0000-0003-0928-2000]{J.~Della~Costa}
\affiliation{Department of Astronomy, San Diego State University, 5500 Campanile Drive, San Diego, CA 92182, USA}
\affiliation{NSF NOIRLab, 950 N. Cherry Ave., Tucson, AZ 85719, USA}

\author{P.~Doel}
\affiliation{Department of Physics \& Astronomy, University College London, Gower Street, London, WC1E 6BT, UK}

\author[0000-0003-4992-7854]{S.~Ferraro}
\affiliation{Lawrence Berkeley National Laboratory, 1 Cyclotron Road, Berkeley, CA 94720, USA}
\affiliation{University of California, Berkeley, 110 Sproul Hall \#5800 Berkeley, CA 94720, USA}

\author[0000-0002-2890-3725]{J.~E.~Forero-Romero}
\affiliation{Departamento de F\'isica, Universidad de los Andes, Cra. 1 No. 18A-10, Edificio Ip, CP 111711, Bogot\'a, Colombia}
\affiliation{Observatorio Astron\'omico, Universidad de los Andes, Cra. 1 No. 18A-10, Edificio H, CP 111711 Bogot\'a, Colombia}

\author[0000-0001-9632-0815]{E.~Gaztañaga}
\affiliation{Institut d'Estudis Espacials de Catalunya (IEEC), c/ Esteve Terradas 1, Edifici RDIT, Campus PMT-UPC, 08860 Castelldefels, Spain}
\affiliation{Institute of Cosmology and Gravitation, University of Portsmouth, Dennis Sciama Building, Portsmouth, PO1 3FX, UK}
\affiliation{Institute of Space Sciences, ICE-CSIC, Campus UAB, Carrer de Can Magrans s/n, 08913 Bellaterra, Barcelona, Spain}

\author[0000-0003-3142-233X]{S.~Gontcho A Gontcho}
\affiliation{Lawrence Berkeley National Laboratory, 1 Cyclotron Road, Berkeley, CA 94720, USA}
\affiliation{University of Virginia, Department of Astronomy, Charlottesville, VA 22904, USA}

\author[0000-0003-4089-6924]{A.~X.~Gonzalez-Morales}
\affiliation{Departamento de F\'{\i}sica, DCI-Campus Le\'{o}n, Universidad de Guanajuato, Loma del Bosque 103, Le\'{o}n, Guanajuato C.~P.~37150, M\'{e}xico}

\author[0000-0002-0676-3661]{D.~Green}
\affiliation{Lawrence Berkeley National Laboratory, 1 Cyclotron Road, Berkeley, CA 94720, USA}

\author{G.~Gutierrez}
\affiliation{Fermi National Accelerator Laboratory, PO Box 500, Batavia, IL 60510, USA}

\author[0000-0003-1197-0902]{C.~Hahn}
\affiliation{Steward Observatory, University of Arizona, 933 N. Cherry Avenue, Tucson, AZ 85721, USA}

\author[0009-0000-8112-765X]{M.~Herbold}
\affiliation{The Ohio State University, Columbus, 43210 OH, USA}

\author[0000-0002-6550-2023]{K.~Honscheid}
\affiliation{Center for Cosmology and AstroParticle Physics, The Ohio State University, 191 West Woodruff Avenue, Columbus, OH 43210, USA}
\affiliation{Department of Physics, The Ohio State University, 191 West Woodruff Avenue, Columbus, OH 43210, USA}
\affiliation{The Ohio State University, Columbus, 43210 OH, USA}

\author[0000-0002-5445-461X]{V.~Ir\v{s}i\v{c}}
\affiliation{Institute for Fundamental Physics of the Universe, via Beirut 2, 34151 Trieste, Italy}
\affiliation{International School for Advanced Studies, Via Bonomea 265, 34136 Trieste, Italy}
\affiliation{Department of Physics, Astronomy and Mathematics, University of Hertfordshire, College Lane Campus, Hatfield, Hertfordshire, AL10 9AB, UK.}
\affiliation{Kavli Institute for Cosmology, University of Cambridge, Madingley Road, Cambridge CB3 0HA, UK}

\author[0000-0002-6024-466X]{M.~Ishak}
\affiliation{Department of Physics, The University of Texas at Dallas, 800 W. Campbell Rd., Richardson, TX 75080, USA}

\author[0000-0003-0201-5241]{R.~Joyce}
\affiliation{NSF NOIRLab, 950 N. Cherry Ave., Tucson, AZ 85719, USA}

\author[0000-0001-7336-8912]{N.~G.~Kara\c{c}ayl{\i}}
\affiliation{Center for Cosmology and AstroParticle Physics, The Ohio State University, 191 West Woodruff Avenue, Columbus, OH 43210, USA}
\affiliation{Department of Astronomy, The Ohio State University, 4055 McPherson Laboratory, 140 W 18th Avenue, Columbus, OH 43210, USA}
\affiliation{Department of Physics, The Ohio State University, 191 West Woodruff Avenue, Columbus, OH 43210, USA}
\affiliation{The Ohio State University, Columbus, 43210 OH, USA}

\author[0000-0002-8828-5463]{D.~Kirkby}
\affiliation{Department of Physics and Astronomy, University of California, Irvine, 92697, USA}

\author[0000-0003-3510-7134]{T.~Kisner}
\affiliation{Lawrence Berkeley National Laboratory, 1 Cyclotron Road, Berkeley, CA 94720, USA}

\author[0000-0001-6356-7424]{A.~Kremin}
\affiliation{Lawrence Berkeley National Laboratory, 1 Cyclotron Road, Berkeley, CA 94720, USA}

\author{O.~Lahav}
\affiliation{Department of Physics \& Astronomy, University College London, Gower Street, London, WC1E 6BT, UK}

\author{A.~Lambert}
\affiliation{Lawrence Berkeley National Laboratory, 1 Cyclotron Road, Berkeley, CA 94720, USA}

\author[0000-0002-6731-9329]{C.~Lamman}
\affiliation{The Ohio State University, Columbus, 43210 OH, USA}

\author[0000-0003-1838-8528]{M.~Landriau}
\affiliation{Lawrence Berkeley National Laboratory, 1 Cyclotron Road, Berkeley, CA 94720, USA}

\author{J.M.~Le~Goff}
\affiliation{IRFU, CEA, Universit\'{e} Paris-Saclay, F-91191 Gif-sur-Yvette, France}

\author[0000-0001-7178-8868]{L.~Le~Guillou}
\affiliation{Sorbonne Universit\'{e}, CNRS/IN2P3, Laboratoire de Physique Nucl\'{e}aire et de Hautes Energies (LPNHE), FR-75005 Paris, France}

\author[0000-0003-1887-1018]{M.~E.~Levi}
\affiliation{Lawrence Berkeley National Laboratory, 1 Cyclotron Road, Berkeley, CA 94720, USA}

\author[0000-0003-4962-8934]{M.~Manera}
\affiliation{Departament de F\'{i}sica, Serra H\'{u}nter, Universitat Aut\`{o}noma de Barcelona, 08193 Bellaterra (Barcelona), Spain}
\affiliation{Institut de F\'{i}sica d’Altes Energies (IFAE), The Barcelona Institute of Science and Technology, Edifici Cn, Campus UAB, 08193, Bellaterra (Barcelona), Spain}

\author[0000-0002-1125-7384]{A.~Meisner}
\affiliation{NSF NOIRLab, 950 N. Cherry Ave., Tucson, AZ 85719, USA}

\author{R.~Miquel}
\affiliation{Instituci\'{o} Catalana de Recerca i Estudis Avan\c{c}ats, Passeig de Llu\'{\i}s Companys, 23, 08010 Barcelona, Spain}
\affiliation{Institut de F\'{i}sica d’Altes Energies (IFAE), The Barcelona Institute of Science and Technology, Edifici Cn, Campus UAB, 08193, Bellaterra (Barcelona), Spain}

\author[0000-0002-2733-4559]{J.~Moustakas}
\affiliation{Department of Physics and Astronomy, Siena College, 515 Loudon Road, Loudonville, NY 12211, USA}

\author{A.~Muñoz-Gutiérrez}
\affiliation{Instituto de F\'{\i}sica, Universidad Nacional Aut\'{o}noma de M\'{e}xico,  Circuito de la Investigaci\'{o}n Cient\'{\i}fica, Ciudad Universitaria, Cd. de M\'{e}xico  C.~P.~04510,  M\'{e}xico}

\author[0000-0001-8684-2222]{J.~ A.~Newman}
\affiliation{Department of Physics \& Astronomy and Pittsburgh Particle Physics, Astrophysics, and Cosmology Center (PITT PACC), University of Pittsburgh, 3941 O'Hara Street, Pittsburgh, PA 15260, USA}

\author[0000-0002-1544-8946]{G.~Niz}
\affiliation{Departamento de F\'{\i}sica, DCI-Campus Le\'{o}n, Universidad de Guanajuato, Loma del Bosque 103, Le\'{o}n, Guanajuato C.~P.~37150, M\'{e}xico}
\affiliation{Instituto Avanzado de Cosmolog\'{\i}a A.~C., San Marcos 11 - Atenas 202. Magdalena Contreras. Ciudad de M\'{e}xico C.~P.~10720, M\'{e}xico}

\author[0000-0003-3188-784X]{N.~Palanque-Delabrouille}
\affiliation{IRFU, CEA, Universit\'{e} Paris-Saclay, F-91191 Gif-sur-Yvette, France}
\affiliation{Lawrence Berkeley National Laboratory, 1 Cyclotron Road, Berkeley, CA 94720, USA}

\author[0000-0002-0644-5727]{W.~J.~Percival}
\affiliation{Department of Physics and Astronomy, University of Waterloo, 200 University Ave W, Waterloo, ON N2L 3G1, Canada}
\affiliation{Perimeter Institute for Theoretical Physics, 31 Caroline St. North, Waterloo, ON N2L 2Y5, Canada}
\affiliation{Waterloo Centre for Astrophysics, University of Waterloo, 200 University Ave W, Waterloo, ON N2L 3G1, Canada}

\author[0000-0003-0247-8991]{Matthew~M.~Pieri}
\affiliation{Aix Marseille Univ, CNRS, CNES, LAM, Marseille, France}

\author{C.~Poppett}
\affiliation{Lawrence Berkeley National Laboratory, 1 Cyclotron Road, Berkeley, CA 94720, USA}
\affiliation{Space Sciences Laboratory, University of California, Berkeley, 7 Gauss Way, Berkeley, CA  94720, USA}
\affiliation{University of California, Berkeley, 110 Sproul Hall \#5800 Berkeley, CA 94720, USA}

\author[0000-0001-7145-8674]{F.~Prada}
\affiliation{Instituto de Astrof\'{i}sica de Andaluc\'{i}a (CSIC), Glorieta de la Astronom\'{i}a, s/n, E-18008 Granada, Spain}

\author[0000-0002-3500-6635]{C.~Ravoux}
\affiliation{Universit\'{e} Clermont-Auvergne, CNRS, LPCA, 63000 Clermont-Ferrand, France}

\author{G.~Rossi}
\affiliation{Department of Physics and Astronomy, Sejong University, 209 Neungdong-ro, Gwangjin-gu, Seoul 05006, Republic of Korea}

\author[0000-0002-9646-8198]{E.~Sanchez}
\affiliation{CIEMAT, Avenida Complutense 40, E-28040 Madrid, Spain}

\author{D.~Schlegel}
\affiliation{Lawrence Berkeley National Laboratory, 1 Cyclotron Road, Berkeley, CA 94720, USA}

\author{M.~Schubnell}
\affiliation{Department of Physics, University of Michigan, 450 Church Street, Ann Arbor, MI 48109, USA}
\affiliation{University of Michigan, 500 S. State Street, Ann Arbor, MI 48109, USA}

\author[0000-0002-6588-3508]{H.~Seo}
\affiliation{Department of Physics \& Astronomy, Ohio University, 139 University Terrace, Athens, OH 45701, USA}

\author[0000-0002-3461-0320]{J.~Silber}
\affiliation{Lawrence Berkeley National Laboratory, 1 Cyclotron Road, Berkeley, CA 94720, USA}

\author[0000-0002-0639-8043]{F.~Sinigaglia}
\affiliation{Departamento de Astrof\'{\i}sica, Universidad de La Laguna (ULL), E-38206, La Laguna, Tenerife, Spain}
\affiliation{Instituto de Astrof\'{\i}sica de Canarias, C/ V\'{\i}a L\'{a}ctea, s/n, E-38205 La Laguna, Tenerife, Spain}

\author{D.~Sprayberry}
\affiliation{NSF NOIRLab, 950 N. Cherry Ave., Tucson, AZ 85719, USA}

\author[0000-0001-8289-1481]{T.~Tan}
\affiliation{IRFU, CEA, Universit\'{e} Paris-Saclay, F-91191 Gif-sur-Yvette, France}

\author[0000-0003-1704-0781]{G.~Tarl\'{e}}
\affiliation{University of Michigan, 500 S. State Street, Ann Arbor, MI 48109, USA}

\author[0000-0002-1748-3745]{M.~Walther}
\affiliation{Excellence Cluster ORIGINS, Boltzmannstrasse 2, D-85748 Garching, Germany}
\affiliation{University Observatory, Faculty of Physics, Ludwig-Maximilians-Universit\"{a}t, Scheinerstr. 1, 81677 M\"{u}nchen, Germany}

\author{B.~A.~Weaver}
\affiliation{NSF NOIRLab, 950 N. Cherry Ave., Tucson, AZ 85719, USA}

\author[0000-0001-5146-8533]{C.~Yèche}
\affiliation{IRFU, CEA, Universit\'{e} Paris-Saclay, F-91191 Gif-sur-Yvette, France}

\author[0000-0001-5381-4372]{R.~Zhou}
\affiliation{Lawrence Berkeley National Laboratory, 1 Cyclotron Road, Berkeley, CA 94720, USA}

\author[0000-0002-6684-3997]{H.~Zou}
\affiliation{National Astronomical Observatories, Chinese Academy of Sciences, A20 Datun Road, Chaoyang District, Beijing, 100101, P.~R.~China}




\begin{abstract}

We perform an analysis of the full shapes of Lyman-$\alpha$ (\lya) forest correlation functions measured from the first data release (DR1) of the Dark Energy Spectroscopic Instrument (DESI). Our analysis focuses on measuring the \alpac\ (AP) effect and the cosmic growth rate times the amplitude of matter fluctuations in spheres of $8$\hMpc, \fsig. We validate our measurements using two different sets of mocks, a series of data splits, and a large set of analysis variations, which were first performed blinded. Our analysis constrains the ratio $D_M/D_H(z_\mathrm{eff})=4.525\pm0.071$, where $D_H=c/H(z)$ is the Hubble distance, $D_M$ is the transverse comoving distance, and the effective redshift is $z_\mathrm{eff}=2.33$. This is a factor of $2.4$ tighter than the Baryon Acoustic Oscillation (BAO) constraint from the same data. When combining with \lya\ BAO constraints from DESI DR2, we obtain the ratios $D_H(z_\mathrm{eff})/r_d=8.646\pm0.077$ and $D_M(z_\mathrm{eff})/r_d=38.90\pm0.38$, where $r_d$ is the sound horizon at the drag epoch.
We also measure $f\sigma_8(z_\mathrm{eff}) = 0.37\; ^{+0.055}_{-0.065} \,(\mathrm{stat})\, \pm 0.033 \,(\mathrm{sys})$, but we do not use it for cosmological inference due to difficulties in its validation with mocks.
In \lcdm, our measurements are consistent with both cosmic microwave background (CMB) and galaxy clustering constraints. Using a nucleosynthesis prior but no CMB anisotropy information, we measure the Hubble constant to be $H_0 = 68.3\pm 1.6\;\kmsMpc$ within \lcdm. Finally, we show that \lyaf\ AP measurements can help improve constraints on the dark energy equation of state, and are expected to play an important role in upcoming DESI analyses.

\end{abstract}

\keywords{}


\section{Introduction} \label{sec:intro}

The physics behind cosmic acceleration represents one of the key unknowns in our current understanding of the nature of the Universe. The simplest explanation, given by the cosmological constant, $\Lambda$, is central to the currently preferred model of cosmology, \lcdm. However, recent Baryon Acoustic Oscillations (BAO) measurements from the Dark Energy Spectroscopic Instrument (DESI; \citealt{DESI2016b.Instr, DESI2022.KP1.Instr}) in combination with cosmic microwave background (CMB) and type Ia supernovae (SNe) have provided strong hints that dark energy might not be described by the cosmological constant \citep{DESI2024.VI.KP7A,DESI.DR2.BAO.cosmo}. Instead, these datasets prefer more complex explanations involving a dynamic equation of state for dark energy. Furthermore, the \lcdm\ model has also come under increasing pressure due to a number of tensions, with perhaps the most discussed being the $>5\sigma$ discrepancy in the value of the Hubble constant between direct measurements based on the cosmic distance ladder \citep[e.g.][]{Breuval:2024,Riess:2022} and indirect probes such as the CMB that rely on \lcdm\ \citep[e.g.][]{Planck:2020}. One of the primary observational goals that can help us untangle these mysteries is to obtain more precise measurements of the expansion rate of the Universe at different stages of its evolution \citep{Weinberg:2013}.

The Lyman-$\alpha$ (\lya) forest currently provides the tightest constraints on the cosmic expansion rate in the redshift interval $2 < z < 4$ \citep{DESI.DR2.BAO.lya,Cuceu:2023b}. The forest traces the \lya\ absorption due to neutral hydrogen in the intergalactic medium (IGM) between us and the background sources, most commonly quasars \citep[e.g.][]{McQuinn:2016}. This gives us a unique view into the large-scale structure (LSS) of the Universe, as it allows continuous mapping of the density field along each line-of-sight. For more than a decade, \lyaf\ observations from the Baryon Oscillation Spectroscopic Survey (BOSS), extended BOSS (eBOSS), and now DESI, have provided increasingly precise constraints on the cosmic expansion rate through measurements of the BAO scale \citep[e.g.][]{Busca:2013,Slosar:2013,duMasdesBourboux:2020,DESI2024.IV.KP6,DESI.DR2.BAO.lya}. These measurements are performed using the three-dimensional (3D) auto-correlation of \lya\ flux, as well as the cross-correlation between \lya\ flux and quasar positions. While the BAO scale provides a very powerful standard ruler for cosmology, the \lyaf\ contains plenty of useful cosmological information beyond BAO.

In this paper, we present an analysis that aims to extract cosmological information using the full shape of the \lyaf\ 3D correlation functions measured from the DESI Data Release 1 (DR1). We perform a template-based analysis in which we focus on two well-understood sources of cosmological information:
the \alpac\ (AP) effect and redshift space distortions (RSD). This is in contrast to direct cosmological analyses (also referred to as \textit{full-modelling}) where cosmological parameters are directly measured within a given model (e.g. \lcdm). In the case of the \lyaf, \cite{Gerardi:2023} used DESI forecasts to show that most of the large-scale 3D information is captured by BAO, AP, and RSD. Therefore, in this work we focus on measuring AP and RSD using the full shapes of the \lyaf\ correlation functions.

The \alpac\ effect introduces an anisotropy in the distribution of LSS when the fiducial cosmology used to transform observed angular and redshift separations into distances is different from the truth \citep{Alcock:1979}. In the context of \lya, AP analyses were first proposed by \cite{Hui:1999,McDonald:1999}. \cite{Cuceu:2021} and \cite{Cuceu:2023b} showed that adding AP information from the full shape of \lya\ correlations leads to significantly tighter cosmic expansion rate constraints from BAO. Therefore, the first goal of this work is to extract the full-shape AP information from DESI DR1 correlations measured by \citet[hereafter \kplya]{DESI2024.IV.KP6}, in order to improve upon their cosmic expansion rate constraints, which are based solely on BAO.

Redshift space distortions are caused by peculiar velocities modifying the observed redshift of quasars, and also the redshift, shape, and amount of \lya\ absorption. When interpreting the redshift as purely cosmological in origin, peculiar velocities lead to distortions in the observed distribution of large-scale structure \citep{Kaiser:1987,Hamilton:1998}. The RSD signal is sensitive to the growth rate of cosmic structure $f$, as well as the amplitude of matter fluctuations, conventionally parameterized by $\sigma_8$, the fluctuation amplitude in spheres of radius $8$ \hMpc. In linear theory, these two parameters are degenerate, and therefore we can only measure their product, \fsig\ \citep{Percival:2009}. As the \lyaf\ transmitted flux is given by a non-linear mapping of the underlying density field, its RSD signal also depends on an unknown velocity divergence bias that multiplies \fsig\ \citep[e.g.][]{McDonald:2003,Seljak:2012,Chen:2021,Ivanov:2024}. This has so far prevented 3D \lyaf\ analyses from extracting useful cosmological information from RSD. On the other hand, galaxy clustering analyses, which do not have to deal with this unknown velocity divergence bias, have successfully used RSD to measure \fsig\ over the last few decades \citep[e.g.][]{Blake:2011,Reid:2012,Alam:2017,Alam:2021,DESI2024.V.KP5}. In this work, we use the quasar RSD signal from large scales in the \lya-QSO cross-correlation to measure \fsig. To obtain this constraint, we perform a joint analysis of the cross-correlation with the \lya\ auto-correlation in order to break the degeneracy between the \lya\ and QSO RSD signals in the cross-correlation \citep{Cuceu:2021}. Therefore, our RSD measurement still depends on constraining the linear \lya\ RSD signal, which has not been as thoroughly studied \citep[see][for the current state-of-the-art]{Chabanier:2024,Belsunce:2025}. This is in contrast to the AP effect, which is geometric in nature and has been used as a cosmological tool for decades.

Our analysis uses the \lyaf\ correlation functions measured from DESI DR1 by \kplya. The DESI survey is in the process of measuring the redshifts of more than 40 million galaxies and quasars over five years \citep{DESI2023a.KP1.SV,Corrector.Miller.2023,FiberSystem.Poppett.2024,Guy:2023,2023Schlafly:SurveyOps}. The first year of DESI data, which comprises the bulk of DESI DR1, contains more galaxy spectra than any previous spectroscopic survey \citep{DESI2016b.Instr,DESI2022.KP1.Instr,DESI2024.I.DR1}. The \lya\ analysis uses more than $420\,000$ \lya\ forests and $700\,000$ quasars, which is roughly double what was used in the previous state-of-the-art analyses from eBOSS. The BAO measurements from this dataset were presented in \kplya, and they have been recently updated with the DESI Data Release 2 (DR2) dataset in \cite{DESI.DR2.BAO.lya}.

We start by introducing the data we use along with our model in \cref{sec:analysis}. The \alpac\ and \fsig\ measurements are presented in \cref{sec:comp_results}, and are followed by our validation tests in \cref{sec:validation}, which encompass analyses on mocks, data splits, and a large set of analysis variations. After that, we discuss our results and their validation in \cref{sec:discussion}. Finally, we present the cosmological implications of our measurements in \cref{sec:cosmo}, and conclude in \cref{sec:summary}.

\section{Analysis} \label{sec:analysis}

Our analysis focuses on modelling \lyaf\ correlation functions in order to extract cosmological information from their full shape. We use the correlation functions measured by \kplya, which are described in \Cref{subsec:corr}. The model we use is based on an extension developed by \cite{Cuceu:2021} of the model used for \lya\ BAO analyses, and is described in \Cref{subsec:model,subsec:scale_pars}. Finally, we describe our blinding and unblinding process in \Cref{subsec:blinding}.

\subsection{\lyaf\ correlation functions} \label{subsec:corr}

We use the four 3D \lyaf\ correlation functions measured by \kplya. The four correlations are computed from three datasets:
\begin{enumerate}
    \item The \lya\ flux overdensity field measured between the \lya\ and \lyb\ emission lines, in the rest-frame interval $1040< \lambda_{RF} < 1205$ \AA, denoted region A. This dataset encompasses 428,403 \lya\ forests, and we refer to it as Ly$\alpha$(A).
    \item The \lya\ flux overdensity field measured between the \lyb\ emission line and the Lyman limit, in the rest-frame interval $920 < \lambda_{RF} < 1020$ \AA, denoted region B. This dataset encompasses 137,432 \lya\ forests, and we refer to it as Ly$\alpha$(B).
    \item A catalog of 709,565 DESI quasars at redshift $z > 1.77$, which are used as discrete tracers.
\end{enumerate}
For a detailed description of these datasets, see \kplya.

The four correlation functions include two auto-correlations of \lya\ flux overdensity, \lyalyalyalya\ and \lyalyalyalyb, and two \lya-QSO cross-correlations, \lyalyaqso\ and \lyalybqso. We will refer to the first pair as the auto-correlation, and the second pair as the cross-correlation. These correlation functions are computed on two-dimensional grids in comoving separation along the line-of-sight, $r_{||}$, and across the line-of-sight, $r_\bot$. For a pair of pixels $(i,j)$ at redshifts $(z_i,z_j)$ and separated by an angle $\Delta\theta$, these comoving coordinates are given by:
\begin{align}
    r_{||} &= [D_\mathrm{c}(z_i) - D_\mathrm{c}(z_j)] \cos{\frac{\Delta \theta}{2}}, \\
    r_\bot &= [D_\mathrm{M}(z_i) + D_\mathrm{M}(z_j)] \sin{\frac{\Delta \theta}{2}},
\end{align}
where $D_c$ is the comoving distance, and $D_M$ is the comoving angular diameter distance. To compute these distances, a fiducial cosmology based on CMB measurements from the Planck satellite \cite{Planck:2020} is used. The parameter values of this fiducial cosmology can be found in Table 2 of \kplya. 

The correlation functions are computed in $4$\hMpc\ wide bins in both $r_{||}$ and $r_\bot$, and are averaged into a single redshift bin with an effective redshift $z_\mathrm{eff}=2.33$. A weighted pair-counting algorithm is used, with the correlation in a comoving coordinate bin $M$ given by:
\begin{equation}
    \xi_M = \frac{\sum_{i,j \in M} w_i w_j \delta_i \delta_j}{\sum_{i,j \in M} w_i w_j},
\end{equation}
where the sums run over pixel-pixel pairs for the two auto-correlations, and over pixel-quasar pairs for the two cross-correlations. In the case of \lyaf\ pixels, $\delta$ corresponds to the measured \lya\ flux overdensity, while for quasars $\delta=1$. The weights for \lya\ pixels are given by the inverse variance, accounting for both noise and intrinsic large-scale structure variance, and also include the redshift evolution of the \lya\ bias. On the other hand, the weights for quasars only account for the redshift evolution of the quasar bias \cite{duMasdesBourboux:2020}. For a detailed description of the continuum fitting process used to measure the \lya\ $\delta$ field, and the weights used in 3D \lyaf\ analyses, see \cite{Ramirez-Perez:2024}.

Finally, \kplya\ compute one covariance matrix for all four correlation functions, which includes their cross-covariance. This is estimated from 1028 measurements of the correlation function in individual \texttt{HEALpix} pixels, corresponding to patches $250\times 250$ $(h^{-1}\rm{Mpc})^2$ in size at $z_\mathrm{eff}=2.33$. This noisy estimate is then smoothed by replacing non-diagonal elements of the correlation matrix, $\text{Corr}_{MN}$, corresponding to the same comoving separation between bins $M$ and $N$ with their average. This method was tested on DESI DR1 mocks by \cite{KP6s6-Cuceu}, and was shown to produce accurate uncertainties on BAO parameters measured from the same mocks. We have performed the same tests in the context of the full-shape analysis below in \Cref{subsec:mocks}.

\subsection{Modelling of correlations} \label{subsec:model}
Our model for \lyaf\ correlation functions closely follows that used in \cite{Cuceu:2023a,Cuceu:2023b}, and is implemented in the \vega\footnote{\url{https://github.com/andreicuceu/vega}} package. It is based on a template approach and at a high level follows these steps:
\begin{enumerate}
    \item A linear isotropic matter power spectrum, $P_\mathrm{fid}(k)$, is computed using CAMB \citep{Lewis:1999} at the effective redshift of our measurements ($z_\mathrm{eff}=2.33$), with a fiducial cosmology based on Planck CMB results \cite{Planck:2020}.\footnote{The template used here is exactly the same as that used for the BAO analysis of the same data set in \kplya.} This is our template.
    \item The isotropic power spectrum is split into a wiggles (or peak) component and no-wiggles (or smooth) component following \cite{Kirkby:2013}. These two components go through the rest of the modelling process independently and are only combined in the final step.
    \item Model \lya\ power spectra, $P_F$, and \lya-QSO cross-spectra, $P_\times$, are computed using the template, the linear Kaiser term \cite{Kaiser:1987}, along with models for small-scale non-linearities, BAO broadening, high column density (HCD) absorbers, and redshift errors. 
    \item The now anisotropic model auto- and cross-power spectra are transformed into 2D model correlations, as a function of $r_{||}$ and $r_\bot$, following \cite{Kirkby:2013}.
    \item The coordinates of the model correlations, $r_{||}$ and $r_\bot$, are re-scaled using free scale parameters that we fit for. These are independent for the smooth and peak components.
    \item The impact of metal absorption is incorporated by computing theoretical models for all \lya-Metal, QSO-Metal, and Metal-Metal correlations. A number of smaller effects discussed below are added at this stage, and also the distortion due to continuum fitting errors is modeled by multiplying the model correlation with a distortion matrix.
    \item Finally, the two model components (peak and smooth) are summed to produce the final model of the correlation functions.
\end{enumerate}

The model anisotropic auto- and cross-spectra are given by:
\begin{align}
    P_F(\mathbf{k}) &= b'^2_F (1 + \beta'_F \mu_k^2)^2\; G(\mathbf{k}) F_{NL}(\mathbf{k}) P_\mathrm{fid}(k), \label{eq:pk1} \\
    P_{F\times Q}(\mathbf{k}) &= b'_F (1 + \beta'_F \mu_k^2) (b_Q + f \mu_k^2)\; G(\mathbf{k}) X_{NL}(\mathbf{k}) P_\mathrm{fid}(k),
    \label{eq:pk2}
\end{align}
where the wavenumber vector $\mathbf{k} = (k,\mu_k) = (k_{||},k_\bot)$ has modulus $k$, line-of-sight and transverse components $(k_{||},k_\bot)$, and $\mu_k=k_{||}/k$. The $G(k,\mu_k)$ term models the binning of the correlation function ($4$\hMpc\ bins), and is given by the Fourier transform of a top-hat function. $F_{NL}$ and $X_{NL}$ model small-scale deviations from linear-theory, and are described in detail below. $b'_F$ and $\beta'_F$ are effective \lya\ linear bias and RSD parameters, which account for both \lya\ flux and HCD absorption, and are described in detail below. $b_Q$ and $f$ are the linear quasar bias and the cosmic growth rate (i.e., the logarithmic derivative of the linear growth factor), respectively. Note that the \lya\ RSD parameter, $\beta_F$, also depends on the growth rate through $\beta_F = b_{\eta,F} f / b_F$. However, due to the presence of the unknown velocity divergence bias $b_{\eta,F}$, we cannot use the linear \lya\ auto-correlation to measure $f$, as the two parameters are degenerate. Therefore, we work with the $\beta_F$ parameter and marginalize over it in our analysis. 

High column density (HCD) absorbers are treated in two different ways depending on their strength. Strong HCDs, represented by Damped \lya\ Absorbers (DLAs), with column densities $\log N_\text{HI}>20.3$, that are found in high signal-to-noise (SNR $> 3$) spectra are masked \citep{Ramirez-Perez:2024}. This is because strong DLAs have damping wings that extend over a significant part of the spectra they are found in, increasing the noise on large scales in the measured \lya\ correlations \citep{Font-Ribera:2012b}. The thresholds above were chosen based on the performance of DLA finder algorithms. On the other hand, HCDs that are either too weak to be detected or in spectra with signal-to-noise below 3 are not masked, and their impact must be modelled at the level of the correlation function. We follow \cite{Font-Ribera:2012b,duMasdesBourboux:2020,DESI2024.IV.KP6}, and model this effect through the effective bias and RSD parameters given by:
\begin{align}
    b'_F &= b_F + b_\mathrm{HCD} F_{HCD}(k_{||}), \\
    b'_F \beta'_F &= b_F \beta_F + b_\mathrm{HCD} \beta_\mathrm{HCD} F_{HCD}(k_{||}),
\end{align}
where $b_F$ and $b_\mathrm{HCD}$ are the linear \lya\ and HCD flux biases, $\beta_F$ and $\beta_\mathrm{HCD}$ are the RSD parameters for \lya\ and HCDs. The $F_\mathrm{HCD}$ function accounts for the profile of the HCDs (i.e., the fact that the absorption is not localized along the line-of-sight). We follow \cite{deSainteAgathe:2019} and use the approximate form $F_\mathrm{HCD}=\exp(-L_\mathrm{HCD} k_{||})$, in line with \kplya. $L_\mathrm{HCD}$ is a free parameter that is allowed to vary in our fits, and can be thought of as the typical scale of undetected HCDs.

For the auto-correlation, we follow previous \lya\ BAO and full-shape analyses and model small-scale deviations from linear theory using the empirical model developed by \cite{Arinyo:2015}. This is given by:
\begin{equation}
    F_{NL}(\mathbf{k}) = \exp \left\{ q_1 \Delta^2(k) \left[1 - \left(\frac{k}{k_v}\right)^{a_v} \mu_k^{b_v}\right] - \left(\frac{k}{k_p}\right)^2 \right\},
    \label{eq:arinyo}
\end{equation}
where $\Delta^2(k)\equiv k^3P_\mathrm{fid}(k)/(2\pi^2)$, and we treat $\{q_1, k_v, a_v, b_v, k_p\}$ as unknown nuisance parameters that we marginalize over.\footnote{Note that the original model developed by \cite{Arinyo:2015}, also contained a higher order $\Delta^4(k)$ term with a corresponding $q_2$ parameter. However, this was found to be very close to zero and has therefore been ignored in subsequent analyses.} This model, which we will refer to as the Arinyo model, represents an empirical fitting formula that was shown to accurately reproduce small-scale \lya\ power spectra in hydrodynamical simulations \citep{Arinyo:2015,Givans2022,Chabanier:2024,ChavesMontero:2025}. In previous \lyaf\ analyses, the parameters of this model have been fixed to the values predicted from simulations \citep{DESI2024.IV.KP6,Cuceu:2023b}, while in this work we allow them to vary. We use wide flat priors for all 5 parameters, with the edges of the priors informed by the range of possible values \cite{Arinyo:2015} measured in simulations (see Appendix \ref{sec:nuisance}).

For the cross-correlation, the largest impact on small scales is due to quasar redshift errors \cite{FontRibera2013}. We follow \kplya\ and model this effect along with quasar non-linear velocities using a simple Lorentzian damping, given by: 
\begin{equation}
    X_{NL}^2 = [1 + (k_{||}\sigma_z)^2]^{-1},
\end{equation}
where $\sigma_z$ is a free parameter. As part of our validation in \cref{subsec:variations}, we will also compare our baseline results to those obtained using a Gaussian damping model instead of the Lorentzian.

In order to account for the broadening of the BAO peak due to non-linear evolution, the model power spectrum for the peak component is also multiplied by $\exp [-(k_{||}^2\Sigma_{||}^2 + k_\bot^2\Sigma_\bot^2)/2]$. The two broadening parameters, are fixed to the theoretical prediction of $\Sigma_\bot=3.24$ \hMpc\ and $\Sigma_{||}=6.37$ \hMpc, following \cite{Eisenstein:2007,duMasdesBourboux:2020}. \citet{DESI.DR2.BAO.lya} has recently tested this choice in the context of the DESI DR2 BAO analysis, and found that these values are consistent with the data, and that allowing these parameters to vary does not impact BAO constraints.

The anisotropic power spectra from \cref{eq:pk1,eq:pk2} are transformed into 2D model correlation functions through the following process. First, a multipole decomposition is performed up to $\ell=6$, followed by a Hankel transform, and then the resulting correlation multipoles are assembled into the 2D correlation model as a function of $r_{||}$ and $r_\bot$.

Besides absorption due to hydrogen, the \lya\ forest regions we use also contain metal absorption. For metals with rest-frame wavelength close to the rest-frame of \lya\ ($1215.67$\AA), the small-scale \lya-Metal (for the auto) and QSO-Metal (for the cross) cross-correlations will appear as extra peaks along the line-of-sight in our correlation functions. This is because we treat all pixels as \lya\ pixels when computing their redshift, and therefore wrongly assign null separation \lya-Metal and QSO-Metal pairs to correlation function bins at large separations \citep{Pieri:2014,Morrison:2024}. The lines found to cause the peaks observed along the line-of-sight in \lya\ correlations are: SiIII(1207), SiII(1193), SiII(1190), and SiII(1260) \citep{Bautista:2017}. Other metal absorption at longer wavelengths only contribute through their auto-correlation. \cite{KP6s5-Guy} used measurements in regions redward of the \lya\ line to show that this type of contamination is dominated by CIV absorption. Therefore, we follow \kplya, and model the impact of these 5 metal lines.

The model for metal absorption involves computing all \lya-Metal and Metal-Metal correlations for the \lya\ auto, and all QSO-Metal correlations for the \lya-QSO cross. The model for each of these correlations is very similar to those of the \lya\ correlations, with each metal having its own linear bias and RSD parameters. The metal biases are free to vary, while the metal RSD parameters are fixed to 0.5, following \kplya.\footnote{This is because we lack the sensitivity to constrain both the bias and RSD parameter for each metal. Given that metal peaks are localized along the line-of-sight, we do not expect this to play a role in our results.} Following \cite{Bautista:2017} and \cite{duMasdesBourboux:2017}, the model metal correlations are transformed from their correct separations $(\Tilde{r}_{||},\Tilde{r}_\bot)$ to the coordinate grid of the measured correlations:
\begin{equation}
    \xi_m^M = \sum M_{MN} \xi_m(\Tilde{r}_{||}(N),\Tilde{r}_\bot(N)),
\end{equation}
where the metal matrices $M_{MN}$ are given by:
\begin{equation}
    M_{MN} = \frac{1}{W_M} \sum_{(i,j)\in M, (i,j)\in N} w_i w_j,
    \label{eq:metal_mat}
\end{equation}
where $(i,j)\in N$ refers to bins computed using the correct redshifts, $(i,j)\in M$ refers to bins computed using the assumed redshifts, and $m$ refers to each of the 5 metal lines described above.

For the auto-correlation, the following components are added besides the main \lyalya\ auto-correlation ($\xi_{F}$):
\begin{equation}
    \xi^t_{F} = \xi_{F} + \sum_m \xi_{F \times m} + \sum_{m_1,m_2} \xi_{m_1 \times m_2} + a_\mathrm{noise}\xi_\mathrm{noise},
\end{equation}
where $F \times m$ refers to \lya-Metal correlations, $m \times m$ refers to Metal-Metal correlations, and the sums are performed over the 5 metal lines above. The $a_\mathrm{noise}\xi_\mathrm{noise}$ term models the correlated sky noise of spectra from fibers of the same spectrograph. \cite{KP6s5-Guy} showed that the dominant contribution is the sky background model noise and that it only affects bins with purely transverse separation (i.e. $r_{||}=0$\hMpc). Following \kplya, $\xi_\mathrm{noise}$ has a fixed shape as a function of $r_\bot$, computed based on the fraction of pairs at each $r_\bot$ that were observed with the same spectrograph, and $a_\mathrm{noise}$ is a free parameter we marginalize over.

For the cross-correlation, the contaminated model is given by:
\begin{equation}
    \xi^t_{F\times Q} = \xi_{F\times Q} + \sum_m \xi_{Q \times m} + \xi_\mathrm{TP},
\end{equation}
where $\xi_{F\times Q}$ is the main \lyaqso\ model, $Q \times m$ refers to QSO-Metal correlations, and $\xi_{TP}$ accounts for the transverse proximity effect. Following \cite{FontRibera2013}, we use the isotropic form:
\begin{equation}
    \xi_\mathrm{TP} = \xi_0^\mathrm{TP} \left(\frac{1 h^{-1} \mathrm{Mpc}}{r} \right)^2 \exp \left(-r / \lambda_\mathrm{UV} \right),
\end{equation}
where the amplitude $\xi_0^\mathrm{TP}$ is a free parameter, and we fix $\lambda_\mathrm{UV}=300$ \hMpc\ \citep{Rudie:2013}. 

Once the contaminated models $\xi^t_{F}$ and $\xi^t_{Q}$ are built, the only missing ingredient is the distortion due to continuum fitting errors. This arises because we fit the amplitude and slope of the continuum for each individual forest in order to account for Quasar diversity \citep[see][]{Ramirez-Perez:2024}. However, this process also fits for any large-scale structure fluctuations of the size of the forest or larger that are present in each forest. These continuum fitting errors lead to distortions in the measured correlation functions. Following \cite{Bautista:2017} and \cite{duMasdesBourboux:2017}, this effect is forward modeled through distortion matrices, which for the auto- and cross-correlation are given by:
\begin{align}
    D_{MN}^\mathrm{auto} &= \frac{1}{W_M} \sum_{i,j \in M} w_i w_j \sum_{i',j' \in N} \eta_{ii'} \eta_{jj'}, \label{eq:dmat1} \\
    D_{MN}^\mathrm{cross} &= \frac{1}{W_M} \sum_{i,j \in M} w_i w_j \sum_{i',j \in N} \eta_{ii'}, \label{eq:dmat2}
\end{align}
with projection matrices, $\eta_{ij}$, given by: 
\begin{equation}
    \eta_{ij} = \delta_{ij}^K - \frac{w_j}{\sum_k w_k} - \frac{w_j\kappa_i\kappa_j}{\sum_k w_k \kappa_k^2},
\end{equation}
where $\delta_{ij}^K$ is the Kronecker delta, and $\kappa_k=\log \lambda_k - \overline{\log \lambda_q}$, with $\lambda_i$ the observed wavelength in pixel $i$, and $\overline{\log \lambda_q}$ the mean of $\log \lambda$ in the forest of quasar $q$. The distortion matrices in \cref{eq:dmat1,eq:dmat2} are computed using model bins $N$ that are $2$ \hMpc\ in size, while the data bins $M$ correspond to the binning size of the measured correlation functions ($4$ \hMpc\ in size). This means the models computed so far use the smaller $2$ \hMpc\ binning for better precision, while the distorted model, $\hat{\xi}_M$, matches the data binning. The distorted model is given by:
\begin{equation}
    \hat{\xi}_M = \sum_N D_{MN} \xi_N^t.
\end{equation}
The computation of the distortion matrices is very intensive, and therefore only a small fraction of pixel pairs are used to measure them. In the baseline analysis, this consists of $1\%$ of pairs, but we also test a variation where that number is doubled in \cref{subsec:variations}.

Finally, the peak and smooth components, $\hat{\xi}_p$ and $\hat{\xi}_s$, which have so far undergone the modelling process separately, are added to produce the full correlation function model:
\begin{equation}
    \xi(r_{||},r_\bot) = \xi_s(q^s_{||} r_{||},q^s_\bot r_\bot) + \xi_s(q^p_{||} r_{||},q^p_\bot r_\bot),
\end{equation}
where $(q^s_{||}, q^s_\bot)$ rescale the coordinates of the smooth component, while $(q^p_{||}, q^p_\bot)$ rescale the coordinates of the peak component. In a BAO analysis, $q^s_{||}=q^s_\bot=1$ and the goal is to fit for the BAO scale parameters $(q^p_{||}, q^p_\bot)$. On the other hand, in this work we wish to extract information from the full shape of the correlation functions, and therefore we also fit for $(q^s_{||}, q^s_\bot)$. In the case of the cross-correlation, the line-of-sight coordinates $r_{||}$, are further modified by an additive free parameter, $\Delta r_{||}$, which accounts for any systematic quasar errors.

\subsection{Fitting \alpac\ and the growth rate}
\label{subsec:scale_pars}

In this section, we introduce the parametrization used to extract cosmological information from \lyaf\ correlation functions. In particular, in this work we focus on the three main sources of information: BAO, the \alpac\ effect, and redshift space distortions. As BAO has already been measured from this data set by \kplya, here we focus on the other two effects.

To isolate the AP effect, we use the parametrization introduced by \cite{Cuceu:2021}:
\begin{equation}
    \phi(z) \equiv \frac{q_{\bot}(z)}{q_{||}(z)} \;\;\text{and}\;\; \alpha(z) \equiv \sqrt{q_{\bot}(z)q_{||}(z)},
\end{equation}
where $\phi$ captures the anisotropy introduced by AP, and is related to the \alpac\ parameter $D_M H(z)$ through:
\begin{equation}
    \phi(z) = \frac{D_M H(z)}{[D_M H(z)]_\mathrm{fid}}, \label{eq:ap_dist}
\end{equation}
where $H(z)$ is the Hubble parameter, and the subscript \textit{fid} indicates quantities computed using the fiducial cosmology. As we have two sets of $(q_{||}, q_\bot)$ parameters, we can measure $\phi$ independently for the peak component and from the broadband. We will use the symbol \phip\ to refer to the measurement of $\phi$ from BAO, while \phis\ will be used for the smooth component. As our end goal is to measure AP from the full shape, we will also use the symbol \phif\ to refer to measurements of AP where the anisotropy in the two components is fitted through a single parameter (i.e., we impose $\phi_\mathrm{p}=\phi_\mathrm{s}$).

On the other hand, $\alpha$ measures an isotropic scale, which for the peak component is related to the comoving acoustic scale $r_d$, through:
\begin{equation}
    \alpha_\mathrm{p}^2(z) = \frac{D_M(z) D_H(z)/r_d^2}{[D_M(z) D_H(z)/r_d^2]_\mathrm{fid}}, \label{eq:bao_dist}
\end{equation}
with $D_H(z) = c / H(z)$, and speed of light, $c$. As before, we use the symbol \alphap\ to refer to the isotropic scale parameter for the peak component, and \alphas\ for the isotropic parameter for the smooth component. The \alphas\ parameter is sensitive to the scale of matter-radiation equality, which has been used as a stand-alone source of cosmological information with galaxy clustering \citep[e.g.][]{Philcox:2021,Zaborowski:2025}. However, in the case of the \lyaf, its measurement has not been validated in the presence of systematics. Given that \alphas\ constraints are significantly weaker than \alphap\ constraints, and would need to be independently validated, we follow \cite{Cuceu:2021} and treat \alphas\ as a nuisance parameter to be marginalized over.

From \Cref{eq:pk2}, we can see that measurements of the anisotropy in the \lya-QSO cross-correlation are sensitive to the growth rate, $f$. However, both \cref{eq:pk1,eq:pk2} have an implicit scaling of $\sigma_8^2$. This means that instead of $\{b'_F, b_Q, f\}$, we are actually sensitive to the products $\{b'_F\sigma_8,b_Q\sigma_8,f\sigma_8\}$. The \lya\ auto-correlation constrains $b'_F\sigma_8$ and $\beta'_F$, which allows us to break the degeneracy between the \lya\ and the QSO part of \cref{eq:pk2}, and therefore measure $b_Q\sigma_8$ and $f\sigma_8$. Also note that all the parameters mentioned above evolve with redshift, and our measurements are made at the effective redshift of our data, $z_\mathrm{eff}=2.33$.

In practice, we fit for $\{b'_F, b_Q, f\}$, and rescale the end result by $\sigma_8$. Because $\sigma_8$ is defined using spheres of $8$ \hMpc, there is an additional dependence on the isotropic scale because we are rescaling the coordinates of the model to match the data. As the RSD information primarily comes from the broadband, the relevant scale is the one measured from the smooth component. Following \cite{GilMarin:2020} and \cite{Bautista:2021}, we account for this effect by using a rescaling of $\sigma_8$ with spheres of $8\alpha_\mathrm{iso}$ \hMpc, where $\alpha_\mathrm{iso} = q_{||}^{1/3} q_\bot^{2/3}$. This effectively keeps the scale $\sigma_8$ is fitted at constant relative to the data. The measurements of \fsig\ presented in this work are rescaled using the $\alpha_\mathrm{iso}$ values derived from the best-fit \alphas\ and \phis.

\subsection{Blinding}
\label{subsec:blinding}

The analysis presented in this work was first performed blinded in order to avoid human biases. Given that we are using the same measured correlation functions as \kplya, we had to work with unblinded correlations and therefore resorted to using the following blinding method. Within the likelihood, we add random values to the parameters of interest, \phis\ and $f$. These random values are kept consistent across all tests we do by fixing the random seed used to generate them. They are generated from Gaussians with null mean, and standard deviations based on the previous eBOSS DR16 measurement in the case of \alpac, $\sigma_{\phi_\mathrm{s}}=0.02$, and the forecasted DESI DR1 uncertainty in the case of the growth rate, $\sigma_f=0.14$\footnote{This is because our analysis is the first to measure \fsig\ from the \lyaf. The forecasted uncertainty is based on \cite{Cuceu:2021}.}.
The actual values of the blinding were $\Delta\phi_\mathrm{s}=-0.020$ and $\Delta f=0.02$, which corresponds to a significant shift in \phis, and a very small shift in $f$.

The decisions related to what constitutes the baseline analysis were taken without knowledge of the final results. Also, all the validation tests presented in \Cref{sec:validation} were first performed blinded. In light of the tests on mocks presented in \Cref{subsec:mocks}, we decided before unblinding that only the \alpac\ result has been fully validated, and will therefore be used for the cosmological inference, as discussed in \cref{sec:discussion}. We still present our measurement of \fsig, along with all of its validation tests, but we do not include it in the cosmological results presented in \Cref{sec:cosmo}.

\section{Compressed results} \label{sec:comp_results}

\begin{figure*}
    \centering
    \includegraphics[width=1.0\textwidth]{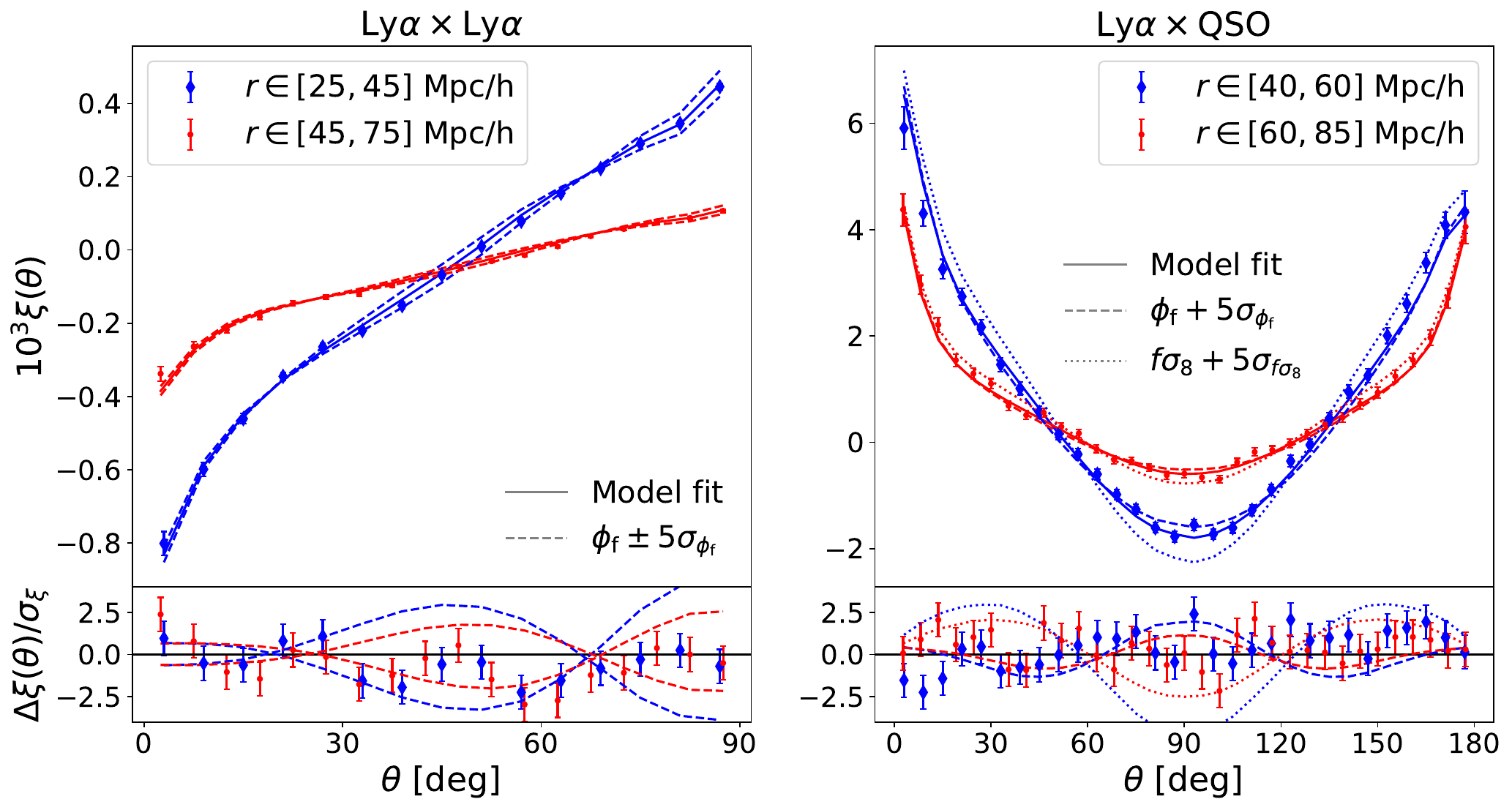}
    \caption{(left) \lyaf\ auto-correlation function (points with error-bars) compressed into shells of separation r, and shown as a function of the line-of-sight angle, $\theta$. The top panel shows the shell compression, while the bottom panel shows the residuals of the shells. The best-fit model is also compressed into the continuous lines and captures well the anisotropy present in the data. The dashed lines indicate models where the AP effect is modified by $5\sigma$ in either direction, illustrating its impact on these shells. (right) Shell compression of the \lya-QSO cross-correlation function. $\theta$ values below $90\degree$ correspond to \lya\ pixels in front of the quasar they are correlated with, whereas values above $90\degree$ correspond to \lya\ pixels behind the quasar (i.e., at larger redshift). Here we also add a model where the value of \fsig\ is modified by $5\sigma$, shown using dotted lines. Note that the separations used to select the shells are different for the auto- and cross-correlation due to differing scale cuts.}
    \label{fig:auto_shells}
\end{figure*}

We model and fit the four correlation functions computed by \kplya\ using the \texttt{Vega} package. We use the \texttt{iminuit} minimizer \citep{iminuit,James:1975dr} to find the best fit, and sample posterior distributions using the Nested Sampler \texttt{PolyChord} \citep{Handley:2015a,Handley:2015b}. We use $4$\hMpc\ bins and compute the auto-correlations up to $200$\hMpc\ in both $r_{||}$ and $r_\bot$, while the cross-correlations are computed from $-200$\hMpc\ to $200$\hMpc\ in $r_{||}$, and from $0$\hMpc\ to $200$\hMpc\ in $r_\bot$. This means the two \lya\ auto-correlations have a total of 2500 data points each, while the two \lya-QSO cross-correlations have 5000 points each, for a total of 15000 data points.

The maximum scale we fit is $r_\mathrm{max}=180$ \hMpc\ for all four correlations. On the other hand, the minimum scale is different for the auto vs the cross due to differences in our confidence of the modelling of small scales. For the auto-correlations, we use a minimum scale of $r_\mathrm{min}=25$ \hMpc, in line with \cite{Cuceu:2023a,Cuceu:2023b}, while for the cross-correlation we choose a more conservative $r_\mathrm{min}=40$ \hMpc. This choice is partly driven by the fact that for the auto-correlation we have an empirical model for small-scale non-linearities \citep{Arinyo:2015}, that has been extensively tested on measurements of the \lya\ power spectrum from hydro-dynamical simulations. However, for the cross-correlation, we do not have an equivalent model, and therefore, we restrict our analysis to larger scales. Furthermore, \cite{Cuceu:2023a} found using eBOSS DR16 mocks that AP measurements are unbiased when using scales larger than $25$ \hMpc, while a significant bias was detected when using scales smaller than that. We find consistent results using the DESI DR1 mocks here (see \cref{subsec:mocks}). Scale cuts are implemented in isotropic separation (i.e., $\sqrt{r_{||}^2+r_\bot^2}$), and after that we are left with a total of 9188 data points.

We show the best-fit model versus data, along with the residuals, in \Cref{fig:auto_shells}. As the quantities we aim to measure come from the anisotropy in correlation functions, we compress the 2D correlations into shells of isotropic separation and show them as a function of the line-of-sight angle $\theta$.\footnote{Where $\theta=0\degree$ corresponds to $r_\bot=0$, and $\theta=90\degree$ corresponds to $r_{||}=0$.} The compression is performed through a weighted average using the covariance matrix. Plots of the same data compressed into wedges instead of shells can be found in Figures 4 and 5 of \kplya. Here we focus on two shells at smaller separations ($r<75$\hMpc\ for \lyalya, and $r<85$\hMpc\ for \lyaqso), as these provide most of the signal for both the AP and RSD measurements. The equivalent plots for the shells at larger separation are in Appendix \ref{sec:large_scales}. \cref{fig:auto_shells} also illustrates the impact of changing \phif\ and \fsig\ through the dashed and dotted lines, respectively. We only show the \fsig\ variation for the cross-correlation, as this is our only source of information for this effect (see \cref{subsec:scale_pars}).  A similar figure showing the impact of several important nuisance parameters can be found in Appendix \ref{sec:nuisance}.

The model gives a good fit to the data, with $\chi^2_\mathrm{min}=9145.1$ for $9163$ degrees of freedom (PTE$=0.55$). We have found that changing the minimum separation used does not significantly degrade the fit quality, which is not surprising given that \kplya\ used $r_\mathrm{min}=10$ \hMpc\ for both the auto and cross-correlation, and still obtained a good fit (PTE$=0.23$). When interpreting the goodness-of-fit from the shell plots shown in \cref{fig:auto_shells}, it is important to note that there is significant correlation between nearby data points. Directly adjacent points have correlation coefficients of $\sim0.5$, and the correlation matrix remains above $0.1$ up to the third or fourth off-diagonal.

\begin{table*}[t]
    \centering
    \begin{tabular}{c|c|c|c}
         & Isotropic BAO & \alpac\ & Growth ($z_\mathrm{eff}=2.33$) \\
         \hline
        BAO (\kplya) & $\alpha_\mathrm{p} = 1.000\pm 0.011$ & $\phi_\mathrm{p} = 1.026 \pm 0.038$ & - \\
         \hline
        Broadband & - & $\phi_\mathrm{s} = 0.987 \pm 0.017$ & $f\sigma_8 = 0.364^{+0.056}_{-0.065}$ \\
         \hline
        Full-shape & $\alpha_\mathrm{p} = 0.999\pm 0.010$ & $\phi_\mathrm{f} = 0.995\pm 0.016$ & $f\sigma_8 = 0.371^{+0.055}_{-0.065}$ \\
    \end{tabular}
    \caption{Constraints on BAO, the \alpac\ effect, and the growth parameter combination \fsig, from BAO-only, broadband-only, and full-shape analyses of the DESI DR1 \lyaf\ correlation functions. The BAO-only results come from \kplya, while the other two analyses are presented here.}
    \label{tab:base_results}
\end{table*}

\begin{figure}
    \centering
    \includegraphics[width=1.0\columnwidth,keepaspectratio]{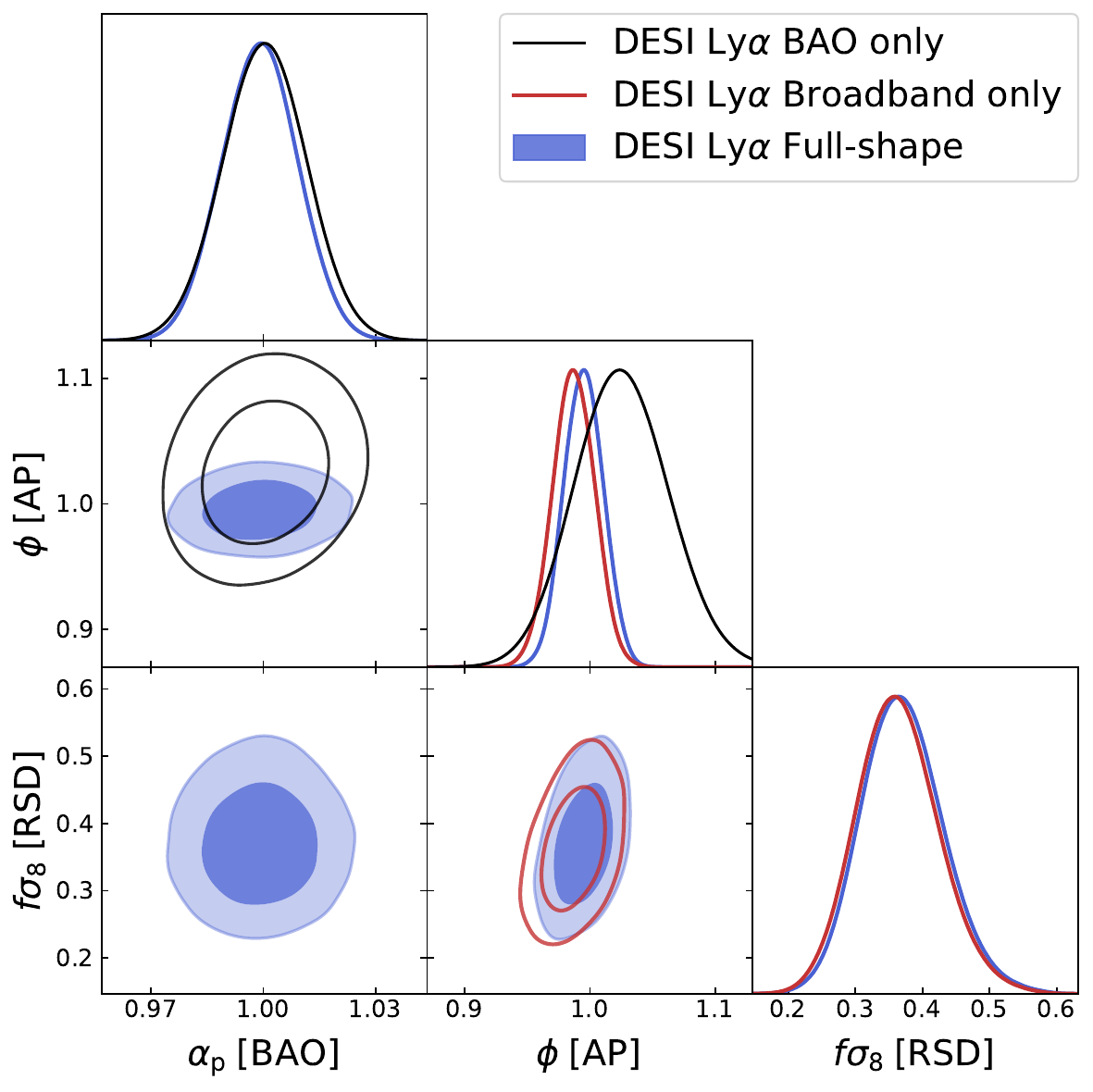}
    \caption{
    Constraints on the isotropic BAO scale, \alphap, the \alpac\ effect, $\phi$, and the growth parameter combination, \fsig. Our main results are shown in blue for the full-shape analysis of the DESI DR1 \lyaf\ correlations, where the AP constraint is given by \phif. For comparison, we also show the BAO measurement measured by \kplya\ from the same dataset in black (AP given by \phip), and results from only the broadband in red (where we marginalize over BAO and AP is given by \phis). The \alpac\ constraint from the broadband is more than a factor of two tighter than the one from BAO.
    }
    \label{fig:base_results}
\end{figure}

Our main measurements are presented in \Cref{fig:base_results} and in \Cref{tab:base_results}, where we focus on the three sources of cosmological information that we use: BAO, the \alpac\ effect, and redshift space distortions. We compare our full-shape results in blue, with the BAO measurement by \kplya\ from the same dataset in black, and the broadband-only constraint in red (where the BAO parameters \alphap\ and \phip\ are marginalized). The isotropic BAO parameter (\alphap) constraint is very similar to that of \kplya, although not identical due to the small correlation between this parameter and the AP effect (correlation coefficient of $\sim0.16$ for \phip, and $\sim0.08$ for \phif). 

The AP measurement from the broadband (\phis) is also consistent with the AP measurement made by \kplya\ (\phip), but represents a $1.7\%$ constraint compared to the previous $3.8\%$ constraint from the BAO peak in the same dataset. Combining the two, the full-shape of the DESI DR1 \lyaf\ correlations gives a $1.6\%$ constraint on the \alpac\ effect (\phif), representing an improvement of about $2.4\times$ in constraining power.

We also obtain the first ever measurement of the growth rate from 3D \lyaf\ correlations, quantified through the combined parameter \fsig. The constraint is much weaker $(15\%-17\%)$ when compared to the AP result, due to the fact that we are only using the quasar RSD signal in the Lya-QSO cross-correlation for this measurement.

\section{Analysis validation} \label{sec:validation}

Our approach for analysis validation closely follows that of \kplya. First, we used a set of 150 mock realizations of DESI DR1 to validate our analysis pipeline, by showing that we can recover unbiased constraints and accurate uncertainties (\Cref{subsec:mocks}). After that, we performed a large set of data splits (\Cref{subsec:datasplits}) and analysis variations (\Cref{subsec:variations}) in order to study the robustness of our measurement. These tests were first performed using the blinding method described above, and informed the choice of baseline model.

As our analysis starts from the correlation functions measured by \kplya, the main purpose of this section is to validate the robustness of the model for these correlation functions when it comes to measurements of \phif\ and \fsig. Therefore, we imposed a threshold of $1/3$ of the DESI statistical uncertainty, $\sigma_\mathrm{DESI}$, for tests of the model to be considered successful. This corresponds to $\sim0.005$ for \phif\, and $\sim0.02$ for \fsig. We will show this threshold in all figures where it was applied. Note that for model variations, this threshold was applied to \phis\ before unblinding, but we did not observe any significant changes when working with \phif\ post-unblinding, and therefore present all our results in terms of \phif.

While our main results in \cref{sec:comp_results} were based on the full posterior distribution as measured by the Nested Sampler \texttt{PolyChord}, for this section we were limited by computational constraints to using the \texttt{iminuit} minimizer with Gaussian approximated uncertainties\footnote{Computed from the second derivative of the likelihood around the best-fit point.}. This was also the case for \kplya, but in their case the posterior distribution of the BAO parameters was very well approximated by a Gaussian. On the other hand, in our case the Gaussian approximation can lead to small differences with respect to the full posterior results as described in Appendix \ref{sec:fitter} \citep[also see][]{Cuceu:2020}. We will discuss the implications of this limitation wherever it is relevant in this section. We will refer to measurements based on the full posterior distribution as sampler results, while those based on the best-fit with Gaussian approximated uncertainties as fitter results.

\subsection{Validation with mocks} \label{subsec:mocks}

We use the set of mocks created by \cite{KP6s6-Cuceu} to validate the DESI DR1 \lyaf\ BAO measurement in \kplya. For a detailed description of how these mocks are created, see \cite{Herrera-Alcantar:2024}. Furthermore, the same type of mocks were also used to validate the \lya\ full-shape analysis of eBOSS DR16 in \cite{Cuceu:2023a}, although that analysis only focused on the \alpac\ effect and did not study RSD. Therefore, here we only give a very brief description of these mocks, focusing on the relevant implications for full-shape analyses.

The mocks were generated using two different sets of fast simulations based on the \lyacolore\ \citep{Farr:2020,Ramirez-Perez:2022} and \saclay\ packages \citep{Etourneau:2023}. Both of these use log-normal density fields and simplified recipes to generate the quasar distribution and paint the \lyaf\ absorption. The recipes were calibrated to approximately match the mean flux, 1D power spectrum, and large-scale bias and RSD parameters measured by eBOSS. These simulations are then used to generate mock realizations of the DESI DR1 quasar catalog with matching synthetic DESI spectra containing simulated \lya\ forests. This is done using the \texttt{desisim} package\footnote{\url{https://github.com/desihub/desisim}}, which simulates the DESI DR1 footprint, magnitude and redshift distributions, spectrograph resolution, noise, and redshift errors. At this stage, astrophysical contaminants are also added, which include Damped \lya\ systems (DLAs), Broad Absorption Lines (BALs), and metal absorption \citep{Herrera-Alcantar:2024}. We use the set of 100 \lyacolore\ and 50 \saclay\ mocks of DESI DR1 generated in \cite{KP6s6-Cuceu} and used to validate the analysis of \kplya. 

The \lyaf\ correlation functions were computed using the same analysis pipeline and methodology that was used to analyze the real data (\cref{sec:analysis}). The only difference with respect to the analysis presented in \cite{KP6s6-Cuceu} is that we now go beyond BAO in our model and focus on the full-shape information as described in \cref{subsec:model}. Following \cite{KP6s6-Cuceu}, the model applied to mocks differs slightly from the one applied to data in the following aspects. First, we include extra Gaussian smoothing to account for the finite cell size in the log-normal simulations. This includes two new free parameters corresponding to the smoothing amplitude along and across the line-of-sight, respectively. Secondly, we ignore correlated sky residuals, CIV contamination, the transverse proximity effect, and BAO broadening due to non-linear evolution since these effects are not included in the mocks. None of these effects are correlated with either \alpac\ or RSD given our scale cuts, and we study their impact below using the real data\footnote{See \cite{Cuceu:2023a} for a discussion of these effects and their impact on full-shape analyses.}. Finally, log-normal mocks do not have correct small-scale clustering, so we do not use the Arinyo model in our baseline mock fits. We have tested including this model and freeing all 5 extra parameters, and we find no significant impact on our measurements. In fact, similar to the measurements on data, we find no preference for deviations from linear theory (i.e., the $q_1$ parameter is consistent with zero).

\begin{figure}
\centering
\includegraphics[width=1.\columnwidth]{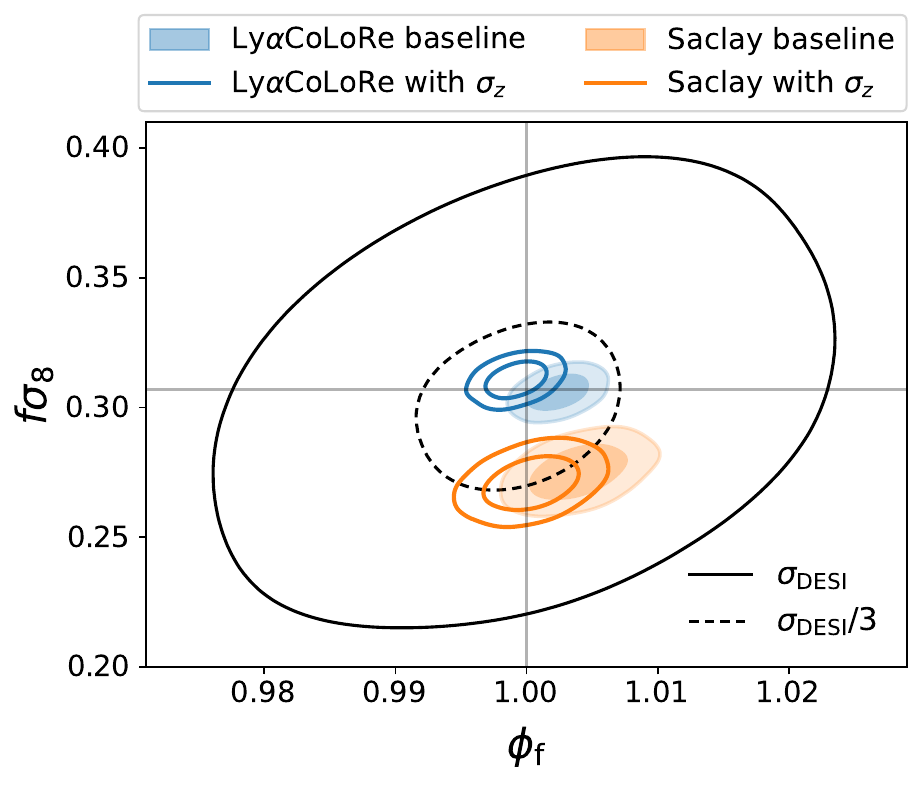}
\caption{$68\%$ and $95\%$ credible regions of AP (\phif) and RSD (\fsig) measurements from the stack of 100 \lyacolore\ (blue) and 50 \saclay\ (orange) mocks. The filled contours show the results for the baseline mock analysis, while the empty contours show results for a variation where redshift errors affect the continuum fitting. The gray cross indicates input mock cosmology, while the dashed black contour shows $1/3$ of the DESI DR1 uncertainty. The AP measurement is unbiased in all cases, while the RSD measurement is unbiased for \lyacolore\ mocks, and shows a small but significant bias for \saclay\ mocks.}
\label{fig:mock_stack}
\end{figure}

We start by fitting the mean of the correlation functions computed from the two sets of mocks, which we will refer to as stacked correlations. \cite{KP6s6-Cuceu} showed that the model described in \cref{subsec:model} provides an excellent fit to these stacked correlations, even when fitting down to $10$ \hMpc\ and fixing the broadband scale parameters. Furthemore, \cite{KP6s6-Cuceu} also found that BAO measurements are unbiased in these mocks. Therefore, our goal here is to study whether we recover unbiased measurements of AP and RSD. Our measurements of \phif\ and \fsig\ from the stack of mocks are shown in \cref{fig:mock_stack}. For comparison, we also show the input mock cosmology marked by the gray cross, along with the DESI contours centered on the input cosmology (continuous black), as well as the contour corresponding to $1/3$ of the DESI uncertainty (dashed black).

\cref{fig:mock_stack} shows two measurements for each of the two sets of mocks. The baseline results are represented by the filled contours, and in the case of the \alpac\ effect they show excellent agreement with the fiducial cosmology used to generate the mocks for both \lyacolore\ and \saclay\ mocks (vertical gray line). On the other hand, the two sets of mocks give slightly different constraints on \fsig. \lyacolore\ mocks are in agreement with the fiducial value (horizontal gray line), while \saclay\ mocks show a small but significant bias. The \fsig\ measurement from \saclay\ mocks is on the edge of the $1/3\,\sigma_\mathrm{DESI}$ contour in 2D. However, in 1D it corresponds to a shift of $\sim0.5\, \sigma_\mathrm{DESI}$ from the truth, and is detected at a significance of more than $4\sigma$. Given that \lyacolore\ mocks are consistent with the fiducial cosmology, it is difficult to draw strong conclusions from this result, as it may be indicative of an issue with one or both sets of mocks rather than a true bias. The redshift-space mapping is not incorporated in the same way for the two sets of mocks, and the two methods rely on different approximations. We discuss this further in \cref{sec:discussion}.

The empty contours in \cref{fig:mock_stack} show results from a different type of analysis of the same sets of mocks. This variation simulates Gaussian redshift errors that affect the continuum fitting process. The baseline analysis does include redshift errors, but only in the quasar catalog used to compute cross-correlations. \cite{Youles:2022} found that redshift errors can also impact the continuum fitting process, leading to spurious correlations appearing in both \lya\ auto and \lya-QSO cross-correlations. \cref{fig:mock_stack} shows that while these spurious correlations do have an impact on our measurements, it is much smaller than the $1/3\,\sigma_\mathrm{DESI}$ threshold, and does not impact our conclusions. While this variation is useful for testing our sensitivity to this effect, the way redshift errors are simulated in these mocks leads to an extreme version of this contamination \citep[see][]{Youles:2022,KP6s6-Cuceu,Gordon:2025}. This is why \cite{KP6s6-Cuceu} did not include this effect in their baseline analysis of these mocks, and we take the same approach here.

Using the mock baseline stacked correlation functions, we also studied how scale cuts affect our ability to recover unbiased constraints on AP and growth. \cite{Cuceu:2023a} previously performed this test for \phis\ using eBOSS mocks (see their Figure 5), and our results are in good agreement with their findings. In particular, we find that using scales below $20$\hMpc\ produces biased constraints on \phif\ at more than $2\sigma$ significance in both \lyacolore\ and \saclay\ mocks. On the other hand, for \fsig\ we find biased results when using cross-correlation scales below $35$\hMpc\ in the case of \lyacolore\ mocks. For Saclay mocks, restricting both the auto- and cross-correlation to scales larger than $50$\hMpc\ does produce \fsig\ constraints consistent with the truth at the $2\sigma$ level. However, that is largely due to the significant increase in uncertainty, rather than a shift in the result.

Next, we perform full-shape fits to individual mocks and study the population statistics of the results. The main goal here is to test the robustness of our uncertainties. Similar to \cite{Cuceu:2023a}, we are limited by computational constraints to using the \iminuit\ minimizer instead of sampling the full posterior distributions. Therefore, our individual mock uncertainties are computed under the assumption that the posterior distributions are Gaussian.

\begin{figure}
    \centering
    \includegraphics[width=1.0\columnwidth,keepaspectratio]{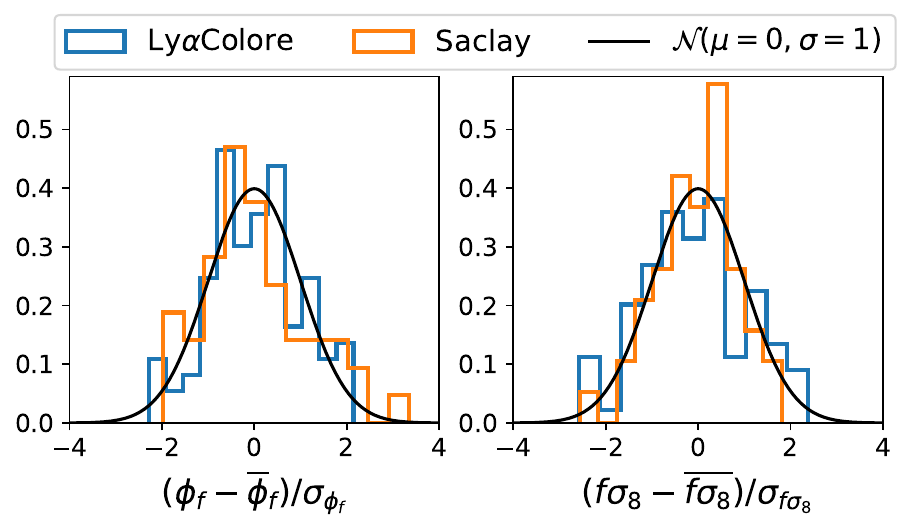}
    \caption{Pull distributions for \phif\ (left) and \fsig\ (right) from the set of 150 mock analyses. The black lines show the expected Gaussian distributions with null mean and unit variance.}
    \label{fig:pull}
\end{figure}

\Cref{fig:pull} shows the pull distributions for \phif\ and \fsig, obtained from the difference between individual results and the mean of all mocks, divided by the measured uncertainty (i.e. $[X - \overline{X}]/\sigma_X$). The results from \lyacolore\ mocks are shown in blue, while those from \saclay\ mocks are shown in orange. We do not find any significant outliers, with the most extreme result being $3.3\sigma$ away from the truth, which is not surprising in a sample of 150 mocks. \Cref{fig:pull} also shows the unit variance Gaussian for comparison. The distribution of pull values for \phif\ matches well the expected Gaussian, while for \fsig\ the histogram shows some deviations. This could be explained by the fact that we are using the Gaussian approximation for the individual mock fits. We present a comparison of sampler and fitter results for the data in Appendix \ref{sec:fitter}, where we show that the Gaussian approximation used to estimate uncertainties in the fitter case produces an uncertainty for \fsig\ of $\pm 0.057$, while the uncertainty from the sampler is asymmetric: $^{+0.055}_{-0.065}$. Therefore, we should not expect a perfect match between the histogram of pull values for \fsig\ and the unit variance Gaussian. However, a full confirmation of the robustness of \fsig\ uncertainties would require running the sampler on the full set of mocks, which we leave for future work.

We also compute the standard deviations of these distributions, which we find to be $1.04 \pm 0.06$ for $\Delta\phi_\mathrm{f}/\sigma_{\phi_\mathrm{f}}$ and $1.08 \pm 0.06$ for  $\Delta f\sigma_8/\sigma_{f\sigma_8}$, with uncertainties obtained through bootstrap. The $68\%$ credible regions\footnote{Computed from half the distance between the 16th and 84th percentiles.} are also very similar, with $1.03 \pm 0.10$ for $\Delta\phi_\mathrm{f}/\sigma_{\phi_\mathrm{f}}$ and $1.07 \pm 0.11$ for $\Delta f\sigma_8/\sigma_{f\sigma_8}$. These results show that the uncertainties for both \phif\ and \fsig\ are well estimated given the uncertainty from the limited number of mocks.

The tests we have performed in this section show that measurements of the AP effect using the full shape of \lyaf\ correlations are unbiased with well-estimated uncertainties. On the other hand, growth measurements are unbiased only when using \lyacolore\ mocks, while \saclay\ mocks show a small but significant bias of about $\sim0.5\,\sigma_\mathrm{DESI}$. Based in part on these results, we decided not to use our growth rate measurements for cosmological inference. We discuss this further in \cref{sec:discussion}.

\subsection{Data splits} \label{subsec:datasplits}

\begin{figure}
\centering
\includegraphics[width=1.\columnwidth,keepaspectratio]{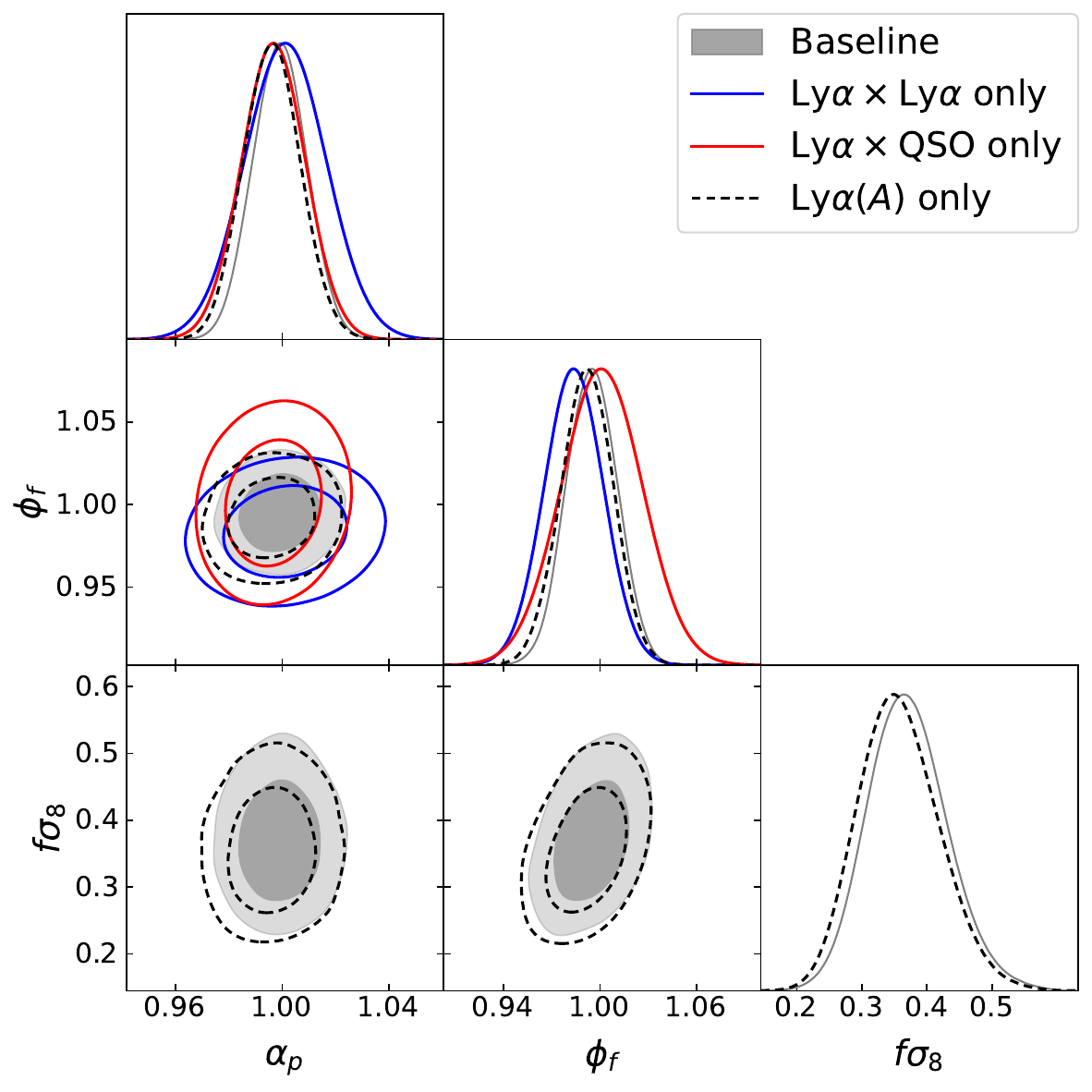}
\caption{Full-shape constraints from the two \lya\ auto-correlations (blue) versus constraints from the two \lya-QSO cross-correlations (red), as well as results from the two correlations that only use the \lya(A) region (black dashed). All three data splits are in good agreement with each other and the baseline results that use all 4 correlations (shown in gray).}
\label{fig:auto_cross}
\end{figure}

The next set of tests involves data splits, where we perform full-shape analyses of different subsets of the data in order to check for potential inconsistencies. We start by fitting the two auto-correlations, \lyalyalyalya\ and \lyalyalyalyb, separately from the two cross-correlations, \lyalyaqso\ and \lyalybqso. As a joint fit of the auto and cross-correlation is needed to break the degeneracy between the \lya\ and quasar RSD terms, for this test we can only measure the \alpac\ effect. We show the constraints on \phif\ and \alphap\ from this data split in \cref{fig:auto_cross}. The results from the auto and cross-correlations are consistent with each other. \cref{fig:auto_cross} shows that while the \alphap\ constraint from the cross is tighter than the one from the auto, the opposite is true for \phif. In the case of \alphap, this is consistent with what \kplya\ found and also with the variance of the correlation functions, which is larger for the auto compared to the cross in DESI DR1. On the other hand, we find that the broadband AP constraint from the cross is much weaker than the one from the auto, with the \phif\ uncertainty from the cross being $\sim40\%$ larger compared to the auto. This is related to the fact that the \lya\ RSD parameter is significantly larger than the QSO one ($\sim1.5$ versus $\sim0.25$), which means that the anisotropy on large scales is stronger in the \lya\ auto-correlation compared to the cross-correlation.

\cref{fig:auto_cross} also shows the results when we only use the \lya(A) region. In this case, we are fitting both the auto and cross-correlation (i.e., \lyalyalyalya\ and \lyalyaqso), and therefore have both AP and RSD constraints. The results are very similar to our baseline constraints using all 4 correlations. This is because the \lya(B) correlations are much noisier due to a few reasons, including that there are fewer forests that contain this region, and that the region is shorter and noisier due to the presence of other Lyman lines. We find that adding the two \lya(B) correlations only improves the constraints by about $3\%$ in both \phif\ and \fsig.

\begin{figure}
\centering
\includegraphics[width=1.\columnwidth,keepaspectratio]{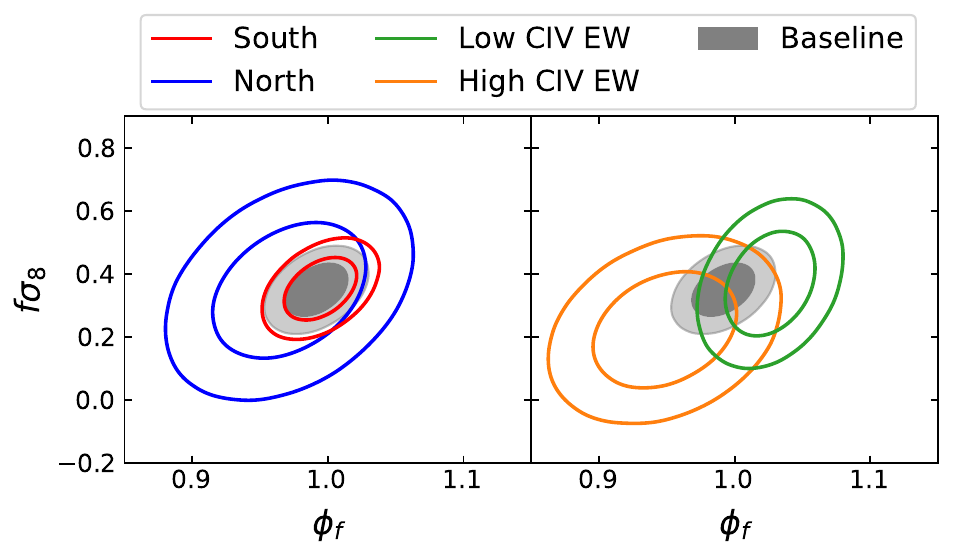}
\caption{AP and RSD constraints from the baseline analysis (gray) and from two different data splits. In the left panel, we show results from the samples using the South (red) vs North (blue) imaging for quasar target selection. In the right panel, we show results from the samples with Low (green) vs High (orange) CIV equivalent width in the quasar spectra.}
\label{fig:data_splits}
\end{figure}

The final two data splits we test are shown in \cref{fig:data_splits}, and are based on measurements by \kplya. In the left panel, we show results from a data split based on the imaging surveys used for quasar target selection. The sample designated as "South" is based on imaging from the DECam camera, and includes the entire South Galactic Cap and the southern part of the North Galactic Cap, at $\delta < 32.375\degree$. The sample designated as "North" is based on imaging from the BASS and MzLS surveys, and only contains $18\%$ of quasars. In the right panel, we show results from a split based on the equivalent width (EW) of the CIV line. This split is done because of the anti-correlation between the quasar continuum luminosity and the EW of the emission lines, known as the Baldwin effect \cite{Baldwin:1977}. We find consistent results for both of these data splits, as shown in \cref{fig:data_splits}.

\subsection{Analysis variations} \label{subsec:variations}

\begin{figure}
    \centering
    \includegraphics[width=1.\columnwidth,keepaspectratio]{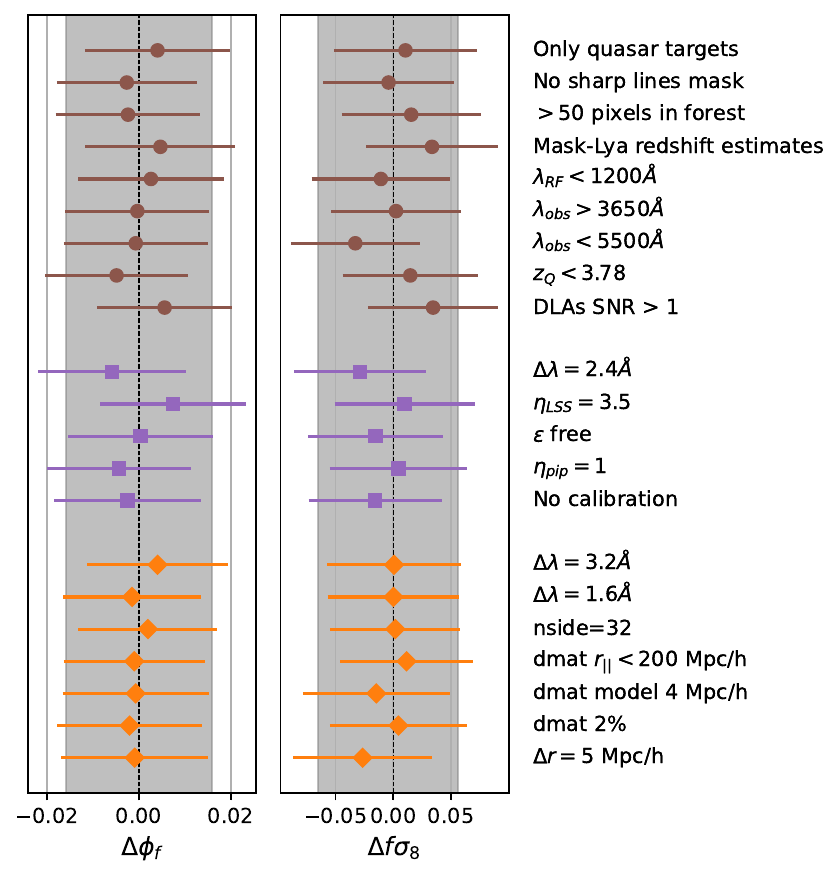}
    \caption{Shifts in the AP and growth measurements from the set of analysis variations performed by \kplya. This includes variations where the dataset is different (brown), variations in the method to estimate the \lya\ overdensities (purple), and variations in the method used to compute correlations, covariances, and distortion matrices (orange). The gray bands show the uncertainty of the baseline measurement. Most of these variations are subject to statistical fluctuations (especially the ones in brown and purple).}
    \label{fig:analysis_variations}
\end{figure}

The next set of tests we perform includes a large set of variations in the analysis methods used up to the measurement of the correlation functions. As the starting point in this work is from the correlation functions measured by \kplya, these tests were all originally performed for the validation of the BAO analysis, and here we only reanalyze those correlation functions in the context of the full-shape analysis. We present the shifts in \phif\ and \fsig\ produced by these variations in \cref{fig:analysis_variations} (points with errorbars), along with the uncertainty from the baseline analysis for comparison (gray shaded bands).

The first subset of analysis variations consists of changes to the dataset. Therefore, we expect these variations to be affected by statistical fluctuations. The shifts in \phif\ and \fsig\ due to these variations are shown in brown in \cref{fig:analysis_variations}, and consist of the following tests:
\begin{itemize}
    \item Only quasar targets: Use only quasars that were originally targeted as quasars.
    \item No sharp lines mask: Do not mask the sharp lines in quasar spectra, as described in \cite{Ramirez-Perez:2024}.
    \item $>50$ pixels in forest: Use lines-of-sight that have at least 50 valid \lya\ pixels, as opposed to the 150 pixel threshold in the baseline analysis.
    \item Mask-\lya\ redshift estimates: Use a different estimator for quasar redshifts, that only takes into accounts wavelengths longer than the \lya\ emission line.
    \item $\lambda_{RF}<1200$\AA: Use only pixels below $1200$\AA\ in rest-frame wavelength ($1205$\AA\ in the baseline analysis).
    \item $\lambda_{obs}>3650$\AA: Use only pixels above $3650$\AA\ in observed wavelength ($3600$\AA\ in the baseline analysis).
    \item $\lambda_{obs}<5500$\AA: Use only pixels below $5500$\AA\ in observed wavelength ($5577$\AA\ in the baseline analysis).
    \item $z_\mathrm{Q} < 3.78$: Use only quasars at redshifts below $3.78$, which corresponds to the highest redshift included in the mocks.
    \item DLAs SNR $>1$: Mask DLAs in spectra that have SNR $>1$, as opposed to SNR $>3$ for the baseline analysis.
\end{itemize}

The next subset of analysis variations includes those that affect the continuum fitting process. These variations can also introduce statistical fluctuations, as they impact the weights used both for continuum fitting and for measuring the correlation functions. The shifts in AP and growth due to these changes are shown in purple in \cref{fig:analysis_variations}, and consist of the following tests:
\begin{itemize}
    \item $\Delta\lambda=2.4$\AA: Perform the pixel re-binning from $0.8$\AA\ to $2.4$\AA\ before the continuum fitting process, as done in \cite{duMasdesBourboux:2020}. For this variation, we set $\sigma^2_\mathrm{mod}=3.1$, following \cite{Ramirez-Perez:2024}.
    \item $\eta_\mathrm{LSS}=3.5$: Reduce the factor that scales the intrinsic \lya\ forest variance from $7$ to $3.5$, thereby reducing its contribution to the weights.
    \item $\epsilon$ free: Include an extra term in the \lya\ weights meant to capture quasar diversity, following \cite{duMasdesBourboux:2020}.
    \item $\eta_\mathrm{pip}=1$: Do not re-calibrate the instrumental noise. See \cite{Ramirez-Perez:2024} for more details.
    \item No calibration: Do not re-calibrate spectra with the CIII region, as described in \cite{Ramirez-Perez:2024}.
\end{itemize}

The final subset of analysis variations includes changes to how correlation functions, covariance matrices, and distortion matrices are computed. These are shown in orange in \cref{fig:analysis_variations}, and consist of the following tests:
\begin{itemize}
    \item $\Delta\lambda=3.2$\AA: Rebin \lya\ pixels in groups of 4 (3 in the baseline analysis).
    \item $\Delta\lambda=1.6$\AA: Rebin \lya\ pixels in groups of 2 (3 in the baseline analysis).
    \item nside $=32$: Measure the covariance matrix from a set of correlations measured in HEALPix pixels with nside $=32$, as opposed to nside $=16$ in the baseline analysis. This results in more HEALPix pixels that are smaller in size.
    \item dmat $r_{||}<200$ Mpc/h: Compute the distortion matrix up to $r_{||}<200$ \hMpc, as opposed to $r_{||}<300$ \hMpc\ in the baseline analysis.
    \item dmat model $4$ Mpc/h: Compute the distortion matrix with $4$ \hMpc\ bins for the model, matching the binning of the data. In the baseline analysis, $2$ \hMpc\ bins are used for the model.
    \item dmat $2\%$: Use $2\%$ of pixels to compute the distortion matrix, as opposed to $1\%$ in the baseline analysis.
    \item $\Delta r=5$ Mpc/h: Use $5$ \hMpc\ bins for the correlation function measurements, as opposed to the $4$ \hMpc\ bins used in the baseline analysis.
\end{itemize}

All these variations produce results consistent with the baseline, with shifts larger than $1/3$ of the DESI uncertainty only observed for variations where we expect statistical fluctuations. As noted earlier, we used the fitter with Gaussian approximated uncertainties for the individual tests, while for the baseline analysis, we computed the full posterior distribution. As shown in Appendix \ref{sec:fitter}, the fitter results can be shifted with respect to the sampler results due to the posterior being asymmetric, and more importantly, due to the difference between conditioning and marginalizing over nuisance parameters. Therefore, we also expect some extra variance in these results due to these differences.

\subsection{Model variations} \label{subsec:model_variations}

We next turn our attention to variations in the modelling process. Unlike the tests in the previous section, most of these variations were designed specifically to test the robustness of the full-shape analysis. This means we perform a set of tests that is fairly different from the one done for the BAO analysis in \kplya.

\begin{figure}
    \centering
    \includegraphics[width=1.\columnwidth,keepaspectratio]{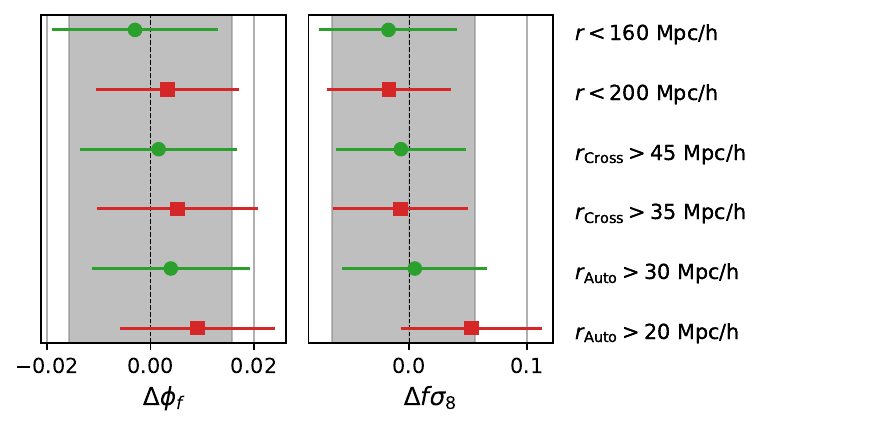}
    \caption{Similar to \cref{fig:analysis_variations}, but showing shifts due to changing the scale cuts used when fitting the model to the measured correlation functions. Changes to the minimum separation are performed independently for the auto and cross-correlation, as they do not use the same limit. Cuts that are more conservative than the baseline are shown using green circles, while less conservative cuts are shown using red squares. These variations are also affected by statistical fluctuations.}
    \label{fig:cuts}
\end{figure}

The first subset of model tests involves changes to the scale cuts, and the resulting shifts in \phif\ and \fsig\ are shown in \cref{fig:cuts}. In the baseline analysis, we use the same maximum fitted scale for all correlations, $r<180$\hMpc, while for the minimum fitted scale we use $r_\mathrm{auto}>25$\hMpc\ for the auto-correlation, and $r_\mathrm{cross}>40$\hMpc\ for the cross-correlation. Here we test changes to these choices that go in both directions, with choices that are more conservative than the baseline shown using green circles, and less conservative choices shown using red squares. These variations also introduce statistical fluctuations because we are changing which data affects our results. This is especially true of changes to the minimum separation, as our measurements come in large part from small scales \citep{Cuceu:2021}.

The results shown in \cref{fig:cuts} are consistent with those obtained using the baseline cuts. The only large shifts are observed when decreasing the minimum scale for the auto-correlation. For this particular case, we also observe a significant increase in the magnitude of the HCD bias, indicating that HCDs start to play an important role at separations below $25$\hMpc. While in the baseline analysis the HCD bias is consistent with zero (see Appendix \ref{sec:nuisance}), in the variation with $r_\mathrm{auto}>20$\hMpc\ we strongly detect a non-zero HCD bias at $\sim12\sigma$.

The final set of variations includes changes to the model used to fit the correlation functions. The shifts in AP and growth produced by these variations are shown in \cref{fig:model_variations}, and consist of the following tests:
\begin{itemize}
    \item Gaussian redshift errors: Use a Gaussian distribution to model redshift errors and peculiar velocities of quasars, as opposed to a Lorentzian distribution in the baseline analysis.
    \item Weak $\Delta r_{||}$ prior: Use a flat prior for the parameter that models the systematic quasar redshift error, $-3 > \Delta r_{||} > 3$ \hMpc. In the baseline analysis, a Gaussian prior is used, with $\mathcal{N}(0, 1)$.
    \item Weak CIV prior: Use a flat prior for the CIV bias parameter, with $-0.5 < b_\mathrm{CIV} < 0$. In the baseline analysis, a Gaussian prior is used, with $\mathcal{N}(-0.0243, 0.0015)$.
    \item UV background fluctuations: Add the impact of fluctuations in the UV background following \cite{GontchoAGontcho:2014} and \cite{Bautista:2017}. 
    \item $L_\mathrm{HCD}=3$ Mpc/h: Fix the value of $L_\mathrm{HCD}$ to $3$ \hMpc. This is allowed to vary in the baseline analysis.

\begin{figure}
    \centering
    \includegraphics[width=1.\columnwidth,keepaspectratio]{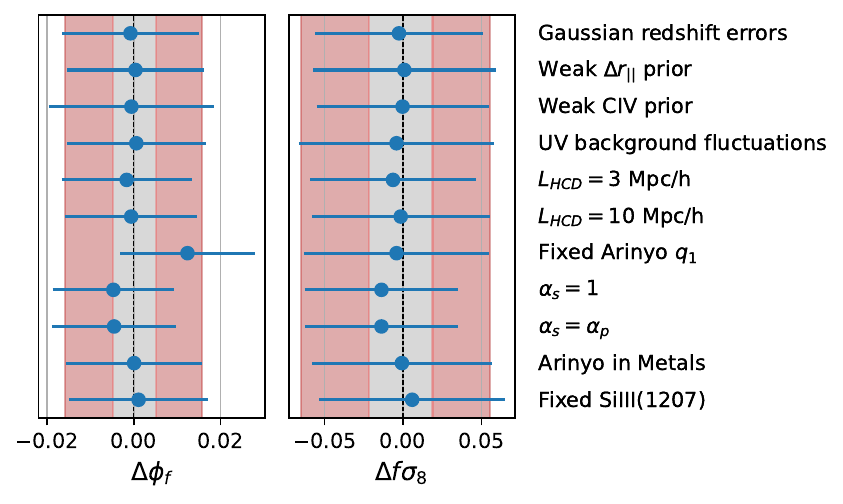}
    \caption{Shifts in the AP and growth measurements when changing the model used to fit the correlation functions. The red bands show the uncertainty from the baseline analysis, while the smaller gray bands show the threshold used for these tests, which represents $1/3$ of the uncertainty from the baseline analysis.}
    \label{fig:model_variations}
\end{figure}

    \item $L_\mathrm{HCD}=10$ Mpc/h: Fix the value of $L_\mathrm{HCD}$ to $10$ \hMpc. This is allowed to vary in the baseline analysis.
    \item Fixed Arinyo $q_1$: Fix the Arinyo $q_1$ parameter to the value used in BAO analyses: $q_1=0.86$. This is based on the measurements from hydro-dynamical simulations of \cite{Arinyo:2015}, interpolated to our effective redshift. In the baseline analysis, this is allowed to vary.
    \item $\alpha_\mathrm{s}=1$: Fix the value of the isotropic scale parameter for the smooth component (\alphas) to $1$. In the baseline analysis, this is allowed to vary.
    \item $\alpha_\mathrm{s}=\alpha_\mathrm{p}$: Impose the condition that the isotropic scale parameters for the peak and smooth components are the same, therefore fitting for the full-shape isotropic scale. In the baseline analysis, they are varied independently of each other.
    \item Arinyo in Metals: Use the Arinyo non-linear correction when modelling the \lya-Metal correlations as well.
    \item Fixed SiIII(1207): Fix the SiIII(1207) bias parameter to the best-fit value obtained by \kplya: $10^3 b_\mathrm{SiIII(1207)}=-9.79$. This parameter is allowed to vary in the baseline analysis. The \lya-SiIII(1207) peak is outside our fitting range ($r_{||}\sim21$ \hMpc), but we are still sensitive to its presence due to the distortion matrix \citep{Cuceu:2023a}.
\end{itemize}

The only variation that produces a shift larger than the $1/3\,\sigma_\mathrm{DESI}$ threshold is the one where the Arinyo $q_1$ parameter is fixed to the value measured from the reference hydrodynamical simulation of \cite{Arinyo:2015}, $q_1=0.86$. However, the data disfavors such a high value for this parameter with a significance of $\sim3\sigma$ ($\Delta\chi^2=9.4$). See Appendix \ref{sec:nuisance} for a discussion of this result. This test has informed our choice of allowing all parameters in the Arinyo model to vary freely in the baseline analysis, instead of imposing constraints based on other measurements. We note that the parameters of this model also have a cosmology dependency, so it would be interesting to explore the implications of their constraints, as recently proposed by \cite{ChavesMontero:2025}. However, since these parameters are treated as a nuisance here, their measurement has not been validated, so a thorough study into these constraints is needed before any cosmological interpretation can take place.

The remaining tests shown in \cref{fig:model_variations} are all consistent with the baseline measurements to within the $1/3\,\sigma_\mathrm{DESI}$ threshold. Also note that the two variations that relate to the isotropic scale parameter, \alphas, both produce significantly smaller uncertainties in both \phif\ and \fsig\ compared to the baseline analysis, and produce shifts that are close to our threshold. For these two constraints, we have validated the results by running the sampler as well. We discuss these further in \cref{sec:discussion}.

\section{Discussion} \label{sec:discussion}

The measurements presented in this work represent the tightest \alpac\ constraint at $z>1$ to date, and the first measurement of \fsig\ from the \lyaf\ cross-correlation with quasars. To validate these results, we performed a large set of tests, including mock analyses, data splits, and analysis and model variations. In this section, we wish to discuss our measurements and the results of the validation tests.

As the starting point of our analysis was from the correlation functions measured by \kplya, we largely focused our validation on the modelling of correlations, and all analysis choices we made were at the level of the model. Therefore, we did not investigate in detail the impact of changes to the analysis methodology prior to the measurement of the correlations. The tests shown in \cref{fig:analysis_variations} are generally consistent with the baseline result, with no shifts larger than $0.5\sigma$ for \phif\ and $0.6\sigma$ for \fsig. Furthermore, for all of the larger shifts, we expect some variation due to statistical fluctuations caused by either changing the dataset or changing the weights used to compute the correlations. Future analyses that also consider changes to the analysis method prior to the measurement of the correlations would also need to study whether the observed shifts are consistent with statistical fluctuations, and if any of them are not, study changes to the analysis method that address those shifts. Also, as the current \lyaf\ analysis framework was built over many iterations with the goal of maximizing BAO signal-to-noise, it would be interesting to consider changes that instead minimize the impact of contaminants on full-shape analyses. For example, 
in-depth tests of the impact of DLA and BAL masking on full-shape measurements \citep[similar to][]{Y3.lya-s2.Brodzeller.2025,KP6s9-Martini} could inform more conservative analysis choices.

Our validation process was also limited by the use of the fitter with Gaussian uncertainties, instead of full posterior sampling. As discussed in Appendix \ref{sec:fitter}, the primary difference is caused by the fitter not fully marginalizing over nuisance parameters. This could introduce extra variance in the set of validation tests, especially for analysis variations where the shape of the posterior changes significantly compared to the baseline. One way to minimize this impact would be to impose Gaussian priors on all parameters that are not well constrained by the data. This is already done with some of the less important parameters ($b_\mathrm{CIV}$, $\beta_\mathrm{HCD}$, $\Delta r_{||}$), but those are all informed by direct measurements from other datasets. The parameters that are unconstrained by the data in our analysis are the Arinyo $a_v$, $b_v$, $k_v$, and $k_p$ parameters, as well as the typical scale of un-masked HCDs, $L_\mathrm{HCD}$. Priors on these parameters could be informed by studies with mocks in the case of $L_\mathrm{HCD}$, or with simulations in the case of the Arinyo parameters. However, these are beyond the scope of the current work.

Another important aspect of our analysis that differentiates it from galaxy clustering analyses, is that we only measure the isotropic scale from the peak component while marginalizing over the isotropic scale of the broadband, \alphas. This was a central part of the framework introduced by \cite{Cuceu:2021}, and it allows us to marginalize over potential systematics that isotropically affect the broadband. However, this also introduces a complication, because we still need this parameter to rescale our \fsig\ measurements as it affects the definition of $\sigma_8$ (\cref{subsec:scale_pars}). Furthermore, this parameter is correlated with both \phif\ ($\rho=0.38$) and with \fsig\ ($\rho=0.53$), meaning that systematics that affect it could still spill over into our primary measurements as well. Because \alphas\ has so far been treated as a nuisance parameter, not much attention was given to validating its constraint. In our analysis we have found that it is very weakly constrained when compared to \alphap\ ($3.5\%$ for \alphas\ versus $1\%$ for \alphap), but still consistent with the fiducial cosmology ($\alpha_\mathrm{s}=1.040\pm0.035$). Therefore, the approach of fitting for a single full-shape isotropic scale may be preferred by future analyses. This does mean that the resulting \alphaf\ measurement would also need to be validated as part of full-shape analyses (similar to what is done for galaxy clustering). However, this measurement would be dominated by the BAO signal given the difference in constraining power on \alphap\ and \alphas, and therefore likely to still be robust. Finally, if this approach was taken for future analyses, the resulting constraints would be roughly $9\%$ tighter for \phif\ and $18\%$ tighter for \fsig\ based on our tests.

One of the most important parts of the validation process is showing that we can recover unbiased constraints using mocks. In \cref{subsec:mocks} we found that \phif\ constraints are unbiased for both \lyacolore\ and \saclay\ mocks, while in the case of \fsig\ only constraints from the \lyacolore\ mocks were consistent with the input cosmology. Both sets of mocks contain all the major contaminants known to affect \lyaf\ correlations (HCDs, metals, BALs, redshift errors). These contaminants are injected in the same way for both types of mocks \citep{Herrera-Alcantar:2024,KP6s6-Cuceu}, and the analysis and modelling are also the same. Therefore, we consider it unlikely that the observed difference in \fsig\ is caused by contaminants. However, this is something that could be tested in future analyses by creating and analysing a set of uncontaminated mocks.

The impact of the fiducial cosmology was tested by \kplya\ for \alphap\ in the context of DESI DR1. For \phif, \citet{Cuceu:2023a} studied this using eBOSS DR16 mocks and found no significant impact. Given that our statistical significance is similar to that of eBOSS DR16, we use the same mock and analysis pipelines, and our \phif\ result is within $1\sigma$ of the fiducial cosmology, we expect this conclusion to hold for DESI DR1 as well. On the other hand, this test has not been done for \fsig. As we were not able to fully validate this measurement with mocks, it is unclear whether a test of the impact of the fiducial cosmology on \fsig\ constraints using the current generation of mocks could be trusted. Therefore, in the case of \fsig, we leave these tests for future work.

Arguably, the most important limitation with the current \lyaf\ mocks is the fact that they rely on log-normal approximation, which means they cannot be used to test the impact of small-scale deviations from linear theory. On one hand, our analysis focuses on large scales where linear theory holds well \citep[e.g.][]{Hadzhiyska2023}. Also, the onset of non-linearities happens at smaller scales for \lya\ when compared to galaxy clustering, both due to the higher redshift and because it traces significantly less dense environments. On the other hand, a number of effects introduce extra sensitivity to small-scale non-linearities, in particular the distortion matrix and contamination due to metals \citep{Bautista:2017,Busca:2025}. The first spreads small-scale information to all scales along the line-of-sight, while the second manifests as small-scale \lya-metal and QSO-metal correlations appearing as extra peaks along the line-of-sight. This means that incorrect modelling of the small scales in mocks can have an impact on the validation.

In the case of the \lya\ auto-correlation, the Arinyo model gives us a way of gauging the importance of deviations from linear theory. We find that the data prefers values of the Arinyo $q_1$ parameter consistent with zero (Appendix \ref{sec:nuisance}), and disfavours large deviations from linear theory over the scales we fit. Furthermore, we do not find significant correlations between any of the Arinyo parameters and \phif\ and \fsig, nor with any of the other nuisance parameters.
On the other hand, this type of test is more difficult for the cross-correlation, because our model for small-scale deviations from linear theory in the cross-correlation is very simplistic, and dominated by the impact of redshift errors. Furthermore, the small-scale quasar clustering in the \lyacolore\ mocks is significantly different than the one measured from data and N-body simulations \citep{Ramirez-Perez:2022,Youles:2022,Y3.lya-s1.Casas.2025}. These facts led us to take a much more conservative approach in our analysis of the cross-correlation, with the minimum scale used being $r > 40$ \hMpc, as opposed to $r > 25$ \hMpc\ for the auto-correlation. Nevertheless, we find that the \fsig\ measurement is somewhat correlated with both $\sigma_z$ ($\rho=0.56$) and $a_{TP}$ ($\rho=-0.39$), indicating it is still fairly sensitive to the modelling of small-scale deviations from linear theory.

When comparing the realism of the mock correlation functions, \cite{KP6s6-Cuceu} found that the stacked cross-correlation of the DESI DR1 \saclay\ mocks matches the DESI data better when compared to the \lyacolore\ one. On the other hand, the \lyacolore\ stacked auto-correlation performs better than the one from \saclay\ mocks. However, in the end both sets of mocks are well fit by our model \citep[see Figures 6 and 7 in][]{KP6s6-Cuceu}. Given this, it is difficult to draw a clear conclusion from our \fsig\ results in \cref{subsec:mocks}. While the \fsig\ signal comes from the cross-correlation, this measurement relies on the effective \lya\ RSD term being calibrated by the auto-correlation. If we only used the cross-correlation alone, the two would be degenerate. Based on this, we conclude that at this stage we cannot fully confirm an unbiased \fsig\ measurement, or provide a well-understood measurement of a systematic uncertainty for this parameter. Further study with more realistic mocks that go beyond the log-normal approximation is needed to fully validate an \fsig\ measurement from the \lya\ forest \citep[e.g.][]{Hadzhiyska2023,Sinigaglia:2023}. Therefore, in the case of the present analysis, we decided before unblinding that we would not use the resulting \fsig\ constraint for cosmology inference.

\begin{figure}
\centering
\includegraphics[width=1.\columnwidth,keepaspectratio]{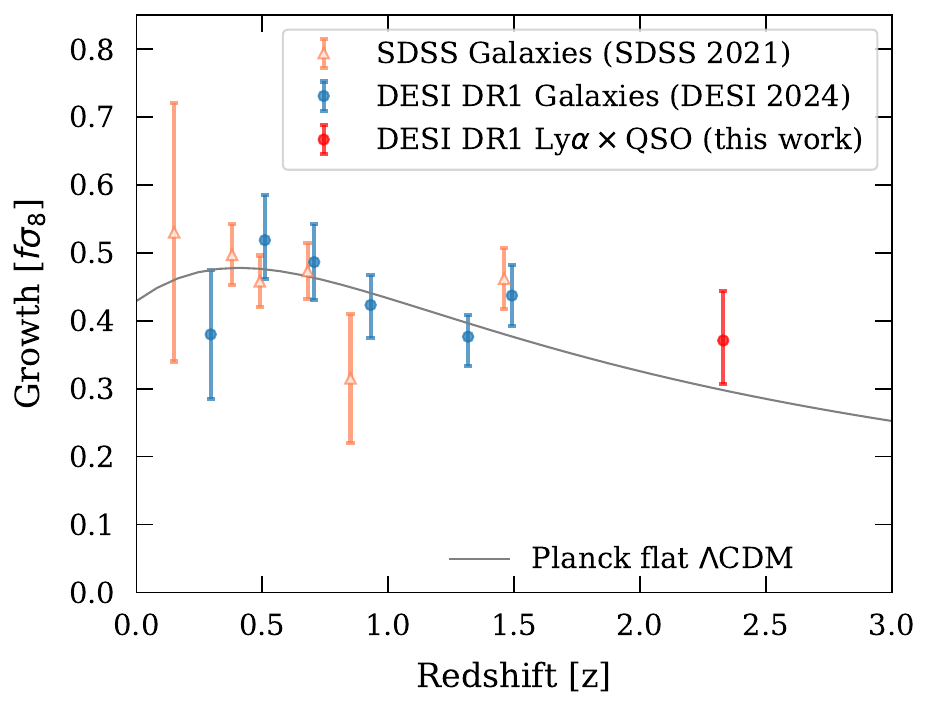}
\caption{Measurements of the growth rate times the amplitude of fluctuations, \fsig, as a function of redshift. The prediction from the Planck CMB data assuming the \lcdm\ model is shown as the gray line. Measurements from galaxy clustering are shown in orange for the Sloan Digital Sky Survey (SDSS) and in blue for DESI. Our constraint using the cross-correlation between the \lyaf\ and quasars is shown in red, and it includes an additional systematic uncertainty based on results from mocks.}
\label{fig:growth}
\end{figure}

To put our \fsig\ measurement in context, in \cref{fig:growth} we compare it to constraints from galaxy clustering at lower redshifts \citep{Alam:2017,Alam:2021,DESI2024.V.KP5}. We include an extra systematic uncertainty based on the shift measured from the stack of \saclay\ mocks, but caution that this is not a well-understood systematic as discussed above. The result we obtain is:
\begin{equation}
    f\sigma_8(z_\mathrm{eff}=2.33) = 0.37\; ^{+0.055}_{-0.065} \,(\mathrm{stat})\, \pm 0.033 \,(\mathrm{sys}).
\end{equation}
To our knowledge, this is the first direct measurement of \fsig\ from large-scale structure at $z>2$.

We now turn our attention back to the validation of the AP measurement. In this case, both sets of mocks show unbiased results. Furthermore, we found that the impact of spurious correlations due to redshift errors \citep{Youles:2022} is much smaller than the current statistical uncertainties, and well within the threshold of $1/3\,\sigma_\mathrm{DESI}$ (\cref{fig:mock_stack}). A model for this contamination was recently introduced by \cite{Gordon:2025}. This was built by propagating the effect of redshift errors from the smearing of the mean quasar continuum to the resulting spurious correlations. Using the same set of mocks we use here, they showed that this model can correct the small shift observed in AP and RSD. As the statistical uncertainty of our measurements will shrink with the upcoming DESI data releases, these spurious correlations are likely to become a dominant source of systematic uncertainty, and this model will likely play an important role in future such analyses. Indeed, it has already become important for the DESI DR2 BAO analysis \citep{Y3.lya-s1.Casas.2025}. However, at the level of the DR1 full-shape analysis, this systematic is still small, and therefore, we did not use the model developed by \cite{Gordon:2025}.

While the mocks we use contain all the major contaminants known to affect \lyaf\ correlations, they do not include a number of the less significant effects \citep[see][]{Cuceu:2023a,KP6s6-Cuceu}. These include CIV contamination, the transverse proximity effect, BAO broadening due to non-linear evolution, UV background fluctuations, and correlated sky residuals. In the case of CIV contamination, the test shown in \cref{fig:model_variations}, labeled \textit{Weak CIV prior}, shows that the impact of this contamination is negligible. Without the Gaussian prior, we find that the CIV bias parameter is consistent with zero (within $1\sigma$), indicating no detection of CIV contamination in the \lyaf\ correlations given current uncertainties. The transverse proximity effect only impacts the cross-correlation, and as discussed in \cref{subsec:datasplits}, the cross is less constraining than the auto when it comes to \phif\ measurements. The results from the auto and cross are also consistent, as shown in \cref{fig:auto_cross}, indicating that this contamination is not a cause for concern in case of AP measurements. The BAO broadening due to non-linear evolution only impacts the peak component, while most of our AP information comes from scales smaller than the BAO scale. It does affect the measurement of \alphap\ and \phip\, but those were validated as part of the BAO analysis by \kplya.

UV background fluctuations introduce a scale-dependent bias \citep{Pontzen:2014,GontchoAGontcho:2014}. This is not present in our baseline analysis as it has not been detected at a significant level so far. In \cref{subsec:model_variations}, we perform a test where we model this effect following \cite{GontchoAGontcho:2014} and \cite{Bautista:2017}. Similar to previous works \citep[e.g.][]{Bautista:2017,Cuceu:2023b,DESI2024.IV.KP6}, we do not detect it at a significant level (within $1\sigma$ of zero). We also find no impact on the AP measurement (\cref{fig:model_variations}).

Correlated sky residuals were studied by \cite{KP6s5-Guy} using a synthetic data set created from the DESI DR1 data by replacing the original \lya\ $\delta$ with realizations of the sky subtraction residuals. The correlation function of this mock data set gives the contamination expected in \lyaf\ correlations. \cite{KP6s5-Guy} then built a model for this contamination and showed that it fits the measurement from the mock data set very well. This model is used both here and in \kplya, and it has a free amplitude parameter, $a_\mathrm{noise}$. This parameter does have a mild correlation with \phif\ ($\rho=0.24$). However, given that this contamination is very well understood and modelled, we do not consider it a significant source of concern. Nevertheless, this effect should be prioritized for inclusion in future mock data sets.

In conclusion, we have found that the measurement of the AP effect is robust, while the measurement of \fsig\ could not be fully validated due to deficiencies with the mock data sets. Therefore, in the following section we only use our measurements of $\phi$ to derive cosmological constraints.

\section{Cosmological interpretation} \label{sec:cosmo}

In this section, we study the cosmological implications of our \alpac\ measurement, alone and in combination with other data sets. We begin in \cref{subsec:cosmo_dist} with the constraints on combinations of cosmic distances corresponding to our measurement. After that, we briefly introduce the external data sets we use in \cref{subsec:cosmo_other}, and then present our results using the \lcdm\ model in \cref{subsec:lcdm}. Finally, we present constraints on the equation of state of dark energy in \cref{subsec:dark_energy}.

\subsection{Distance measurements} \label{subsec:cosmo_dist}

\begin{figure}
\centering
\includegraphics[width=1.\columnwidth,keepaspectratio]{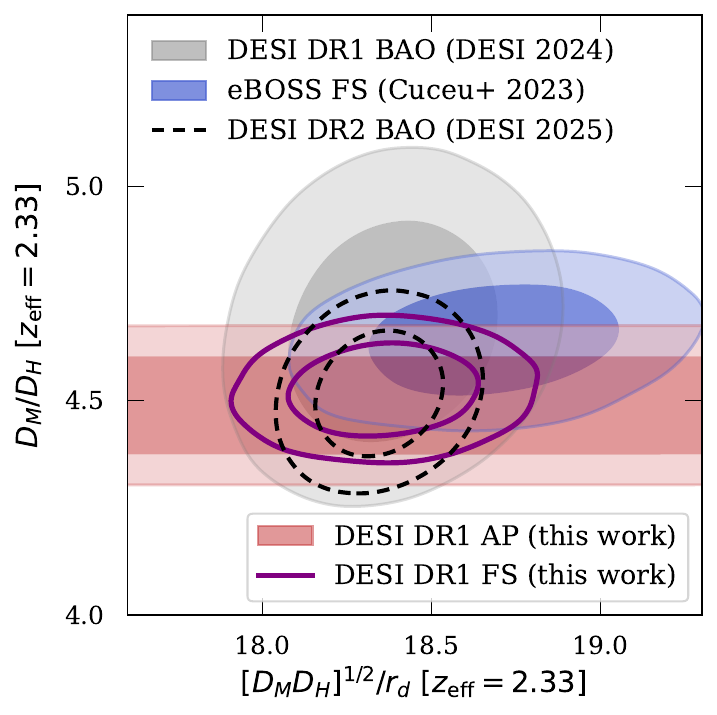}
\caption{Constraints on the \alpac\ parameter given by the distance ratio $D_M/D_H$, and the distance combination corresponding to our isotropic BAO measurement, $\left[D_M D_H(z_\mathrm{eff})\right]^{1/2} / r_d$. DESI \lya\ BAO results are shown in grey for DR1 (\kplya) and in black dashed for DR2 \citep{DESI.DR2.BAO.lya}, while the eBOSS \lya\ full-shape results from \cite{Cuceu:2023b} are shown in blue. Our broadband-only AP constraint is given by the horizontal bands in red, while the full-shape result, which includes broadband AP and BAO information, is shown in purple.}
\label{fig:dmdh}
\end{figure}

We first focus on the relevant distance combinations for our AP and isotropic BAO measurements, which are given by \cref{eq:ap_dist,eq:bao_dist}, respectively. The measurements corresponding to our full-shape results are:
\begin{equation}
\left\{
\begin{array}{ll}
    \mathbf{AP}:&\; D_M / D_H (z_\mathrm{eff}) = 4.525 \pm 0.071, \\
    \mathbf{BAO}:&\; \left[D_M D_H(z_\mathrm{eff})\right]^{1/2} / r_d = 18.36 \pm 0.18,
\end{array}
\right.
\end{equation}
with the effective redshift of our measurements equal to $z_\mathrm{eff}=2.33$, and the correlation coefficient between the two distance combinations equal to $0.09$. We also compute the distance ratio corresponding to our AP measurement from the broadband (BB) alone:
\begin{equation}
    \mathbf{broadband\;AP}:\; D_M / D_H (z_\mathrm{eff}) = 4.489 \pm 0.076.
\end{equation}
In this case, we marginalize over the scale parameters of the peak component. We will use this constraint when combining our measurement with DESI DR2 \lya\ BAO.

In \cref{fig:dmdh}, we compare our results with the DR1 and DR2 DESI \lya\ BAO measurements \citep{DESI2024.IV.KP6,DESI.DR2.BAO.lya}, and with eBOSS \lya\ full-shape \citep{Cuceu:2023b}. Our broadband-only AP constraint is shown as the horizontal shaded area in red. The DESI full-shape constraint is shown in purple, and it contains information from both the broadband through the AP effect and from BAO. Adding the AP information from the broadband improves upon the BAO constraint by an impressive factor of $2.4\times$ (purple versus grey). Furthermore, the full-shape \alpac\ result from DESI DR1 is $37\%$ tighter than the DESI DR2 BAO constraint (purple versus black), with the two measurements being in excellent agreement. While the BAO information is highly correlated between DR1 and DR2, the broadband AP measurement is independent of both DESI DR1 and DR2 BAO as discussed in Appendix \ref{sec:independence}.

When compared to the results from the eBOSS full-shape analysis (blue in \cref{fig:dmdh}), our constraints prefer slightly lower values in both the isotropic and the AP direction. We have computed the correlation coefficient between our results and those from eBOSS using the same method as described in Appendix F of \kplya, and in Appendix \ref{sec:independence} below. We find a correlation coefficient of $13\%$, which is very similar to the $10\%$ correlation \kplya\ found between the DESI DR1 BAO and eBOSS BAO results. Given that the correlation is small, we conclude that the eBOSS and DESI full-shape results are consistent.

We also provide our constraints of the more common distance combinations, $D_M/r_d$ and $D_H/r_d$. The results from the full-shape \phif\ and \alphap\ measurements are:
\begin{equation}
\begin{array}{ll}
\text{DESI DR1}& \\
\text{full-shape}&
\end{array}
\left\{
\begin{array}{ll}
    D_H(z_\mathrm{eff})/r_d &= 8.632 \pm 0.105, \\
    D_M(z_\mathrm{eff})/r_d &= 39.05 \pm 0.52, \\
    \rho(D_M/r_d, D_H/r_d) &= -0.244,
\end{array}
\right. \label{eq:dmdh_dr1}
\end{equation}
where $\rho$ gives the correlation coefficient between $D_H/r_d$ and $D_M/r_d$. For comparison, the DESI DR2 BAO constraints from \cite{DESI.DR2.BAO.lya} are $D_H(z_\mathrm{eff})/r_d = 8.632 \pm 0.101$ and $D_M(z_\mathrm{eff})/r_d = 38.99 \pm 0.53$ ($\rho=-0.43$), again showing the excellent agreement between the two. 
Both our analysis and that of \cite{DESI.DR2.BAO.lya} were performed blinded, and there were no modifications that impact the position of either constraint post-unblinding. Our analysis was unblinded about five months prior to the unblinding of the DESI DR2 BAO measurement. When compared to the distances measured by \kplya\ using DESI DR1 BAO, our uncertainties are $35\%$ smaller for $D_H/r_d$ and $43\%$ smaller for $D_M/r_d$.

As our broadband \alpac\ constraint is independent from the DESI DR2 BAO measurement, we also combine the two, and obtain the following distance ratios:
\begin{equation}
\begin{array}{ll}
\text{DR2 BAO}& \\
\text{+ DR1 AP}&
\end{array}
\left\{
\begin{array}{ll}
    D_H(z_\mathrm{eff})/r_d &= 8.646 \pm 0.077, \\
    D_M(z_\mathrm{eff})/r_d &= 38.90 \pm 0.38, \\
    \rho(D_M/r_d, D_H/r_d) &= -0.016.
\end{array}
\right. \label{eq:dmdh_dr1_dr2}
\end{equation}
This corresponds to a $0.9\%$ measurement of $D_H/r_d$, and a $1\%$ measurement of $D_M/r_d$.

In the following sections, we will use both the DESI DR1 full-shape constraints from \cref{eq:dmdh_dr1}, and the combination of the DR1 broadband AP and DR2 BAO constraints from \cref{eq:dmdh_dr1_dr2}. We will refer to constraints that use \cref{eq:dmdh_dr1} as \textit{\lya-FS} , those that use \cref{eq:dmdh_dr1_dr2} as \textit{Best-\lya}, and those that only use the \lya\ broadband AP constraint as \textit{\lya-AP}. 

\subsection{Other datasets} \label{subsec:cosmo_other}

For the cosmological analysis, we combine our measurement with results from a number of external probes in order to obtain the tightest constraints. These closely match the probes used in \cite{DESI.DR2.BAO.cosmo}, and they also describe the motivation behind these choices in more detail.

Besides the DESI DR2 \lyaf\ BAO measurement, described above, we also use DESI BAO measurements from galaxy clustering at lower redshifts. We use the latest results from DESI DR2 presented in \cite{DESI.DR2.BAO.cosmo}. These consist of $D_M/r_d$ and $D_H/r_d$ measurements in 5 redshift bins that span the range $0.5 < z < 1.5$, and one isotropic BAO constraint at $z=0.295$.

In order to calibrate the scale of the sound horizon, $r_d$, BAO data needs to be combined with an external measurement of the baryon density $\Omega_b h^2$. This allows us to constrain the Hubble constant $H_0$, by breaking its inherent degeneracy with $r_d$. We use the $\Omega_b h^2$ estimate from \cite{Schoneberg:2024}, which is based on Big Bang Nucleosynthesis (BBN) and measurements of the primordial deuterium abundance \citep{Cooke:2018}, and is given by:
\begin{equation}
    \Omega_b h^2 = 0.02218 \pm 0.00055.
    \label{eq:bbn}
\end{equation}
This is based on the \texttt{PRyMordial} code \citep{Burns:2024}, and includes marginalisation over uncertainties in nuclear reaction rates.

We use CMB measurements from Planck \citep{Planck:2020gen,Planck:2020} and the Atacama Cosmology Telescope \citep[ACT; ][]{Madhavacheril:2024,Qu:2024,MacCrann:2024}. This includes the Planck temperature (TT), polarization (EE) and cross (TE) power spectra using the \texttt{simall}, \texttt{Commander} (for $\ell<30$) and \texttt{CamSpec} likelihoods \citep[for $\ell\geq30$; ][]{Efstathiou:2021,Rosenberg:2022}, and the combination of Planck and ACT DR6 CMB lensing from \cite{Madhavacheril:2024}. \cite{DESI.DR2.BAO.cosmo} found that using some of the other CMB likelihoods available in the literature \citep[e.g.][]{Tristram:2024} does not significantly impact dark energy constraints. Therefore, in this work we only focus on combinations with the likelihoods mentioned above, and refer the reader to Appendix A of \cite{DESI.DR2.BAO.cosmo} and to \cite{GarciaQuintero:2025}, for discussions of the impact of different CMB likelihoods on current dark energy constraints.

Finally, for constraints on the equation-of-state of dark energy, we also combine our measurement with results from Type Ia supernovae (SNe) analyses. Following \cite{DESI2024.VI.KP7A} and \cite{DESI.DR2.BAO.cosmo}, we study combinations with each of the three SNe samples from Pantheon+ \citep{Scolnic:2022,Brout:2022}, Union3 \citep{Rubin:2023}, and the Dark Energy Survey Year 5 \citep[DESY5; ][]{DES:2024}. The Pantheon+ and Union3 samples consist of spectroscopically-classified SNe drawn from multiple observational programs. They have a significant fraction of objects in common, but use different calibration, modelling and treatment of systematic uncertainties. On the other hand, DESY5 consists of a large sample of photometrically-classified SNe at $z>0.1$, with uniform calibration based on a single survey, and a small set of SNe at $z<0.1$, which are drawn from historical sources. We use the publicly available Cobaya likelihoods for the three SNe samples.\footnote{\url{https://github.com/CobayaSampler/sn_data}}

\subsection{\lcdm\ constraints} \label{subsec:lcdm}

We start with cosmological constraints in the \lcdm\ model, focusing on the measurements from \lya\ alone (\cref{subsec:cosmo_dist}) and in combination with DESI DR2 BAO (\cref{subsec:cosmo_other}).  We use the \texttt{Cobaya} package \citep{Torrado:2019,Torrado:2021} and the associated Markov Chain Monte Carlo Metropolis sampler \citep{Lewis:2002,Lewis:2013} for cosmological inference.  Chains are run until the convergence criterion based on the Gelman-Rubin statistic \citep{Gelman:1992} satisfies $R-1<0.01$. Theory models are computed using \texttt{CAMB} and \texttt{Cobaya} likelihoods, and we use the same priors as \cite{DESI.DR2.BAO.cosmo} for our cosmological analyses.

In \lcdm, measurements of $D_M/r_d$ and $D_H/r_d$ are sensitive only to two cosmological parameters: $\Omega_m$ and the degenerate combination $hr_\mathrm{d}$, where $h=H_0/(100 \text{ km }\text{s}^{-1}\; \text{Mpc}^{-1})$. Therefore, we directly sample these two parameters. The constraints from DESI DR1 full-shape are:
\begin{equation}
\left.
\begin{aligned}
    \Omega_{\mathrm{m}} &=  0.309^{+0.024}_{-0.028} \\
    hr_\mathrm{d} &= (100.1\pm 3.4) \text{ Mpc}
\end{aligned}
\right\} \text{DESI DR1 Ly$\alpha$-FS},
\end{equation}\\
with a correlation coefficient $\rho=-0.95$. This is consistent with the individual DESI DR2 BAO measurements, as well as the combined measurement presented in \cite{DESI.DR2.BAO.cosmo}.

The constraints from the combination of DESI DR2 \lya\ BAO and DR1 \lya\ AP, which we denote \textit{Best-\lya}, are:
\begin{equation}
\left.
\begin{aligned}
    \Omega_{\mathrm{m}} &=  0.299^{+0.019}_{-0.023} \\
    hr_\mathrm{d} &= (101.3\pm 2.8) \text{ Mpc}
\end{aligned}
\right\} \text{DESI Best-Ly$\alpha$},
\end{equation}
with $\rho=-0.97$. DR1 \lya\ full-shape gives an $\sim8\%$ constraint on the matter fraction, while the Best-\lya\ measurement gives a $\sim7\%$ constraint.

For comparison, the three SNe samples introduced above give $\Omega_m$ constraints of $\sim5\%$ (DESY5), $\sim7\%$ (Union3), and $\sim5\%$ (Pantheon+), respectively. This means the \lyaf\ measurements from DESI already provide constraints that are competitive with current SNe measurements. However, \lya\ prefers smaller $\Omega_m$ values when compared to the three SNe samples. Our Best-\lya\ measurement is $1.3\sigma$ lower than that from Pantheon+, with the difference increasing to $1.7\sigma$ for Union3, and $2\sigma$ for DESY5. While these differences are not statistically significant, they follow the trend of DESI BAO measurements \citep{DESI2024.VI.KP7A,DESI.DR2.BAO.cosmo}, which also prefer smaller $\Omega_m$ values compared to SNe.

\begin{figure}
    \centering
    \includegraphics[width=1.\columnwidth,keepaspectratio]{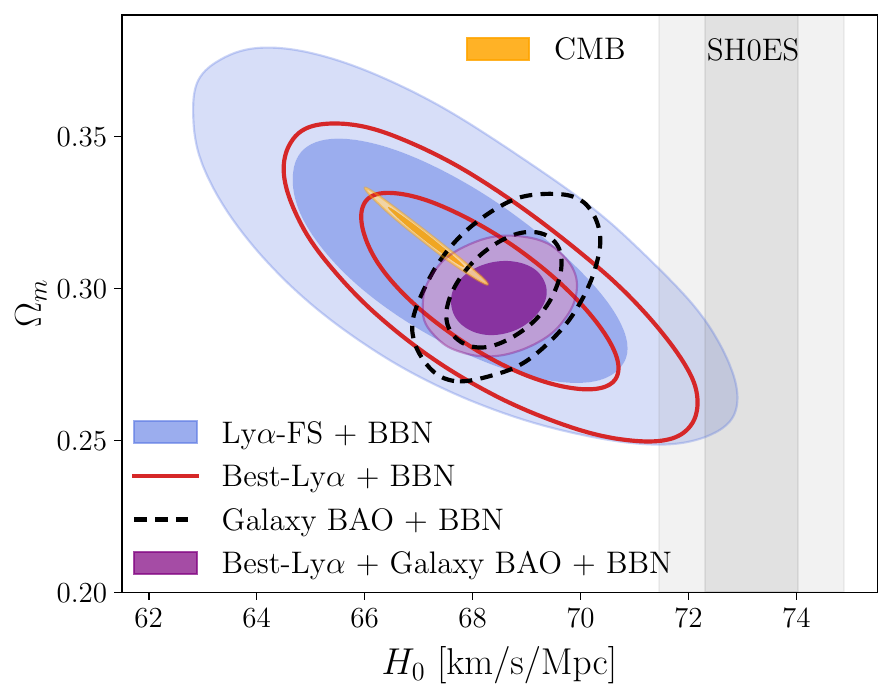}
    \caption{Constraints on the Hubble constant and the matter fraction within the \lcdm\ model from DESI measurements combined with an $\Omega_b h^2$ prior from BBN, which helps break the degeneracy in $hr_\mathrm{d}$. The results from DR1 \lya\ full-shape are shown in blue, while the Best-\lya\ combination is shown in red, and includes DR1 \lya\ AP and DR2 \lya\ BAO. The \lya\ constraints agree well with both the DESI DR2 galaxy BAO results (black) and the CMB (orange), but are in tension with SH0ES (grey). The combined result in purple includes all DESI DR2 BAO measurements from both galaxies and \lya\, and the DR1 \lya\ AP constraint.}
    \label{fig:lcdm}
\end{figure}

Our measurements are fully consistent with the CMB constraints on $\Omega_m$ and $hr_\mathrm{d}$, and therefore cannot currently provide insight into the small tension reported in \cite{DESI.DR2.BAO.cosmo} between CMB and DESI BAO constraints. We discuss this later in this section in the context of \cref{fig:lcdm}.

We next combine our broadband AP measurement (denoted \textit{\lya-AP}) with all of the DESI DR2 BAO results (now including galaxy BAO). We will refer to this combination as either "\lya-AP$+$BAO" or "Best-\lya$+$Galaxy BAO", but the two are equivalent. The result we obtain is:
\begin{equation}
\left.
\begin{aligned}
    \Omega_{\mathrm{m}} &=  0.2972\pm 0.0081, \\
    hr_\mathrm{d} &= (101.55\pm 0.70) \text{ Mpc}
\end{aligned}
\right\}
\begin{array}{ll}
\;\;\;\;\;\;\;\text{Best-\lya}& \\
\;\text{+ Galaxy BAO},&
\end{array}
\end{equation}
with $\rho=-0.92$. This corresponds to a $6\%$ tighter constraint on $\Omega_m$, and $4\%$ tighter on $hr_\mathrm{d}$ when compared to DESI DR2 BAO alone. Given that we are only adding the DR1 \lya\ AP constraint on top of the combined DESI DR2 BAO measurement, we consider that even this small improvement is fairly impressive, and shows the potential of \lya\ AP measurements with future DESI data releases. However, given there is no significant shift in the result when adding \lya\ AP, the tensions between this constraint and SNe $\Omega_m$ results remain the same as the ones reported in \cite{DESI.DR2.BAO.cosmo}.

In order to break the degeneracy in $hr_\mathrm{d}$, we need to calibrate the relative BAO distance measurements. This can be achieved by adding an $\Omega_b h^2$ prior from BBN, which, along with our $\Omega_m$ measurement, helps constrain $r_\mathrm{d}$, and results in a measurement of the Hubble constant $H_0$. The constraint from DR1 \lya\ FS is:
\begin{equation}
    H_0 = (67.7\pm 2.1)\;\kmsMpc\;\text{(\lya-FS + BBN)},
\end{equation}
while the constraint from the Best-\lya\ measurement is:
\begin{equation}
    H_0 = (68.3\pm 1.6)\;\kmsMpc\;\text{(Best-\lya\ + BBN)}.
\end{equation}
These represent $3.1\%$ and $2.3\%$ constraints on $H_0$, respectively.

We compare our measurements with those from the CMB, DESI DR2 galaxy BAO, and SH0ES \citep{Breuval:2024}, in \cref{fig:lcdm}. The \lya\ constraints (blue and red) are in good agreement with both the CMB and DESI galaxy BAO measurements. This plot shows that the $2.3\sigma$ tension reported in \cite{DESI.DR2.BAO.cosmo} between DESI BAO and the CMB is driven by DESI galaxies. Interestingly, the \lya\ results align well with the direction of the tension due to the high redshift of the \lyaf, and because the \lya\ isotropic BAO constraint (\alphap) is in excellent agreement with the CMB prediction in \lcdm. Significant improvements in the \lya\ AP constraint (e.g. from future DESI data releases) would shrink the major axis of the \lya\ ellipses in \cref{fig:lcdm}, and could shed light on this tension. We discuss this further in \cref{subsec:future}.

While the \lya\ measurements agree with the CMB constraints, they are significantly lower than the cosmic distance ladder results from SH0ES \citep{Breuval:2024}, which are shown using the vertical grey bands in \cref{fig:lcdm}. The two \lya\ constraints are in $2.4\sigma$ (\lya FS) and $2.7\sigma$ (Best-\lya) tension with SH0ES.

Finally, when combining our \lya\ AP constraint with all DESI DR2 BAO measurements, we obtain an $H_0$ value of: 
\begin{equation}
\begin{aligned}
    &\;\;\;\;\; H_0 = (68.49\pm 0.58)\;\kmsMpc\; \\
    &\text{(DESI DR2 BAO + DR1 Ly$\alpha$-AP + BBN)}.
\end{aligned}
\end{equation}
This is almost identical with the constraint from DESI DR2 BAO alone, which means both the tension with the CMB ($2.3\sigma$) and with SH0ES ($4.5\sigma$) remain the same as reported in \cite{DESI.DR2.BAO.cosmo}. 

Our results show that a weaker Hubble tension of the same sign persists within \lcdm\ using only DESI \lyaf\ data ($2.7\sigma$) or only DESI BAO plus \lya-AP ($4.5\sigma$), with no CMB data. These constraints do still assume standard pre-recombination physics to compute $r_d$, and they rely on the BBN prior of \cref{eq:bbn}. Therefore, early dark energy or other departures from \lcdm\ that affect the sound horizon $r_d$ would change the level of $H_0$ tension for both DESI results and for the CMB.

\subsection{Dark energy} \label{subsec:dark_energy}

Next, we turn our attention to constraints on the nature of dark energy, focusing on measurements of its equation-of-state parameter: $w(z)=P(z)/\rho(z)$, where $P(z)$ is the dark energy pressure, and $\rho(z)$ is its energy density. We use the common Chevallier-Polarksy-Linder parametric model \citep{Chevallier:2001,Linder:2003}, where the time-evolution of the equation-of-state as a function of the expansion factor $a=(1+z)^{-1}$, is given by:
\begin{equation}
    w(a) = w_0 + w_a(1-a),
\end{equation}
where $w_0$ is the equation-of-state today and $w_a$ describes its rate of change, with $w\sim w_0 + w_a$ in the distant past. Following \cite{DESI.DR2.BAO.cosmo}, we use wide flat priors with $w_0\in\mathcal{U}[-3, 1]$, $w_a\in\mathcal{U}[-3, 2]$, and also impose the condition $w_0+w_a<0$ to enforce early matter domination. To quantify the statistical significance of the preference for evolving dark energy, we use the $\Delta\chi^2_\mathrm{MAP}$ between the best fit \lcdm\ and $w_0w_a$CDM models for a particular combination of datasets.

As the \lya\ measurements are at $z>2$, combinations with the CMB do not lead to significant constraints on $w_0-w_a$. This is because both results come from before the epoch of dark energy domination, which leads to similar parameter degeneracies. Instead, we need to combine our results with low-redshift probes, so \lya\ can play the role of the high-redshift anchor for the dark energy evolution (similar to the role played by the CMB). Here we consider two low-redshift probes: DESI Galaxy BAO and Type Ia Supernovae. We present $w_0-w_a$ constraints for various combinations of the four probes (\lya, Galaxy BAO, Supernovae, CMB), focusing on those where the \lya\ measurements play a significant role.

As our best measurement is given by the combination of DR1 \lya\ AP and DR2 \lya\ BAO (\textit{Best-\lya}), we will use this result for our dark energy constraints. We have also performed the same analysis using only the DR1 \lya\ full-shape constraints and found consistent but less constraining results.

\begin{figure}
    \centering
    \includegraphics[width=1.\columnwidth,keepaspectratio]{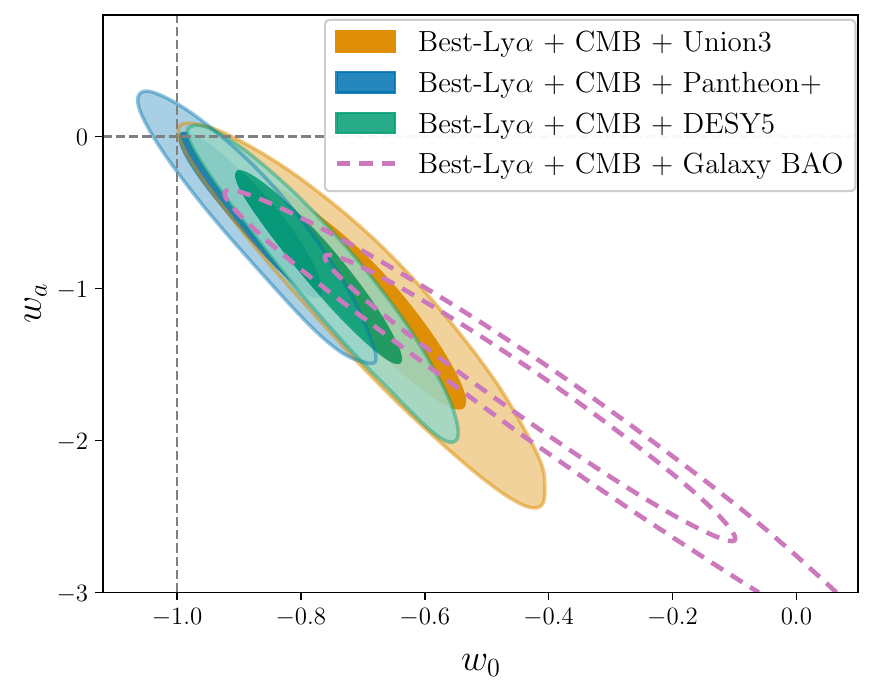}
    \caption{Posterior distributions of the dark energy equation-of-state parameters $w_0$ and $w_a$ when combining the Best-\lya\ measurement from DESI with the CMB and either of the three SNe datasets (filled contours), or with DESI Galaxy BAO constraints (dashed empty contour). The intersection of the gray dashed lines indicates the \lcdm\ limit ($w_0=-1$, $w_a=0$). The significance of rejection of \lcdm\ is not significant for combinations with SNe, ranging between $1.1\sigma$ and $2.4\sigma$. On the other hand, the combination with DESI Galaxy BAO shows a more significant $3.1\sigma$ deviation from \lcdm, at the same level found by \cite{DESI.DR2.BAO.cosmo}.}
    \label{fig:w0wa_1}
\end{figure}

\begin{table}[t]
    \centering
    \begin{tabular}{l|c|c}
        Datasets & $\Delta\chi^2_\mathrm{MAP}$ & Significance \\
         \hline \hline
        Best-\lya+Pantheon+ & $-1.8$ & $0.8 \sigma$ \\
        Best-\lya+Union3 & $-5.6$ & $1.9 \sigma$ \\
        Best-\lya+DESY5 & $-6.6$ & $2.1 \sigma$ \\
        \hline
        Best-\lya+CMB+Pantheon+ & $-2.5$ & $1.1 \sigma$ \\
        Best-\lya+CMB+Union3 & $-6.3$ & $2.0 \sigma$ \\
        Best-\lya+CMB+DESY5 & $-8.4$ & $2.4 \sigma$ \\
        \hline \hline
        \lya-AP+BAO & $-4.6$ & $1.6 \sigma$ \\
        \lya-AP+BAO+CMB & $-12.3$ & $3.1 \sigma$ \\
        \hline
        \lya-AP+BAO+CMB+Pantheon+ & $-10.8$ & $2.8 \sigma$ \\
        \lya-AP+BAO+CMB+Union3 & $-17.6$ & $3.8 \sigma$ \\
        \lya-AP+BAO+CMB+DESY5 & $-21.2$ & $4.2 \sigma$ \\
        \hline\hline
    \end{tabular}
    \caption{Difference in the effective $\chi^2_\mathrm{MAP}$ value of the best-fit $w_0w_a$CDM model and the best-fit \lcdm\ model for various combinations of datasets, along with the corresponding significance levels given two extra free parameters. Combinations labeled \lya-AP$+$BAO are equivalent to Best-\lya$+$Galaxy BAO.}
    \label{tab:de_significance}
\end{table}

In \cref{tab:de_significance}, we show $\Delta\chi^2_\mathrm{MAP}$ values and the associated frequentist significance of the preference for $w_0w_a$CDM over \lcdm\ given combinations between our measurement and various other data sets (described in \cref{subsec:cosmo_other}). The top half of the table shows results when combining the Best-\lya\ constraint with non-DESI probes, while the second half shows results when adding our broadband AP measurement to DESI DR2 BAO measurements (i.e., also including Galaxy BAO). We find that data combinations without DESI Galaxy BAO do not significantly exclude \lcdm, with the preference for $w_0w_a$CDM ranging between $0.8\sigma$ and $2.1\sigma$ when combining \lya\ with SNe, and slightly increasing to $1.1\sigma-2.4\sigma$ when also adding CMB data. The marginalized posterior  distributions of the equation-of-state parameters, $w_0$ and $w_a$, from the combination of \lya, CMB, and SNe are shown as filled contours in \cref{fig:w0wa_1}, and are given by:
\begin{equation}
\left.
\begin{aligned}
    w_{0} &= -0.874\pm 0.078 \\
    w_{a} &= -0.53^{+0.41}_{-0.32}
\end{aligned}
\right\} 
\begin{array}{ll}
\text{Best-\lya+CMB}& \\
\text{+Pantheon+},&
\end{array}
\end{equation}\\
when combined with Pantheon+,
\begin{equation}
\left.
\begin{aligned}
    w_{0} &= -0.70^{+0.11}_{-0.13} \\
    w_{a} &= -1.06^{+0.58}_{-0.43}
\end{aligned}
\right\} 
\begin{array}{ll}
\text{Best-\lya+CMB}& \\
\text{+Union3},&
\end{array}
\end{equation}\\
with Union3, and
\begin{equation}
\left.
\begin{aligned}
    w_{0} &= -0.764^{+0.084}_{-0.094} \\
    w_{a} &= -0.88^{+0.47}_{-0.37}
\end{aligned}
\right\} 
\begin{array}{ll}
\text{Best-\lya+CMB}& \\
\text{+DESY5},&
\end{array}
\end{equation}\\
for the combination with DESY5.

On the other hand, combinations that include Galaxy BAO (second half of \cref{tab:de_significance}) show more significant deviations from \lcdm, in line with the results presented in \cite{DESI.DR2.BAO.cosmo}. In fact the results presented in the second half of \cref{tab:de_significance} are almost identical to those based on DR2 BAO alone \citep{DESI.DR2.BAO.cosmo}. This is not surprising given that DESI DR2 BAO includes the BAO measurement from \lya, which provides a similar constraining power on AP as our DR1 broadband AP measurement here (see \cref{subsec:cosmo_dist}). 

The constraint from the combination of \lya, Galaxy BAO, and CMB is also shown in \cref{fig:w0wa_1} (empty contour). This constraint is not as tight as the ones using SNe, but shows a more pronounced deviation from \lcdm, at $3.1\sigma$ significance. The marginalized posterior for this combination is:
\begin{equation}
\left.
\begin{aligned}
    w_{0} &= -0.43\pm 0.21 \\
    w_{a} &= -1.72\pm 0.58
\end{aligned}
\right\} 
\begin{array}{ll}
\text{Best-\lya+CMB}& \\
\text{+Galaxy BAO},&
\end{array}
\end{equation}\\
which is very similar to the constraint from DESI DR2 BAO and CMB presented in \cite{DESI.DR2.BAO.cosmo}.

All the combinations shown in \cref{fig:w0wa_1} prefer a similar region in the $w_0-w_a$ space, with $w_0>-1$ and $w_a<0$, and have similar degeneracy directions, although not identical and not all pointed exactly at \lcdm. The regions where the empty contour intersects each of the three filled contours very roughly correspond to the joint posteriors obtained when combining \lya, Galaxy BAO, CMB, and SNe. These combinations give results very similar to those presented in \cite{DESI.DR2.BAO.cosmo}, with the same significance levels of the preference for $w_0w_a$CDM over \lcdm, ranging from $2.8\sigma$ to $4.2\sigma$ (\cref{tab:de_significance}).

\begin{figure}
    \centering
    \includegraphics[width=1.\columnwidth,keepaspectratio]{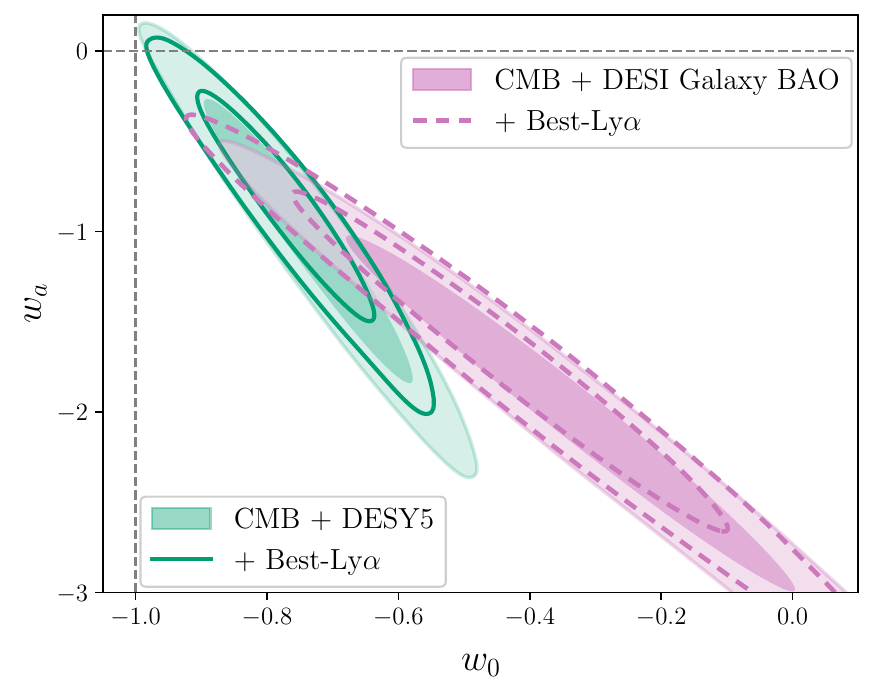}
    \caption{Similar to \cref{fig:w0wa_1}, but showing the impact of adding the DESI Best-\lya\ constraint to the combination of CMB and DESY5 SNe (green filled versus empty contours), and to the combination of CMB and DESI Galaxy BAO measurements (pink filled versus empty contours). In the first case, adding \lya\ leads to tighter constraints that are closer to \lcdm. For the second combination, adding \lya\ does not shrink the constraints, but does shift the posterior towards \lcdm, slightly reducing the tension from $3.3\sigma$ to $3.1\sigma$.}
    \label{fig:w0wa_2}
\end{figure}

To better understand the role played by \lya, in \cref{fig:w0wa_2} we show results from two of the combinations presented above, with and without the Best-\lya\ constraint. First, the green contours show the constraints from the combination of CMB and DESY5 SNe without \lya\ (filled) and with \lya\ (empty). In this case, adding \lya\ leads to tighter constraints, corresponding to $\sim22\%$ larger figure-of-merit, and an increase in the tension with \lcdm\ from $2.2\sigma$ to $2.4\sigma$. On the other hand, adding \lya\ to the combination of CMB and DESI Galaxies (pink filled versus empty contours) does not improve the constraints, but it does shift the posterior towards \lcdm, slightly reducing the tension from $3.3\sigma$ to $3.1\sigma$. \cref{fig:w0wa_2} also illustrates that shifts along the degeneracy direction for either of these combinations cannot fully reconcile the combined DESI, CMB, and DESY5 SNe constraint with \lcdm.

\subsection{Other constraints} \label{subsec:cosmo_other_res}

In this section, we present several other cosmological constraints that are improved by the addition of our \lya\ AP measurement. These include neutrino mass and curvature constraints.

Our AP measurement is not directly sensitive to massive neutrinos. However, it provides a geometrical measurement on $\Omega_m$, which can help improve CMB constraints on the neutrino mass. This is because the neutrino constraints from the CMB are highly correlated with several other parameters, including $\Omega_m$ \citep{Planck:2020,Loverde:2024}. This is exactly the same way that BAO measurements contribute to neutrino mass constraints \citep{DESI2024.VI.KP7A,DESI.DR2.BAO.cosmo}. 

We sample the sum of the neutrino masses, $\sum m_\nu$, assuming three degenerate mass eigenstates, and closely follow the analysis setup described in \cite{DESI.DR2.BAO.cosmo}. Within \lcdm, combining our Best-\lya\ constraint with the CMB produces the following upper bound:
\begin{equation}
    \sum m_\nu < 0.156\;\text{eV} \;\text{(95\%, Best-\lya+CMB)}.
\end{equation}
When adding our \lya-AP measurement to the combination of CMB and DESI DR2 BAO, we obtain:
\begin{equation}
    \sum m_\nu < 0.0638\;\text{eV} \;\text{(95\%, \lya-AP+BAO+CMB)},
\end{equation}
which is similar but slightly tighter than the constraint presented in \cite{DESI.DR2.BAO.cosmo}. When extending to $w_0w_a$CDM, our \lya-AP measurement does not change the constraints from BAO+CMB.
We note that these neutrino mass constraints do depend on the likelihood used for the CMB. Here we only use the \texttt{CamSpec} likelihood, but \cite{DESI.DR2.BAO.cosmo} and \cite{Y3.cpe-s2.Elbers.2025} present results using all three likelihoods currently available.

When allowing curvature to vary, \alpac\ measurements constrain a degenerate combination of $\Omega_m$ and the curvature fraction $\Omega_k$ \citep[see][]{Cuceu:2023b}. Combining our constraint with other AP and BAO measurements at lower redshifts helps break this degeneracy, leading to constraints on curvature. The result from the combination of \lya-AP and DESI DR2 BAO is given by:
\begin{equation}
    \Omega_k = 0.017\pm 0.036\;\text{(\lya-AP+BAO)}
\end{equation}
This is consistent with a flat Universe, and represents a $14\%$ tighter constraint compared to that from BAO alone. However, when combined with BAO+CMB, the \lya\ constraint does not change the result presented in \cite{DESI.DR2.BAO.cosmo}.

\subsection{Cosmology Discussion} \label{subsec:future}

We conclude this part with a discussion of our cosmological constraints, with an eye on the value added by the improved \lya\ forest \alpac\ measurement presented here, and the potential value of future such analyses. Within \lcdm, our measurement is consistent with both the CMB and DESI Galaxy BAO results, which means it cannot currently provide insight into the tension observed between the two. However, the disagreement is exactly along the direction constrained by the AP effect at high redshift, which means \lya\ full-shape analyses with future DESI data releases could shed light on this tension.

Within \lcdm, adding our DR1 \lya\ broadband AP to DESI DR2 BAO results in a $6\%$ reduction in the $\Omega_m$ uncertainty. The AP measurement also helps tighten upper bounds on the neutrino mass and DESI constraints on curvature. However, given that here we are combining the DR1 \lya\ AP constraint with DR2 BAO, the improvements are fairly small. We expect the improvements to be more significant with the next iteration of this analysis using DESI DR2.

\begin{figure*}
    \centering
    \includegraphics[width=0.8\textwidth,keepaspectratio]{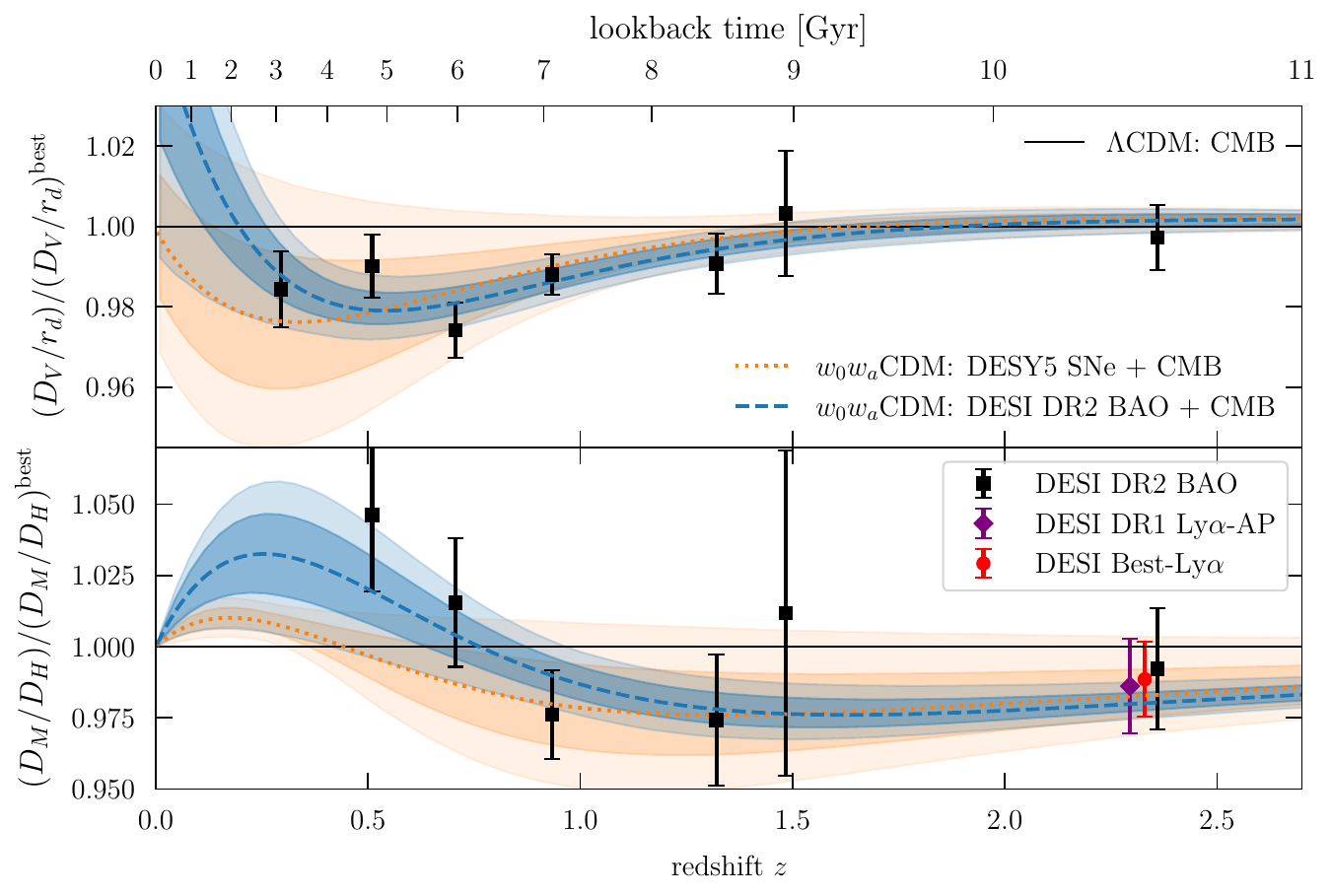}
    \caption{
        DESI constraints on the isotropic BAO distance $D_V/r_d$ (top), and on the \alpac\ parameter $D_M/D_H$ (bottom), relative to the CMB best-fit \lcdm\ model (black line). The DESI DR2 BAO measurements are shown in black, our DR1 \lya\ broadband AP constraint is shown in purple, and the Best-\lya\ constraint on AP is shown in red. For visualization purposes, the \lya\ constraints are shown at slightly different redshifts. The constraints within the $w_0w_a$CDM model from the combination of DESI BAO and CMB are shown in blue, with the dashed line indicating the best-fit model. We also show constraints from the combination of DESY5 SNe and CMB within $w_0w_a$CDM in orange (best-fit represented by the dotted line). This shows that within $w_0w_a$CDM, current data prefer AP values that are $\sim2\%$ lower than the CMB \lcdm\ predictions in the range $1 < z < 2.5$. Our Best-\lya\ combination gives a $1.3\%$ constraint on AP at $z=2.33$, but this falls right in between the \lcdm\ and $w_0w_a$CDM best-fit models. \cite{Cuceu:2021} forecasted \lya\ broadband AP constraints of $0.3\%-0.9\%$ with a 5-year DESI survey, meaning \lya\ could play an important role in future dark energy constraints from DESI.
    }
    \label{fig:dv_fap_evol}
\end{figure*}

When it comes to constraints on the dark energy equation-of-state parameters, we found that our \lya\ AP constraint is particularly useful in combination with Supernovae alone and when also including the CMB. These combinations produce fairly tight constraints on $w_0$ and $w_a$, rivaling those from the joint BAO and CMB analysis. On the other hand, adding our \lya\ AP constraint to combinations that contain DESI DR2 BAO does not lead to any significant changes in the posterior. This is in part because DESI DR2 BAO contains a \lya\ BAO measurement at the same redshift as our DR1 broadband AP constraint, and with a similar uncertainty ($1.7\%$ for DR1 broadband AP, and $2.1\%$ for DR2 BAO). However, even when we compared with the results from CMB and Galaxy BAO alone (i.e., without \lya), we found that adding \lya\ does not result in tighter constraints, but does produce a small shift towards \lcdm.

To gain a better understanding of these results, we project the $w_0w_a$CDM posterior distributions into the space of our measurements in \cref{fig:dv_fap_evol}. The figure shows the DESI constraints on the isotropic BAO distance $D_V/r_d$ in the top panel and on the \alpac\ parameter in the bottom panel, both normalized relative to the best-fit \lcdm\ model from the CMB. The actual measurements are represented by points with error bars, while the inferred posteriors within $w_0w_a$CDM are shown using the colored bands, with the dashed and dotted lines representing the best-fit models. The blue bands show results from DESI DR2 BAO and the CMB, while the orange bands show results from DESY5 SNe and the CMB.

The top panel of \cref{fig:dv_fap_evol} shows that DESI Galaxies at $z<1$ are the primary driver of the tension with the CMB \lcdm\ model and of the preference for $w_0w_a$CDM. This is because the uncertainties of the BAO measurements in this region are similar to the uncertainty in the $w_0w_a$CDM posterior (black points versus blue bands), and are significantly tighter than the uncertainty in the SNe$+$CMB posterior. On the other hand, at $z>1.5$ the $w_0w_a$CDM posteriors agree well with each other and with the CMB \lcdm\ model, as this region is constrained very tightly by the CMB acoustic scale. This means future improvements in the isotropic BAO measurements at high redshift will not be very helpful to dark energy constraints that include CMB information.

When it comes to the \alpac\ constraints, shown in the lower panel of \cref{fig:dv_fap_evol}, the conclusions are quite different. At $z>1$, both $w_0w_a$CDM posteriors prefer AP values that are $\sim2\%$ smaller than the CMB prediction in \lcdm. In this region, our measurement from DESI DR1 \lya\ full-shape (purple) gives the tightest constraint to date ($1.7\%$; shown in purple), and its combination with DR2 \lya\ BAO produces a $1.3\%$ constraint, which is shown in red in \cref{fig:dv_fap_evol}. The inferred constraints from SNe$+$CMB are similar in this region ($\sim0.9\%$), while the ones from BAO$+$CMB are significantly tighter ($\sim0.3\%$). This explains why adding \lya\ to the former helped improve the constraints, while adding it to the latter did not. Furthermore, the \lya\ constraints are currently consistent with both the CMB \lcdm\ model and the best-fit $w_0w_a$CDM models, meaning they do not strongly pull in either direction.

The fact that AP uncertainties are significantly larger than the BAO$+$CMB inferred constraints in $w_0w_a$CDM (blue bands) means they do not currently play as significant a role as low-redshift isotropic BAO constraints. However, these constraints are expected to improve with full-shape analyses of DESI DR2 on both the \lya\ and Galaxy clustering side. \cref{fig:dv_fap_evol} shows that future full-shape analyses with emission line galaxies in the range $1 < z< 1.5$, and with \lya\ at $z>2$, could help improve upon current constraints. The \lya\ forest is particularly powerful when it comes to AP constraints, as we have seen here. With DESI DR1, it led to a $2.4\times$ tighter constraint when compared to BAO. \cite{Cuceu:2021} forecasted \lya\ broadband AP uncertainties in the range $0.3\%-0.9\%$ for the full 5-year DESI survey, with the main contributing factor being the minimum scale used in the analysis. Given that the difference between \lcdm\ and $w_0w_a$CDM is $\sim2\%$ in this region, future DESI \lyaf\ analyses could play an important role in deciphering the nature of dark energy.

More generally, the DESI DR2 analysis suggests that the dark energy density $\rho_{\rm DE}(z)$ peaks at $z \sim 0.3-1$ and declines significantly towards higher redshift. The $w_0w_a$CDM model provides a compact parameterization that allows this behavior, but if the basic finding of evolving dark energy does hold up, there is no reason to expect that $w_0w_a$CDM provides a full description. The correct physical model might even involve exchanges of energy between the dark energy, matter, and radiation components, rather than dark energy evolution on its own. In this context, measurements of $D_H(z)$, which directly probe the energy density at $z>2$ are especially valuable, whereas $D_M(z)$ constrains an integral over $H(z)$ and has limited room to depart from the trend measured by galaxies at lower redshift and $\theta_*$ at $z\sim1100$ (see, e.g., Eqs. 3 and 6 of \citealt{DESI.DR2.BAO.cosmo}). \lyaf\ BAO is already quite good at measuring $D_H$ because the strong redshift-space distortions amplify the BAO signal in the line-of-sight direction. However, a full-shape \lya\ AP measurement in combination with BAO gives greater leverage on the energy density, improving $D_H/r_d$ constraints by a factor of $\sim1.5$ relative to BAO alone \citep{Cuceu:2021}. Even moderate gains in precision may play a critical role in distinguishing physical models of evolving dark energy.

Finally, while the discussion here focuses on a dynamic dark energy equation-of-state as the way to reconcile BAO and CMB results, several other potential explanations have recently appeared in the literature \citep[e.g.,][]{Chen:2025,Sailer:2025}. If alternative explanations lead to different predictions for the behaviour of the high-redshift AP parameter compared to $w_0w_a$CDM, the \lyaf\ would also help test these alternatives. A concrete example is non-zero curvature, as proposed by \cite{Chen:2025}, for which the best-fit BAO$+$CMB constraints predict AP values roughly consistent with \lcdm\ at $z\sim2$ (see their Figure 6). Therefore, future DESI \lyaf\ full-shape analyses could also shed light on whether curvature or a dynamic dark energy equation-of-state offer better explanations for the current tension.

\section{Summary} \label{sec:summary}

We have presented a full-shape analysis of \lyaf\ 3D correlation functions measured from the first data release (DR1) of DESI. We used both the \lya\ auto-correlation and its cross-correlation with quasars, which were measured by \cite{DESI2024.IV.KP6}. The first goal of this analysis was to measure the \alpac\ (AP) effect from a broad range of scales in order to improve upon the expansion rate constraints based on BAO measurements from the same data set \citep{DESI2024.IV.KP6}. The second goal was to use RSD in the \lya-QSO cross-correlation to measure the product of the growth rate and the amplitude of fluctuations in spheres of $8$\hMpc, \fsig.

Our analysis, introduced in \cref{sec:analysis}, largely follows the methods previously used for the \lya\ full-shape analysis of eBOSS \citep{Cuceu:2023a,Cuceu:2023b}, and for the first DESI \lya\ BAO measurement \citep{DESI2024.IV.KP6}. The DESI \lyaf\ analysis leading to the measurement of correlation functions was presented in \cite{Ramirez-Perez:2024,Gordon:2023,DESI2024.IV.KP6}. The model used to fit the \lya\ correlations and extract full-shape information is described in \cref{subsec:model,subsec:scale_pars}. Our analysis was first done blinded as described in \cref{subsec:blinding}.

Our main results are presented in \cref{sec:comp_results}, and include a $1.6\%$ constraint on the \alpac\ effect and a $\sim16\%$ constraint on \fsig, at an effective redshift $z_\mathrm{eff}=2.33$. The AP measurement represents a factor of $2.4\times$ improvement compared to the BAO analysis of the same dataset, and is almost $40\%$ better than the BAO constraint from the second DESI data release \citep[DR2;][]{DESI.DR2.BAO.lya}. The \fsig\ measurement is the first of its kind from the \lya-QSO cross-correlation, and to our knowledge, the first direct growth rate constraint at $z>2$.

In \cref{sec:validation}, we performed a large set of tests to validate our measurements, including tests with two different sets of mocks, data splits, and analysis and modelling variations. The tests on mocks, presented in \cref{subsec:mocks}, show that we can recover unbiased AP constraints in the presence of all the major \lyaf\ contaminants. Furthermore, we found no evidence of significant systematic shifts in AP with the tests on data, as detailed in \cref{subsec:datasplits,subsec:variations,subsec:model_variations}.

On the other hand, for \fsig, one of the two sets of mocks shows a $\sim0.5\sigma$ bias, which we were not able to track down to any particular systematic effect. Therefore, we decided not to use the \fsig\ constraint for cosmological inference as discussed in \cref{sec:discussion}. Our final growth rate measurement, including a systematic uncertainty based on the shift observed in mocks, is: $f\sigma_8(z_\mathrm{eff}) = 0.37\; ^{+0.055}_{-0.065} \,(\mathrm{stat})\, \pm 0.033 \,(\mathrm{sys})$. 

We present the cosmological interpretation of our AP measurement alone, and in combination with other data sets, in \cref{sec:cosmo}. The \lya\ full-shape analysis from DESI DR1 measures the ratios $D_H(z_\mathrm{eff})/r_d=8.632\pm0.105$ and $D_M(z_\mathrm{eff})/r_d=39.05\pm0.52$, where $D_H$ is the Hubble distance, $D_M$ is the transverse comoving distance, and $r_d$ is the sound horizon at the drag epoch. We also combined our DR1 broadband AP constraint with the \lya\ BAO measurement from DESI DR2 \citep{DESI.DR2.BAO.lya}, and obtained the ratios $D_H(z_\mathrm{eff})/r_d=8.646\pm0.077$ and $D_M(z_\mathrm{eff})/r_d=38.90\pm0.38$.

Within \lcdm\ our measurements are consistent with both CMB and galaxy clustering BAO constraints, as discussed in \cref{subsec:lcdm}. Using a BBN prior on the baryon density, we measure the Hubble constant to be $H_0 = 68.3\pm 1.6\;\kmsMpc$ in \lcdm. When allowing for evolving dark energy equation-of-state parameters (\cref{subsec:dark_energy}), we find that \lya\ helps improve constraints from the combination of Supernovae and CMB, and slightly increases the tension with \lcdm. On the other hand, for combinations that include CMB and Galaxy BAO results, adding \lya\ leads to a small shift towards \lcdm, but does not improve constraints.

Finally, in \cref{subsec:future} we discuss our cosmology results and the role future DESI \lyaf\ analyses are expected to play when it comes to studying the nature of dark energy. We show that further incremental improvements in isotropic BAO constraints at $z>1.5$ will not help because this quantity is constrained very precisely by the CMB within $w_0w_a$CDM. On the other hand, best-fit models using \lcdm\ or $w_0w_a$CDM currently give predictions of the \alpac\ effect that are $\sim2\%$ different at $z>1$. \lya\ full-shape analyses using future DESI data releases are expected to produce sub-percent constraints of the AP effect \citep{Cuceu:2021}, and could therefore play a key role in deciphering the nature of dark energy.

\section{Data Availability}

The data used in this analysis is public as part of DESI Data Release 1 (details in \url{https://data.desi.lbl.gov/doc/releases/}). The data points corresponding to the figures from this paper will be available in a Zenodo repository.

\section{Acknowledgements}

AC acknowledges support provided by NASA through the NASA Hubble Fellowship grant HST-HF2-51526.001-A awarded by the Space Telescope Science Institute, which is operated by the Association of Universities for Research in Astronomy, Incorporated, under NASA contract NAS5-26555. HKHA acknowledges support from the French National Research Agency (ANR) under grant ANR-22-CE31-0009 (HZ-3D-MAP project) and grant ANR-22-CE92-0037 (DESI-Lya project). CG, CRP, and AFR acknowledge support from the European Union’s Horizon Europe research and innovation programme (COSMO-LYA, grant agreement 101044612), as well as support by the Spanish Ministry of Science, Innovation and Universities (MICIU) under grants SEV-2016-0588, PGC-2018-094773-B-C31, RYC-2018-025210, PGC2021-123012NB-C41, PID2024-159420NB-C41 and CEX2024001442-S, some of which include ERDF funds from the European Union. IFAE is partially funded by the CERCA program of the Generalitat de Catalunya.

This material is based upon work supported by the U.S. Department of Energy (DOE), Office of Science, Office of High-Energy Physics, under Contract No. DE–AC02–05CH11231, and by the National Energy Research Scientific Computing Center, a DOE Office of Science User Facility under the same contract. Additional support for DESI was provided by the U.S. National Science Foundation (NSF), Division of Astronomical Sciences under Contract No. AST-0950945 to the NSF’s National Optical-Infrared Astronomy Research Laboratory; the Science and Technology Facilities Council of the United Kingdom; the Gordon and Betty Moore Foundation; the Heising-Simons Foundation; the French Alternative Energies and Atomic Energy Commission (CEA); the National Council of Humanities, Science and Technology of Mexico (CONAHCYT); the Ministry of Science, Innovation and Universities of Spain (MICIU/AEI/10.13039/501100011033), and by the DESI Member Institutions: \url{https://www.desi.lbl.gov/collaborating-institutions}. Any opinions, findings, and conclusions or recommendations expressed in this material are those of the author(s) and do not necessarily reflect the views of the U. S. National Science Foundation, the U. S. Department of Energy, or any of the listed funding agencies.

The authors are honored to be permitted to conduct scientific research on I'oligam Du'ag (Kitt Peak), a mountain with particular significance to the Tohono O’odham Nation.


%






\bibliography{main,DESI_supporting_papers}{}
\bibliographystyle{aasjournal}

\appendix

\section{Large-scale fits} \label{sec:large_scales}

\begin{figure}
    \centering
    \includegraphics[width=1.0\textwidth,keepaspectratio]{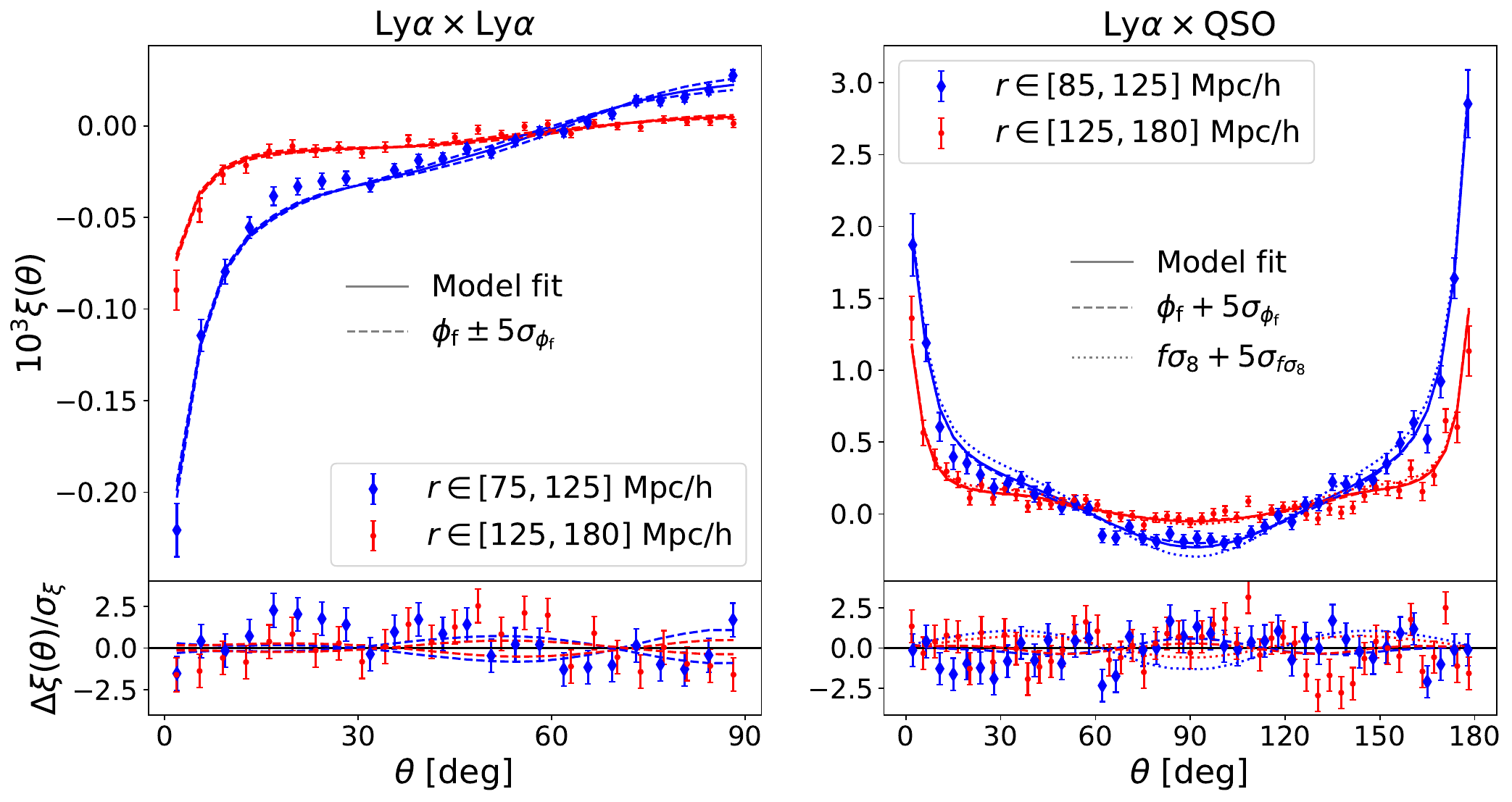}
    \caption{Similar to figure \ref{fig:auto_shells}, but showing the shell compression for larger scales. The impact of changing the AP and RSD parameters by $5\sigma$ is lower than our statistical uncertainty for these shells, indicating that large scales do not contribute very significantly to our constraints.}
    \label{fig:cross_shells}
\end{figure}

In \cref{sec:comp_results}, we showed a comparison between the best-fit model and the data correlation functions when compressing into shells at smaller separations. Here we show the equivalent plots for the larger separations. These are illustrated in \cref{fig:cross_shells}, where the blue shells roughly capture the information from the BAO region (left for the auto-correlation, and right for the cross-correlation), while the red shells show the broadband information at scales larger than BAO. Similar to \cref{fig:auto_shells}, the dashed and dotted lines indicate models where we vary AP and \fsig, respectively. In this case, changing \phif\ and \fsig\ by $5\sigma$ produces a change in the model that is smaller than our uncertainties. This indicates that large scales do not contribute very significantly to our constraints.

\section{Nuisance parameters} \label{sec:nuisance}

\begin{table}
\centering
\begin{tabular}{c|c}
Parameter                                 & Priors                         \\
\hline
$\left\{\alpha, \phi \right\}$  & $\mathcal{U}[0.01, 2.00]$    \\
$f$                            & $\mathcal{U}[0.00, 2.00]$    \\
$b_{F}$                            & $\mathcal{U}[-2.00, 0.00]$    \\
$\beta_{F}$                        & $\mathcal{U}[0.00, 5.00]$     \\
$b_{Q}$                                 & $\mathcal{U}[0.00, 10.00]$    \\
$b_{HCD}$                                 & $\mathcal{U}[-0.20, 0.00]$  \\
$\beta_{HCD}$                             & $\mathcal{N}(0.500, 0.090)$   \\
$L_{\rm HCD} (h^{-1} {\rm Mpc})$          & $\mathcal{U}[0.00, 15.00]$     \\
$10^3 b_{\rm SiII(1190)}$                 & $\mathcal{U}[-500.00, 0.00]$     \\
$10^3 b_{\rm SiII(1193)}$                 & $\mathcal{U}[-500.00, 0.00]$     \\
$10^3 b_{\rm SiII(1260)}$                 & $\mathcal{U}[-500.00, 0.00]$    \\
$10^3 b_{\rm SiIII(1207)}$                & $\mathcal{U}[-500.00, 0.00]$  \\
$10^3 b_{\rm CIV(eff)}$                   & $\mathcal{N}(-24.3, 1.5)$     \\
$\Delta r_{\parallel} (h^{-1} {\rm Mpc})$ & $\mathcal{N}(0.0, 1.0)$    \\
$\sigma_v (h^{-1} {\rm Mpc})$             & $\mathcal{U}[0.00, 15.00]$   \\
$\xi_0^{TP}$                              & $\mathcal{U}[0.00, 2.00]$   \\
$10^4 a_{\rm noise}$                      & $\mathcal{U}[0.00, 100.00]$   \\
$q_1$                                     & $\mathcal{U}[0.00, 2.00]$   \\
$k_v$                                     & $\mathcal{U}[0.30, 2.00]$   \\
$a_v$                                     & $\mathcal{U}[0.10, 1.00]$   \\
$b_v$                                     & $\mathcal{U}[1.00, 2.00]$   \\
$k_p$                                     & $\mathcal{U}[5.00, 22.00]$   \\

\end{tabular}
\caption{
Parameter priors used in our analysis. $\mathcal{U}[\mathrm{min}, \mathrm{max}]$ indicates a flat prior while $\mathcal{N}(\mu, \sigma)$ indicates a Gaussian with mean $\mu$ and standard deviation $\sigma$.
}
\label{tab:priors}
\end{table}

Our analysis has 21 nuisance parameters, introduced in \cref{subsec:model}. These include the \lya\ bias and RSD parameters, 5 parameters related to the Arinyo model, 13 parameters related to contaminants and systematics, and the isotropic scale of the broadband, \alphas. Our priors are presented in \cref{tab:priors}, and largely follow the priors used by \kplya.\footnote{As mentioned in \cref{subsec:scale_pars}, we sample the growth rate, $f$, and treat \fsig\ as a derived parameter.} The important exceptions are the systematic quasar redshift error, $\Delta r_{||}$, which had a flat prior in \kplya, and the Arinyo parameters, $\{q_1,k_v,a_v,b_v,k_p\}$, which were fixed in \kplya. As we use a significantly larger minimum separation for the cross-correlation compared to \kplya\ ($40$\hMpc\ versus $10$\hMpc), we found that $\Delta r_{||}$ was poorly constrained and decided to impose a wide Gaussian prior to discourage extreme values. $\Delta r_{||}$ has been measured by \cite{KP6s4-Bault} for the DESI DR1 quasar sample and found to be significantly smaller than $1$\hMpc\ \citep[also see][]{RedrockQSO.Brodzeller.2023}. The prior has no impact on our results as shown in \cref{subsec:model_variations}. This was also recently done in \cite{DESI.DR2.BAO.lya}, and we use the same prior. In the case of the Arinyo parameters, we chose wide flat priors based on the fits to hydrodynamical simulations from \cite{Arinyo:2015} and \cite{Chabanier:2024}. Only one of these parameters, $q_1$, is constrained by the data. The other four parameters, $\{k_v,a_v,b_v,k_p\}$, produce posteriors consistent with the flat priors we imposed and show no significant correlation with our main parameters (\phif\ and \fsig).

\begin{figure}
    \centering
    \includegraphics[width=1.0\textwidth,keepaspectratio]{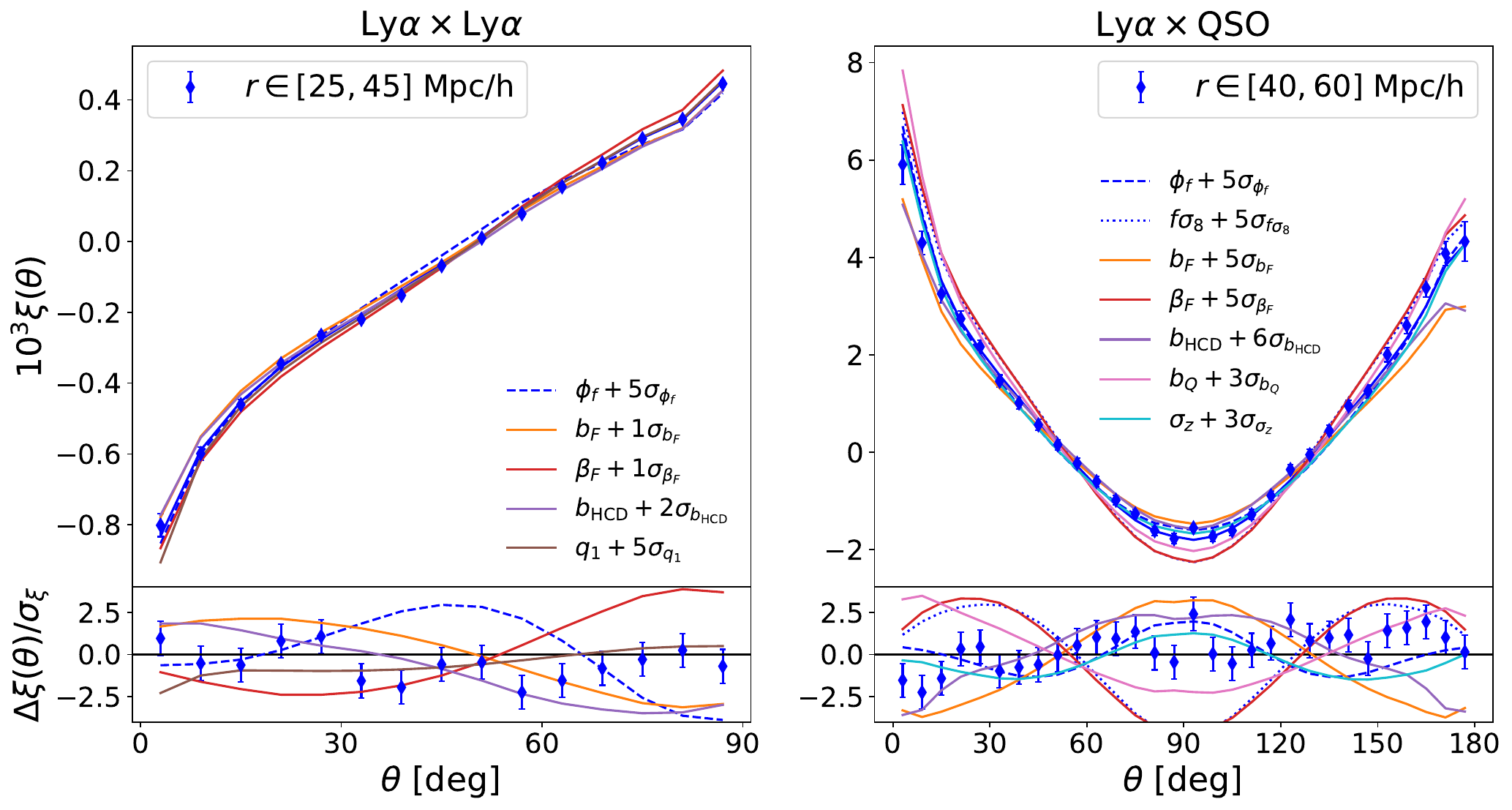}
    \caption{Similar to \cref{fig:auto_shells}, but showing only one shell at the smallest separations used in our analysis, and including the effect of changing several of the important nuisance parameters. For the auto-correlation (left), none of the nuisance parameters produce changes similar to those produced by changing AP. On the other hand, for the cross-correlation (right), the impact of \fsig\ is very similar to that of $\beta_{F}$, as the \lya\ and QSO RSD terms are degenerate on large scales. This is why the auto-correlation is needed for our \fsig\ constraints, as it helps constrain $\beta_{F}$.  While AP looks quite different from RSD in this compression, its impact has some similarity to that of redshift errors (dashed blue versus cyan).}
    \label{fig:nuisance_shells}
\end{figure}

In \cref{fig:nuisance_shells}, we show the impact of changing several of the important nuisance parameters on the lowest separation shell for both the auto-correlation (left) and the cross-correlation (right). The magnitude of the variation in each parameter was chosen such that the change in the model can be visualized in the lower panels of \cref{fig:nuisance_shells}. For the auto-correlation, we show the impact of changing the \lya\ bias and RSD parameters, along with the HCD bias and the main parameter in the Arinyo model ($q_1$). The other HCD and Arinyo parameters are not included here, as they do not produce significant changes. None of the parameters shown produces changes similar to those produced by AP. Both biases and the RSD parameter produce changes in the slope of the anisotropy, while AP affects the curvature of the anisotropy. On the other hand, for the cross-correlation, the \lya\ and QSO RSD terms are degenerate (see \cref{subsec:model}), which means the impact of the \lya\ RSD parameter is very similar to the impact of \fsig. However, $\beta_F$ is tightly constrained by the auto-correlation, thus allowing us to break this degeneracy. In the case of AP, the impact of the smearing due to redshift errors, modelled through the $\sigma_z$ parameter, shows some similarity to the impact of \phif. However, as we will see in \cref{fig:nuisance} below, there is no significant correlation between the posteriors of the two parameters. This is partly because a majority of the \phif\ information comes from the auto-correlation. Nevertheless, the complete lack of correlation between the two parameters indicates that we are able to distinguish the two effects.

\begin{figure}
\centering
\includegraphics[width=1.\columnwidth,keepaspectratio]{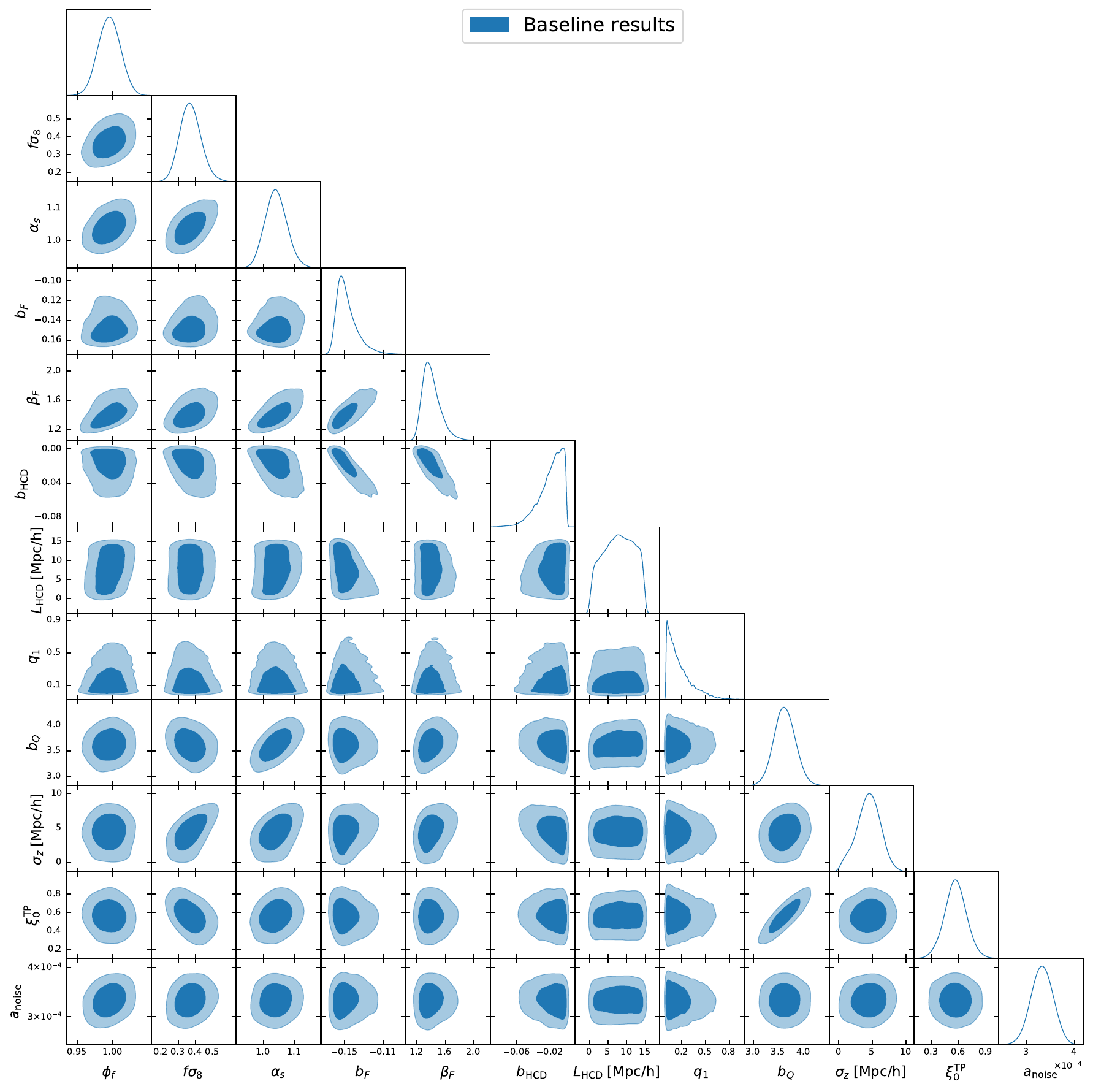}
\caption{Posterior distribution for \phif\ and \fsig, along with a selection of nuisance parameters that show some degree of correlation with either of the two main parameters.
}
\label{fig:nuisance}
\end{figure}

We present the marginalized posterior distribution for the main parameters and a selection of nuisance parameters in \cref{fig:nuisance}. We selected the subset of nuisance parameters that show any visible correlation with either of the two main parameters. The $L_\mathrm{HCD}$ parameter, which represents the typical length scale of unmasked HCDs, is completely unconstrained as we just recover the flat prior. This is not surprising given the scale cuts we use, with both \cite{Cuceu:2023b} and \cite{DESI.DR2.BAO.lya} finding similar behaviour. \cite{DESI.DR2.BAO.lya} decided to impose a Gaussian prior on this parameter, as they found this had no impact on their BAO constraints. However, in our case, we found that imposing constraints on this parameter has a small impact on \phif\ due to the small correlation between the two parameters (see \cref{fig:nuisance}), and decided to keep using the flat prior.

Besides the unconstrained parameters described above, we also have two parameters that only have upper or lower bounds. These are the HCD bias, $b_\mathrm{HCD}$, which only has a lower bound, and $q_1$, which only has an upper bound. In the case of $b_\mathrm{HCD}$, this behaviour was encountered before \citep{Cuceu:2023b,DESI.DR2.BAO.lya} when using scale cuts $r>25$\hMpc, as we do here. For the Arinyo $q_1$ parameter, this is the first ever measurement on real data, with all previous measurements taking place on hydro-simulations. The constraint we obtain, $q_1 < 0.51$ at $95\%$ CR, is smaller than the value used in previous \lya\ BAO analyses, $q_1=0.86$, which was based on the result from the reference simulation of \cite{Arinyo:2015}, interpolated to our effective redshift. However, both \cite{Arinyo:2015} and more recent analyses \citep[e.g.][]{Chabanier:2024,ChavesMontero:2025} have found a wide range of values for this parameter depending on the exact simulation and model setup, some of which are consistent with our results. Nevertheless, we caution against over-interpreting our results in the context of future analyses on hydro-simulations due to two main reasons. First, given that we could not test the Arinyo model in mocks, we do not exclude the possibility that this parameter is also capturing some subdominant systematic effects (e.g., from imperfect modelling of HCDs, metals, or the distortion matrix). Secondly, this parameter was treated as a nuisance parameter in this analysis, and the robustness of its measurement to analysis and modelling variations has not been tested. It would be interesting to perform an analysis that specifically aims to measure the Arinyo parameters for the purpose of comparing with results from hydro-simulations, but that is beyond the scope of this work.

\section{Comparison of sampler and fitter results} \label{sec:fitter}

The main results presented in this paper are given by marginalized posterior distributions over the parameters of interest (\phif, \fsig), obtained from the full posterior distribution over all parameters, which is sampled using the \texttt{PolyChord} Nested Sampler. However, due to computational constraints, a significant part of our validation results were obtained using the \iminuit\ minimizer with Gaussian approximated uncertainties. These include the fits to individual mocks, and most of the analysis and modelling variations presented in \cref{sec:validation}. Here, we discuss the differences between the two types of analyses and their impact on our validation.

The first difference relevant to our analysis is due to parameter uncertainties being approximated as Gaussian when using the fitter. The main results affected by this approximation are the tests of the robustness of uncertainties using mocks (\cref{subsec:mocks}), which include the pull distributions presented in \cref{fig:pull}. This is because the uncertainties in the individual mock fits are approximated as Gaussian, while the distribution of parameter results in mocks is not Gaussian. Nevertheless, we found that the Gaussian approximation works well and produces accurate uncertainties, especially for \phif. Finally, this approximation does not impact our analysis and modelling variations, because in that case, we are only testing for shifts in the best-fit parameter values.

The other main difference relevant to our analysis arises from the fact that when using the sampler, we marginalize over nuisance parameters, whereas the fitter results for \phif\ and \fsig\ are conditioned on the best-fit values of the nuisance parameters. This is relevant because a number of nuisance parameters are either completely or partly unconstrained by the data (see Appendix \ref{sec:nuisance}). We illustrate this in \cref{fig:sampler_fitter}, where we compare sampler and fitter constraints using our baseline model. We also show the best-fit parameter values from the fitter using the dashed red lines, and the maximum a posteriori results from the sampler using the dashed blue lines. We found that the parameter primarily responsible for this difference is $L_\mathrm{HCD}$.\footnote{This conclusion was reached after performing tests where unconstrained parameters are fixed one-by-one. We found that fixing $L_\mathrm{HCD}$ brings the sampler and fitter results in very good agreement.} This is because this parameter is completely unconstrained and has a small correlation with \phif\ ($\rho\sim0.2$). Sampler results marginalize over all values of this parameter within the flat prior boundary $0 < L_\mathrm{HCD}<15$\hMpc, while the fitter tries to find a best-fit value, which in most cases ends up being either the upper or lower bound of the flat prior (in the baseline analysis it is the upper bound). Based on tests with mocks, we expect this parameter to have values between $3$\hMpc\ and $7$\hMpc, with the larger values obtained when no DLAs are masked (see Appendix B of \kplya). The flat prior bounds were chosen to be larger than this range, in order to be conservative. In \cref{subsec:model_variations} we found that fixing this parameter to either $3$\hMpc\ or $10$\hMpc\ does not have a significant impact on \phif\ constraints, producing shifts of at most $0.1\sigma$. Nevertheless, because the fitter tends to choose either the upper or lower bound of the flat prior on $L_\mathrm{HCD}$, our individual mock results and analysis variations likely have some extra variance. Therefore, future \lya\ full-shape analyses could choose a more restrictive prior for this parameter, given that it has a small impact on results. For example, a Gaussian prior could be imposed based on the range of results found in mocks, as was recently done by \cite{DESI.DR2.BAO.lya}.

\begin{figure}
\centering
\includegraphics[width=0.5\columnwidth,keepaspectratio]{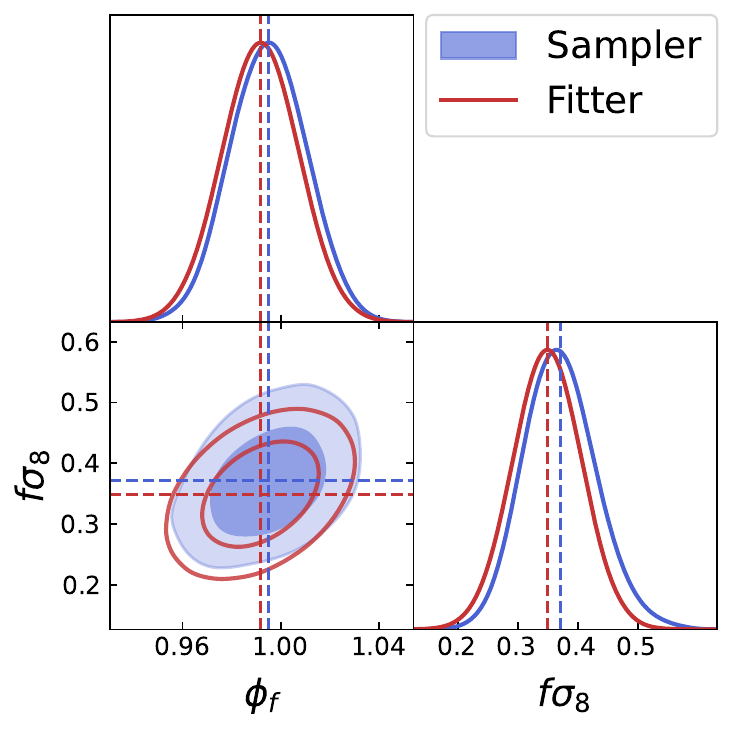}
\caption{Comparison of AP and growth results using a sampler and marginalization versus a fitter and minimization. Both results use the baseline model described in \cref{subsec:model}. The dashed lines indicate the best-fit values obtained by the fitter (red) and the maximum a posteriori values from the sampler (blue). The fitter minimizes the $\chi^2$ over all parameters, and estimates the Gaussian uncertainty using the second derivative of the $\chi^2$ around the best fit. On the other hand, the sampler computes the full posterior distribution, which allows us to marginalize over nuisance parameters. The primary difference arises because the fitter results are conditioned on the best-fit value of each nuisance parameter, whereas the sampler results are marginalized over nuisance parameters.
}
\label{fig:sampler_fitter}
\end{figure}

\section{Covariance between DESI DR1 broadband AP and DR2 BAO} \label{sec:independence}

To quantify the covariance between our DESI DR1 broadband AP measurements and BAO results from DESI DR2, we use the method described in Appendix F of \kplya. First, we compute the joint covariance matrix of DESI DR1 and DR2 using only the \lya(A) region. This is done using the same method for computing the covariance matrix of the four DR1 correlations as described in \cref{subsec:corr}, but replacing the DR1 \lya(B) region correlations with the DR2 \lya(A) region correlations \citep{DESI.DR2.BAO.lya}. This covariance matrix now includes the cross-covariance between the DR1 \lya(A) correlations and the DR2 \lya(A) correlations. Similar to \kplya, we ignore \lya(B) correlations, as they contribute a very small amount of information (see \cref{sec:comp_results}).

We next produce a set of 4096 Monte Carlo realisations of the \lya(A) auto and QSO cross-correlations for both DESI DR1 and DR2 using the best-fit model from \cref{sec:comp_results}, with noise generated from the covariance matrix described above. This means the auto and cross-correlation pairs generated for DR1 and DR2 have realistic noise and cross-covariance. After that, we perform full-shape fits on the set of DR1 correlations using the model described in \cref{subsec:model} with independent \phis\ and \phip\ parameters. In parallel, we also perform BAO fits on the set of DR2 correlations using the model described in \cite{DESI.DR2.BAO.lya}. This results in a set of 4096 broadband AP constraints based on DESI DR1, and a set of 4096 BAO constraints based on DESI DR2.

Using these two sets of best fit parameters, we can compute the cross-covariance between DR1 AP and DR2 BAO. The correlation coefficients we obtain between DR1 \phis\ and the DR2 BAO parameters, \apar\ and \atrans, are:
\begin{align}
    \rho(\phi_\mathrm{s}^\mathrm{DR1}, \alpha_{||}^\mathrm{DR2}) &= -0.015 \pm 0.016, \\
    \rho(\phi_\mathrm{s}^\mathrm{DR1}, \alpha_\bot^\mathrm{DR2}) &= 0.001 \pm 0.016,
\end{align}
where the uncertainties are computed through bootstrap. Therefore, we conclude that the cross-covariance between DESI DR1 broadband AP and DR2 BAO is consistent with zero, and the two measurements can be treated as independent.




\end{document}